\documentclass[11pt,twoside]{article}
\newcommand{\etal}{\textit{et al}.}

\newcommand{\eg}{\textit{e}.\textit{g}.}
\usepackage{graphicx}
\usepackage{xcolor}
\usepackage{soul}
\graphicspath{{./figs/}}
\usepackage{ap-article}
\def\labart{yourLabel}      
\shortauthors{J. Chen et al.}
\shorttitle{Mutation Operators for Simulink Models}
\title{%
Timed Model-Based Mutation Operators for Simulink Models
}
\author{%
Jian Chen{*}\,\up{1}
Manar H. Alalfi\,\up{2}
Thomas R. Dean\, \up{3}
}
\addresses{%
\up{1}
Department of Electrical and Computer Engineering, Queen's University\\
Kingston, ON, Canada
\\ \vspace*{0.04truein}
E-mail: jian.chen@queensu.ca
\\ \vspace*{0.05truein}
\up{2}
Department of Computer Science, Ryerson University\\
Toronto, ON, Canada \\ 
\vspace*{0.04truein}
E-mail: manar.alalfi@cs.ryerson.ca
\\ \vspace*{0.05truein}
\up{3}
Department of Electrical and Computer Engineering, Queen's University\\
Kingston, ON, Canada
\\ \vspace*{0.04truein}
E-mail: tom.dean@queensu.ca
\\ \vspace*{0.05truein}
}
\begin{document}
\label{\labart-FirstPage}

\maketitle
\abstracts{%
    Model-based mutation analysis is a recent research area, and real-time system testing can benefit from using model mutants. Model-based mutation testing (MBMT) is a particular branch of model-based testing. It generates faulty versions of a model using mutation operators to evaluate and improve test cases. Mutation testing is an effective way to ensure software correctness and has been applied to various application areas. Simulink is a vital modeling language for real-time systems. This paper introduces Simulink model mutation analysis to improve Model-in-the-loop (MIL) testing. We propose a set of Simulink mutation operators based on  AUTOSAR, which reflects the temporal correctness when a Simulink model is mapped to Operating System tasks. We implement a mutation framework that generates mutants for implicit clock Simulink models. Finally, we demonstrate how this framework generates mutants to reveal task interference issues in the simulation. Our work integrates the Simulink model with the timed systems to better support mutation testing automation.
}

\medskip

\vspace*{7pt}\textlineskip
\begin{multicols}{2}

\section{Introduction}

    Today, cars come equipped with advanced technologies that did not exist before, such as Automatic Emergency Braking (AEB), Adaptive Cruise Control (ACC), Lane Departure Warning/Lane Keeping, and Autonomous driving. All of these features rely on software to realize sophisticated control algorithms. Generally, such software is developed within the timed system context, in which the system correctness not only relies on the software implemented functions correctness but also depends on the system to meet time constraints. Many factors can contribute to the execution time of a system running on a target platform. Issues such as task interference may cause delays during task execution. Software quality plays a crucial role in such safety-critical applications. 
	
    Model-Based Testing (MBT) is a promising technique for the automated testing of timed systems. A model represents the behavior of software, and the model is usually abstracted from real-time specifications. However, some modeling environments support this feature in the Hardware-in-the-loop (HIL) simulation testing instead of the MIL. For example, Matlab/Simulink (ML/SL) simulations assume block behaviors are completed in nearly zero execution time, while real execution requires a finite execution time, which may cause a failure. ML/SL models are based on the Synchronous Reactive (SR) model \cite{LeeEtAl:14:SR} that may assume the task execution times are zero. Errors in the model may not be apparent without an explicit real-time execution in the MIL phase. Usually, a Simulink model can be well simulated in the MIL, but it may have errors in the real-time context.
	
    Hence, MBT needs an extension to accommodate the real-time context, which includes modeling the system through a timed formalism, and checking the implementation conforms to its specification. Traditionally, this is done via conformance checks \cite{Schmaltz2008}. Recently, several tools have been proposed to simulate the real-time execution effects for ML/SL models in MIL, such as TrueTime \cite{Henriksson2003}, TRES \cite{Cremona2015}, Timing-aware blocks \cite{Naderlinger2017}, and SimSched \cite{chen2018modeling}. SimSched uses a model transformation to integrate scheduling into the model to validate the real-time context during simulation. To evaluate SimSched, we turn to mutation testing using mutation analysis to assist the evaluation of the SimSched tool.
	
    In this paper, we propose a set of mutation operators with a timed task model, which is based on the AUTomotive Open System ARchitecture (AUTOSAR), that reflects the temporal correctness when a Simulink model is mapped to Real-Time Operating System (RTOS) tasks in a real-time context.
    
    This paper is organized as follows: Section 2 introduces background information. Section 3 presents the set of proposed timed mutation operators for Simulink models. Section 4 explains the usage of the timed mutation operators. Section 5 presents validation experiments and results. Section 6 summarizes related studies in MBT. Finally, Section 7 presents the conclusions of our work and outlines future work.

    \section{Background}
    This section gives an overview of the background information on the material needed to explain our work. We begin with a basic introduction to mutation testing, Simulink, and AUTOSAR; then, we present our timed task model.
	
    \subsection{Mutation testing}
    Mutation testing was introduced in the 1970s \cite{Hamlet1977,DeMillo1978,Jia2011} and proved to be an effective way to reveal software faults \cite{Papadakis2019}. It is a fault-based software testing technique, which has been extensively studied and used for decades. It contributes a range of methods, tools, and reliable results for software testing. Mutation testing is designed to find valid test cases and discover real errors in the program. 
	
	Model-Based Mutation Testing (MBMT) takes the advantages of both model-based testing and mutation testing and has been widely applied to multiple types of models such as feature models \cite{Henard2013}, statechart-based models \cite{Trakhtenbrot2010,Aichernig2015a},  timed automata \cite{Aichernig2013,Aichernig2015b}, and Simulink \cite{Brillout2010,Matinnejad2016,Stephan2014a}. However, in real-time system development, both logical and temporal correctness is crucial to the correct system functionality. The temporal correctness depends on timing assumptions for each task. Timed Automata (TA) \cite{Alur1994} is a common formalism to model and verify real-time systems to see whether designs meet temporal requirements. Aichernig \etal \cite{Aichernig2013} propose an MBMT technique for timed automata that applies to \emph{input/output} timed automata (TAIO) model. Nilsson \emph{et al.} \cite{Nilsson2004} add an extension to the TA formalism with a task model, and their mutation operators focus on timeliness. Simulink\footnote{https://www.mathworks.com/products/simulink.html} is widely used for model-driven development of software within the automotive sector. Most of the mutation operators proposed for Simulink models are from a property point of view either run-time or design-time such as signal modification, arithmetic alternation, or block change \cite{Hanh2012,Stephan2014a,Stephan2014b,Zhan2005}. Some of the proposed mutation testings are targeted at test case generation for Simulink models \cite{Brillout2010,5981757}. However, there is no mutation operator with an explicit clock model for Simulink. 
	
	\subsection{Simulink}
	Simulink is one of the most popular modeling languages for modeling dynamical systems, and MATLAB provides a graphical programming environment to perform system simulations. Simulink models are graphical blocks and lines, and they are connected by signals between input and output ports. The Simulink simulation engine determines the execution order of blocks based on the data dependencies among the blocks before a simulation execution. Simulink defines two types of blocks, direct feedthrough, and non-direct feedthrough, to assure the correct data dependencies in the simulation. Simulink uses the following two basic rules \cite{Simulinkguide} to determine the sorted execution order: A block must be executed before any of the blocks whose direct-feedthrough ports it drives; Blocks without direct feedthrough inputs can execute in arbitrary order as long as they precede any block whose direct-feedthrough inputs they drive. All blocks are scheduled in sorted order and executed in sequential execution order. The Simulink engine maintains a virtual clock to execute each ordered block at each virtual time. 
	
	Simulink Coder\footnote{https://www.mathworks.com/products/simulink-coder.html} supports code generation and offers a framework to execute the generated code in a real-time environment. Simulink Coder can generate code for the periodic task, either using a single task or a multi-task. Single-task implementations can preserve the semantics during the simulation because the generated code is invoked by a simple scheduler in a single thread without preemptions. For multi-task implementations, the generated code is invoked by a rate monotonic (RM) \cite{Lehoczky1989} scheduler in a multithreaded RTOS environment, where each task is assigned a priority and preemptions occur between tasks. As a consequence of preemption and scheduling, the implementation semantic can conflict with the model semantic in a multi-rate system. 
	
	\subsection{AUTOSAR}
	AUTOSAR is an open industry standard to meet the needs of future car development. AUTOSAR defines three main layers: the application, the runtime environment (RTE), and the basic software (BSW) layer \cite{Autosar}. The functions in the application layer are implemented by SW-Cs, which encapsulate part or all of the automotive electronic functions, as shown in Figure \ref{fig:autosar}. The components communicate via a Virtual Functional Bus (VFB), which is an abstraction of all the communication mechanisms of AUTOSAR. Engineers abstract the communication details of software components employing VFBs. A set of runnables represents the SW-Cs internal behaviors, and a runnable is the smallest executable code that can be individually scheduled, either by a timer or an event. Lastly, runnables are required to map to a set of tasks for a target platform, and the mapping has to preserve ordering relations and causal dependencies. Simulink has supported AUTOSAR compliant code generation since version R2006a\footnote{https://www.mathworks.com/products/simulink.html}.  All AUTOSAR concepts can be represented by Simulink blocks and the existing Simulink blocks can be easily used in the AUTOSAR development process. Some of AUTOSAR concepts and Simulink concepts mapping relation is shown in Table \ref{tbl:mapping} \cite{Applysingsimulink}. 
	
	\begin{Figure}
		\includegraphics[width=.99\columnwidth]{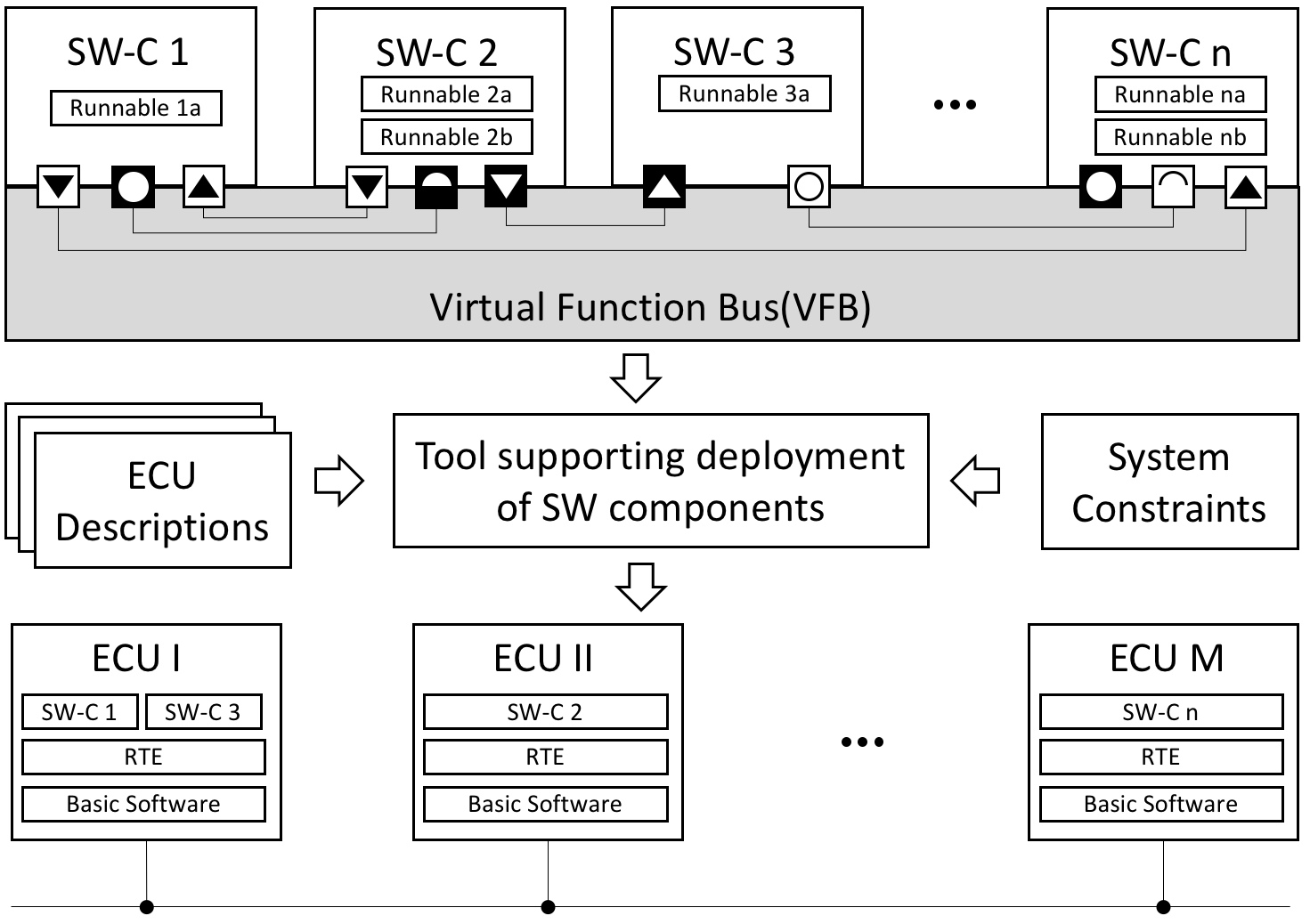}
		\fcaption{AUTOSAR components, interfaces, and runnables. (Adapted from \cite{Autosarparternship})}
		\label{fig:autosar}       
	\end{Figure}
	
	\begin{Table} 
		\tcap{Examples of ML/SL and AUTOSAR Concepts Mapping.}
		\begin{tabular}{ l l}
			\hline
			ML/SL & AUTOSAR \\ \hline
			Subsystem   &   Atomic Software\\
			 & Component \\
			Function-call subsystem & Runnable \\
			Function calls &  RTEEvents \\
			\hline
		\end{tabular}
		\label{tbl:mapping}
	\end{Table}
	
	\subsection{Task model}
	In automotive software, Simulink models are often drawn from real-time specifications and are realized as a set of tasks running on an RTOS. In order to better test this kind of software in the MIL phase,  model-based testing needs to be scaled to the real-time context, which includes a timed formalism to model the system under test conforming with the real-time requirements. We define a task model to model the timing properties of tasks in the Simulink environment and the application is modeled as a set of periodic tasks.
	
	A task model, $T$, is represented by a tuple $\{\phi, \rho, c, \gamma, prect, precr, prio, jitter\}$, where $\phi$ is an offset of the task, $\rho$ is the period of the task, $c$ is the  Worst Case Execution Time (WCET) of the task,  $\gamma$ is a list of runnables that belong to the task, $prect$ is the precedence constraint of the task, $precr$ is the precedence constraint of the runnables within the task, $prio$ is the priority associated with the task, and $jitter$ is the deviation of a task from the periodic release times. Every task has an implicit deadline which means the deadline of a task is equal to $\rho$. An offset $\phi$ refers to the time delay between the arrival of the first instance of a periodic task and its release time. A WCET is the summation of each runnable execution time. A precedence constraint $prect$ is a list of tasks that specifies the task execution order, and $precr$ is a list of runnables within a task.
	
	\begin{Figure}
		\includegraphics[width=.99\columnwidth]{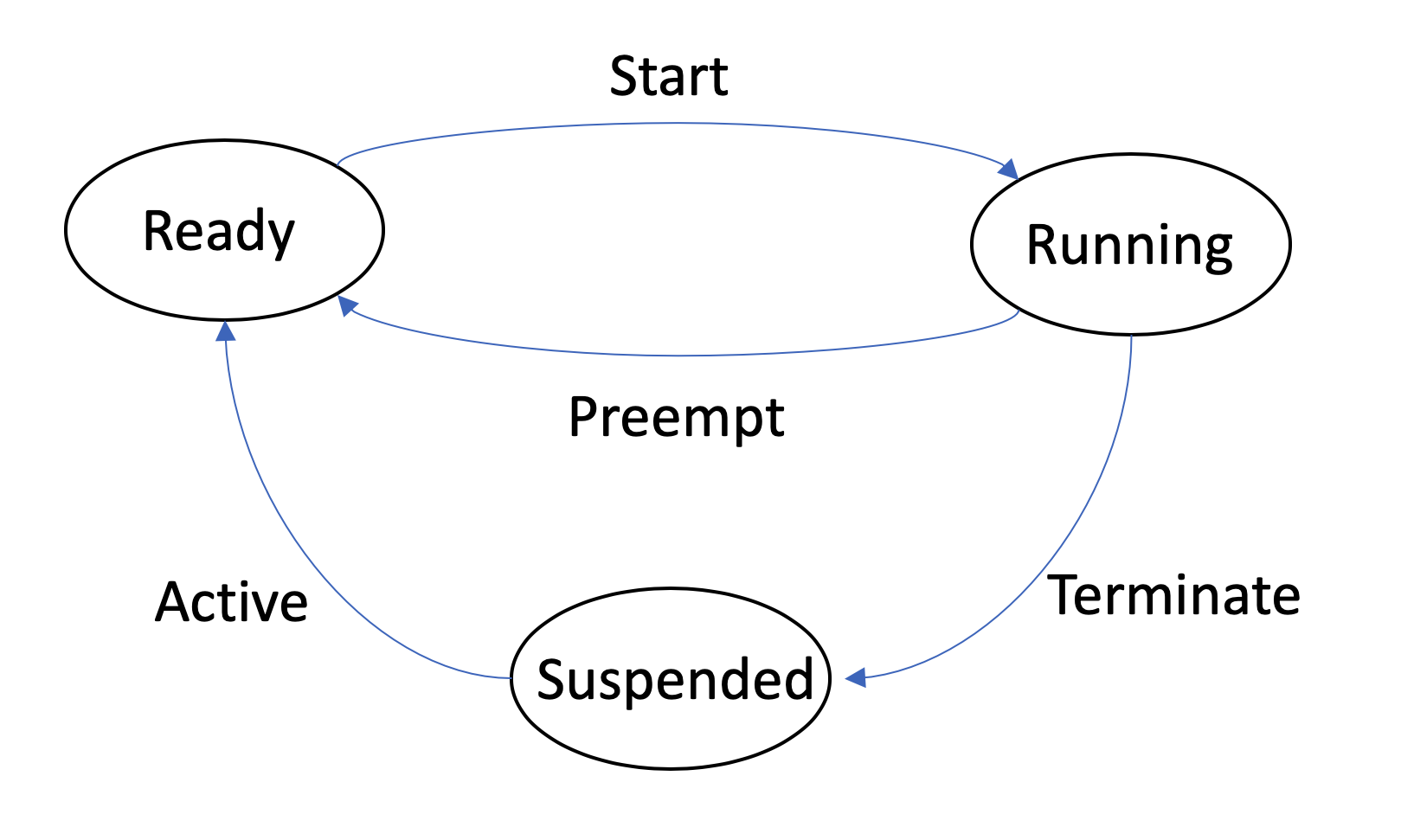}
		\fcaption{Task states and transitions of task model.}
		\label{fig:taskState}       
	\end{Figure}
	
	Figure \ref{fig:taskState} shows the task-state and transition diagrams of the task model that is based on OSEK's basic task-state model. The task model includes three states: suspended, ready, and running, and four transitions: Active, Stare, Preempt, and Terminate. The transitions represent the actions to activate, start, preempt, or terminate a task.
	\begin{Figure}
		\includegraphics[width=.99\columnwidth]{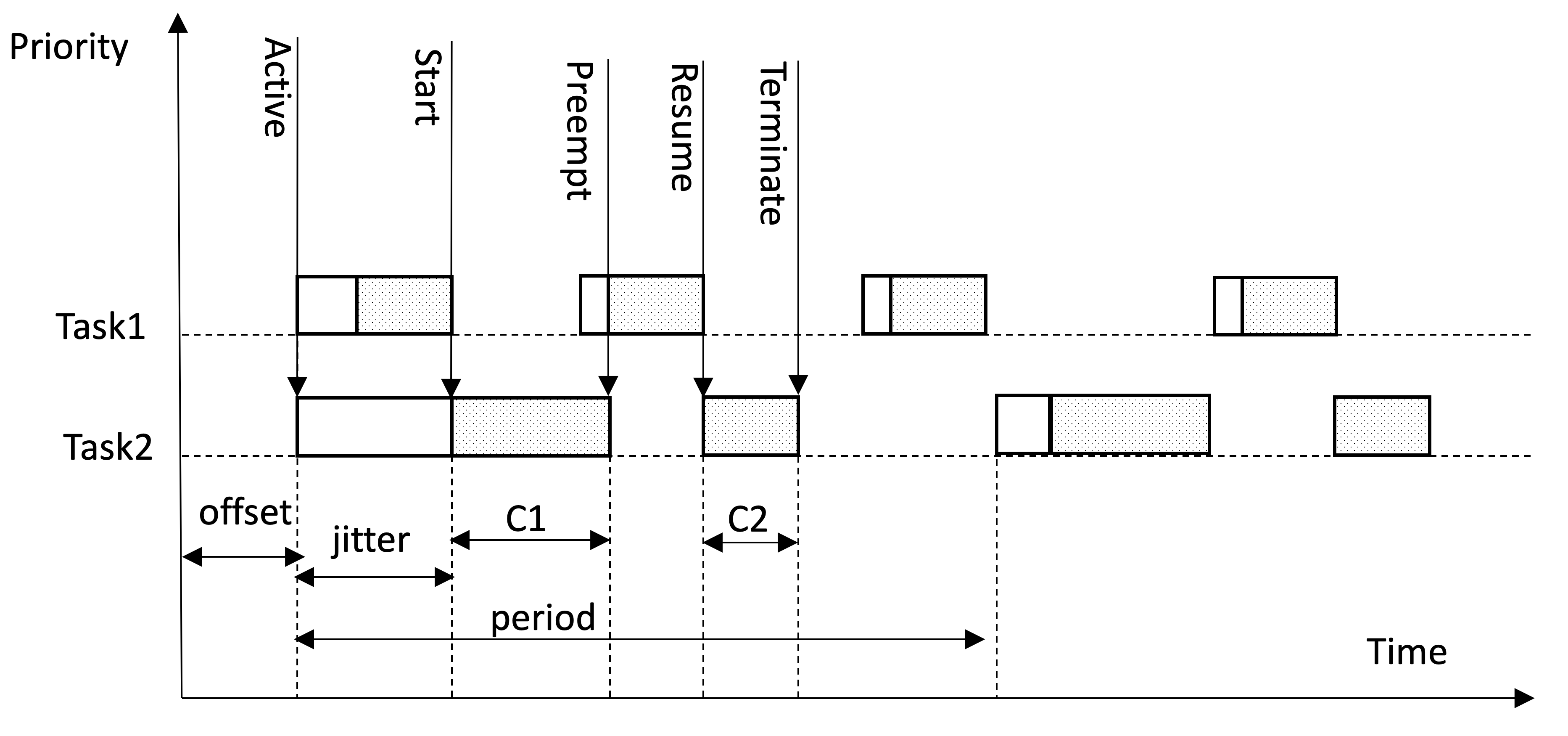} 
		\fcaption{Task timing parameters shown in Gantt chart (all related to Task2).}
		\label{fig:taskTimingParameter}
	\end{Figure}
	Figure \ref{fig:taskTimingParameter} shows the timing parameters of a task model and  different timing parameters can alter the application's real-time behavior within a system.

	\section{Mutation Operators for Simulink Model}
	This section introduces a mutation analysis approach to validate real-time context during the simulation. Mutation operators are the key elements of mutation testing. Model-based mutation testing is a method of injecting faults into models to check whether the tests are passed or failed, thus validating the software. The injecting faults are the mutation operators. Before we apply mutation operators to the model, we need to identify them which is to understand what kind of errors can cause failure. We have proposed the following task-related mutation operators.

    \subsection{Offset mutation operators}
    The task release offset is one of the factors that affect the computation result in terms of task interference. In order to take the offset into account for analysis, we introduced an offset mutation operator. For a known offset $\phi$, a task can now execute after $\phi $ time units with respect to the start of its period. The execution time of the task is unchanged at $c$ time units before the next period starts. 

    \subsubsection{mITO} This operator adds $\delta$ time to the current offset. For a given task $\tau_i\{\phi_i, \rho_i, c_i, \gamma_i, prect_i, precr_i, prio_i, jitter_i\}\in T$, this mutation operator changes the offset $\phi_i$ to $\phi_i+\delta$.

    \subsubsection{mDTO} This operator subtracts $\delta$ time to the current offset. For a given task $\tau_i\{\phi_i, \rho_i, c_i, \gamma_i, prect_i, precr_i, prio_i, jitter_i\}\in T$, this mutation operator changes the offset $\phi_i$ to $\phi_i-\delta$.

    \subsection{Period mutation operators}
    An RTOS usually applies a preemptive multitasking scheduling algorithm to determine the execution order of tasks, and the most picked algorithm is fixed-priority scheduling (FPS). The algorithm assigns each task a static priority level. The RTOS scheduler executes the highest priority task from the ready task queue. Simulink Coder supports an RM scheduler, where the priority of a task is associated with its period, if a task has a smaller period, then it has a higher priority. Furthermore, a lower-priority task can be preempted by a more top-priority task during the execution. 

    \subsubsection{mITPER} This operator increases the period of a task, which changes the task to a slower rate. For a given task $\tau_i\{\phi_i, \rho_i, c_i, \gamma_i, prect_i, precr_i, prio_i, jitter_i\}\in T$, this mutation operator changes the period of the task $\_i$ to $\rho_i+\delta$.

    \subsubsection{mDTPER} This operator decreases the period of a task, which changes the task to a faster rate. For a given task $\tau_i\{\phi_i, \rho_i, c_i, \gamma_i, prect_i, precr_i, prio_i, jitter_i\}\in T$, this mutation operator changes the period of the task $\_i$ to $\rho_i-\delta$.

    \subsection{Execution time mutation operators}
    The correctness of a real-time system is determined on one hand by the computation results of the logical program, and on the other hand, is strictly related to the time at which the results are produced. Hence, execution time analysis is essential during the process of designing and verifying embedded systems. For this reason, we propose execution time operators, which can adjust the execution time of each task at the runnable level to simulate the run time execution on different processor speeds. The longer execution time of a task may lead to a scenario where a lower-rate task blocks a higher-rate task so that it misses its deadline. 

\subsubsection{mITET} This operator adds  $\delta$ time to the current execution time of each runnable, which increases the total execution time. For a given task $\tau_i\{\phi_i, \rho_i, c_i, \gamma_i, prect_i, precr_i, prio_i, jitter_i \}\in T$, this mutation operator changes the execution time $c_i$ to $c_i+\delta$.

\subsubsection{mDTET} This operator subtracts  $\delta$ time from the current execution time of each runnable, which decreases the total execution time. For a given task $\tau_i\{\phi_i, \rho_i, c_i, \gamma_i, prect_i, precr_i, prio_i, jitter_i\}\in T$, this mutation operator changes the execution time $c_i$ to $c_i-\delta$.
\subsection{Execution precedence mutation operators}
The RTOS scheduler selects tasks to execute according to the priority level of the task. However, the spawn order determines the execution order of tasks with the same priority. Whichever task is spawned first is realized and gets the CPU first to run. This results in situations where a pair of tasks have a precedence relation in the implementation that does not exist in the design phase lost an existing precedence relation in the implementation.  The incorrect precedence can cause a wrong execution order of tasks. Hence, we proposed the precedence mutation operators which can specify a precedence relation between a pair of tasks and runnables. This operator creates mutants by assigning  a specific execution order to a set of tasks or runnable to reflect the precedence relation. 

\subsubsection{mATPREC} For a given task $\tau_i\{\phi_i, \rho_i, c_i, \gamma_i, prect_i, precr_i, prio_i, \\ jitter_i\}\in T$, for each task $\tau_j \in T$ $(j \neq i)$, if $\tau_j\notin prect_i$, this mutation operator adds  $\tau_j$ to $prect_i$.

\subsubsection{mRTPREC} For a given task $\tau_i\{\phi_i, \rho_i, c_i,  \gamma_i,prect_i, precr_i, prio_i, \\ jitter_i\}\in T$, for each task $\tau_j \in T$ $(j \neq i)$, if $\tau_j\in prect_i$, this mutation operator removes  $\tau_j$ from $prect_i$.

\subsubsection{mARPREC}  For a given task $\tau_i\{\phi_i, \rho_i, c_i, \gamma_i, prect_i, precr_i, prio_i, \\ jitter_i\}\in T$, for each runnable $\gamma_{im} \in\tau_i$, if $\gamma_{im}\notin precr_i$, this mutation operator adds  $\gamma_{im}$ to $precr_i$.
\subsubsection{mRRPREC} For a given task $\tau_i\{\phi_i, \rho_i, c_i, \gamma_i, prect_i, precr_i, prio_i, \\ jitter_i\}\in T$, for each runnable $\gamma_{im} \in\tau_i$, if $\gamma_{im}\in precr_i$, this mutation operator removes  $\gamma_{im}$ from $precr_i$.

\subsection{Priority mutation operators}
In an RTOS, each task is assigned a relative priority, which is a static integer to identify the degree of importance of tasks. The highest priority task always gets the CPU when it becomes ready to run. The most common RTOS scheduling algorithm is preemptive scheduling. 

\subsubsection{mITPRI} This operator increases the priority of a task. For a given task $\tau_i\{\phi_i, \rho_i, c_i, \gamma_i, prect_i, precr_i, prio_i, jitter_i\}\in T$, this mutation operator changes the priority of the task $prio_i$ to $proi_i+\delta$.

\subsubsection{mDTPRI} This operator decreases the priority of a task. For a given task $\tau_i\{\phi_i, \rho_i, c_i, \gamma_i, prect_i, precr_i, prio_i, jitter_i\}\in T$, this mutation operator changes the period of the task $\_i$ to $proi_i-\delta$.

\subsection{Jitter mutation operators}
Timing jitter exists in the RTOS, and it is the delay between subsequent periods of time for a given task. 
\subsubsection{mITJ} This operator increases the jitter time of a task. For a given task $\tau_i\{\phi_i, \rho_i, c_i, \gamma_i, prect_i, precr_i, prio_i, jitter_i\}\in T$, this mutation operator changes the priority of the task $jitter_i$ to $jitter_i+\delta$.

\subsubsection{mDTJ} This operator decreases the jitter time of a task. For a given task $\tau_i\{\phi_i, \rho_i, c_i, \gamma_i, prect_i, precr_i, prio_i, jitter_i\}\in T$, this mutation operator changes the period of the task $jitter_i$ to $jitter_i-\delta$.

\subsection{Shared memory mutation operators} 
It is common that RTOS tasks exchange data or information via shared memory(\eg, global variable, memory buffer, hardware register). The shared memory can easily cause access conflict if the logical software design is neglected in any corner case. Here we introduce a set of variable mutation operators.
\subsubsection{mDSM} This operator defines a new value to a global variable in a task. If a task reads this global variable, then we define a new value right before the reads occurred. 
\subsubsection{mUDSM} This operator un-defines a global variable in a task.  If a task writes this global variable, then ignore this writes operation. 
\subsubsection{mRDSM} This operator removes the definition of a global variable. If a global variable is initialized in a task then do not initialize it.  
\subsubsection{mRSM} This operator adds a reference to a global variable.
\subsubsection{mRMSMR} This operator removes reference to a global variable.
\subsubsection{mRSMR} This operator replaces a reference of a global variable with a different global variable.

\begin{Table} 
	\tcap{Simuilnk Mutation Operators}
	\begin{tabular}{ l  l }
		\hline
		Mutation Key & Title \\ \hline
		mITO   & Increase Task Offset \\
		mDTO & Decrease Task Offset \\\hline
		mITPER &  Increase Task Period \\
		mDTPER &  Decrease Task Period \\ \hline
		mITET & Increase Task Execution Time \\
		mDTET & Decrease Task Execution Time \\ \hline
		mATPREC & Add Task Precedence \\ 
		mRTPREC & Remove Task Precedence\\
		mARPREC & Add Runnable Precedence\\
		mRRPREC & Remove Runnable Precedence\\ \hline
		mITPRI & Increase Task Priority\\
		mDTPRI & Decrease Task Priority\\ \hline
		mITJ & Increase Task Jitter\\
		mDTJ & Decrease Task Jitter\\ \hline
		mDSM & Define Shared Memory\\
		mUDSM & Un-define Shared Memory\\
		mRDSM & Remove Definition Shared Memory\\
		mRSM & Reference a Shared Memory\\
		mRMSMR & Remove a Shared Memory Reference\\
		mRSMR & Replace a Shared Memory Reference\\
		
		\hline
	\end{tabular}
	\label{tbl:mutationoperator}
\end{Table}

\section{Mutation operators demonstration}
We have introduced twenty mutation operators categorized into seven classes and explained each mutation class. The mutation operators are summarized in Table \ref{tbl:mutationoperator}. We use simple examples to demonstrate the use of each mutation operator. To demonstrate our mutation operators, we use the tool SimSched to alter the properties of software applications realized as Simulink models. From Table \ref{tbl:mapping}, we know that each function-call subsystem represents an AUTOSAR runnable. The function-call subsystem can be executed conditionally when a function-call event signal arrives. Both an S-function block and a Stateflow block can provide such a function-call event. SimSched applies the function-call invocation mechanism to use an S-function to generate a function-call event to schedule each runnable (function-call subsystem). Figure \ref{fig:msParameter} shows the SimSched parameters dialogue that we can utilize it to adjust the timing properties to implement the mutation operator for Simulink models. 

    \begin{Figure}
		\includegraphics[width=0.99\textwidth]{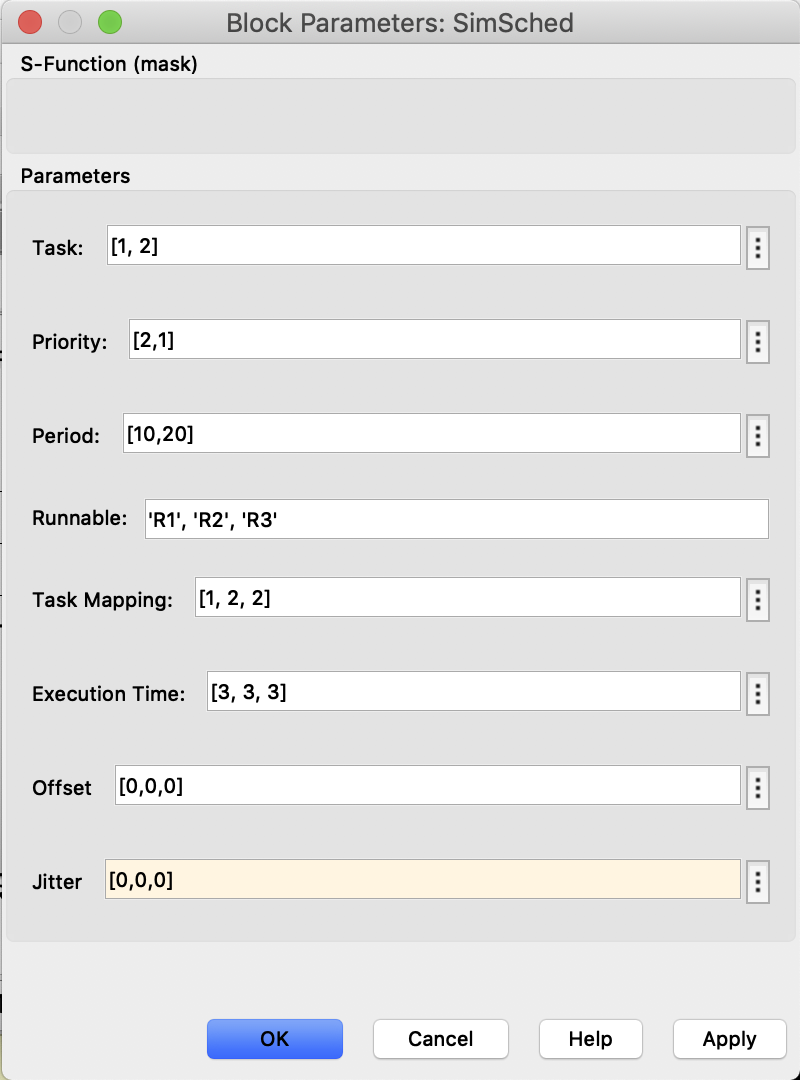}
		\fcaption{SimSched Parameter setting dialogue.}
		\label{fig:msParameter}
	\end{Figure}

In this section, we use several simple examples to exhibit the mutants generated by our mutation operators. Figure \ref{fig:Scheduler_DataRace} illustrates the use of SimSched to schedule a Simulink model. In this example, SimSched is at the top left corner, which schedules three runnables ($R_1$, $R_2$, $R_3$), and they are mapped to two tasks ($T_1$, $T_2$). Runnable $R_1$ is mapped to $T_1$, the period of $T_1$ is $10ms$, priority is $2$, and the execution time of $R_1$ is $3ms$. $R_2$ and $R_3$ are mapped to $T_2$. The period of $T_2$ is $20ms$, priority is $1$, and the execution time of $R_2$ and $R_3$ are $3ms$ and $3ms$ accordingly. The detailed parameter settings are listed in Table \ref{tbl:simpleparameter_M}. There is a Data Store Memory block in this example, named $A$, which defines a shared data store that is a memory area used by Data Store Read and Data Store Write block with the same data store name. $R_1$ writes a constant value to a global variable $A$. $R_2$ reads $A$ first then writes the summation of $A$ and its delay value to $A$. $R_3$ reads $A$ then subtracts its delay value from $A$, and outputs the result.

\begin{Table}
	\centering
	\tcap{The simple example settings}
	\label{tbl:simpleparameter_M}
	\begin{tabular}{ c c c c c}
		\hline
		Task & Period & Execution & Priority & Runnable \\ 
		 &  ($ms$) & Time($ms$) &  &  \\ \hline
		$T_1$   & 10 & 3 & 2 & $R_1$ \\
		$T_2$   & 20 & 3 & 1 & $R_2$ \\
		$T_2$   & 20 & 3 & 1 & $R_3$ \\
		\hline
	\end{tabular}
\end{Table}

\begin{Figure}
    \includegraphics[width=.99\columnwidth]{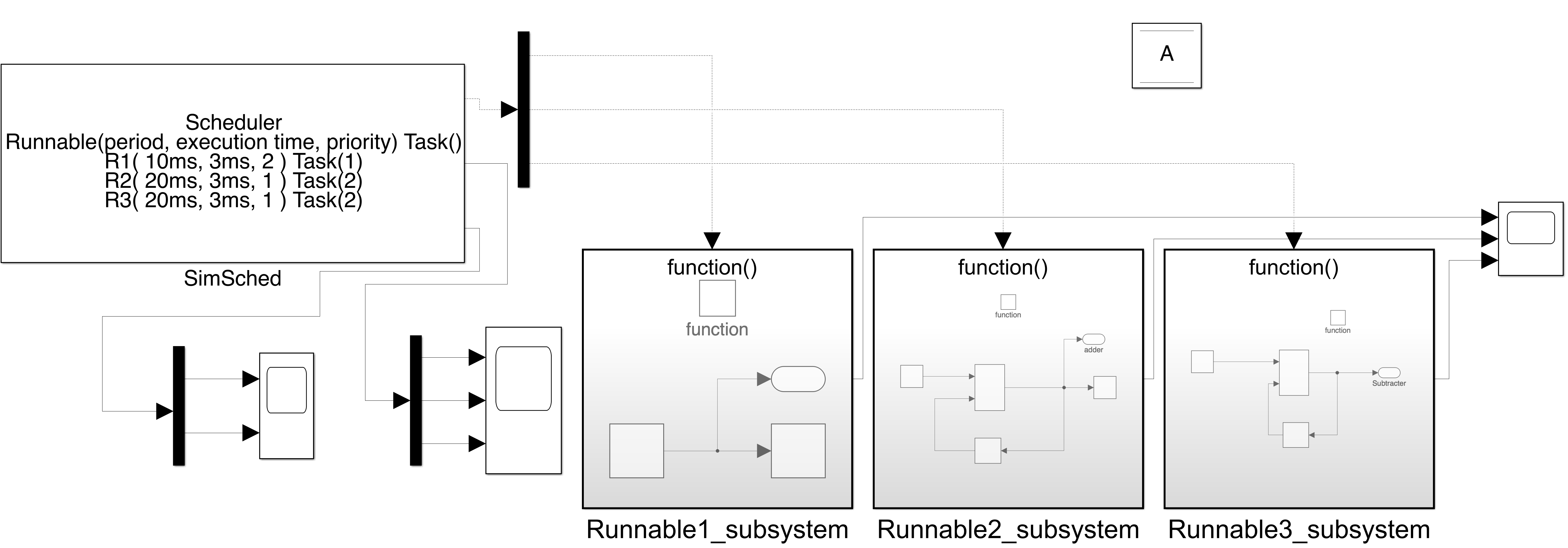} 
	\fcaption{A simple example of using SimSched to schedule AUTOSAR SW-Cs.}
	\label{fig:Scheduler_DataRace}
\end{Figure}

\begin{Figure}
	\includegraphics[width=.99\columnwidth]{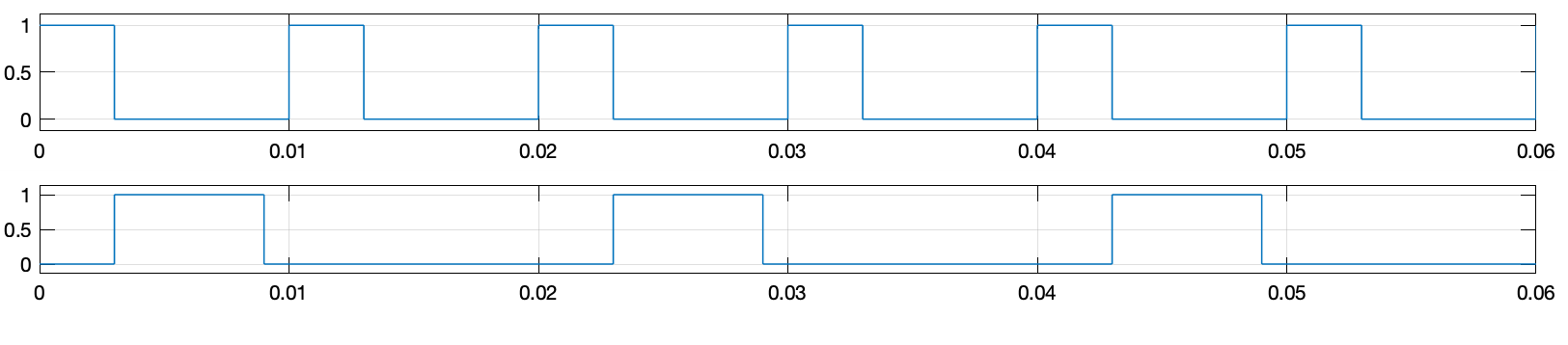}
	\fcap{Task executions Gantt chart of the running example.}
	\label{fig:taskexecution}
\end{Figure}

\begin{Figure}
	\includegraphics[width=.99\columnwidth]{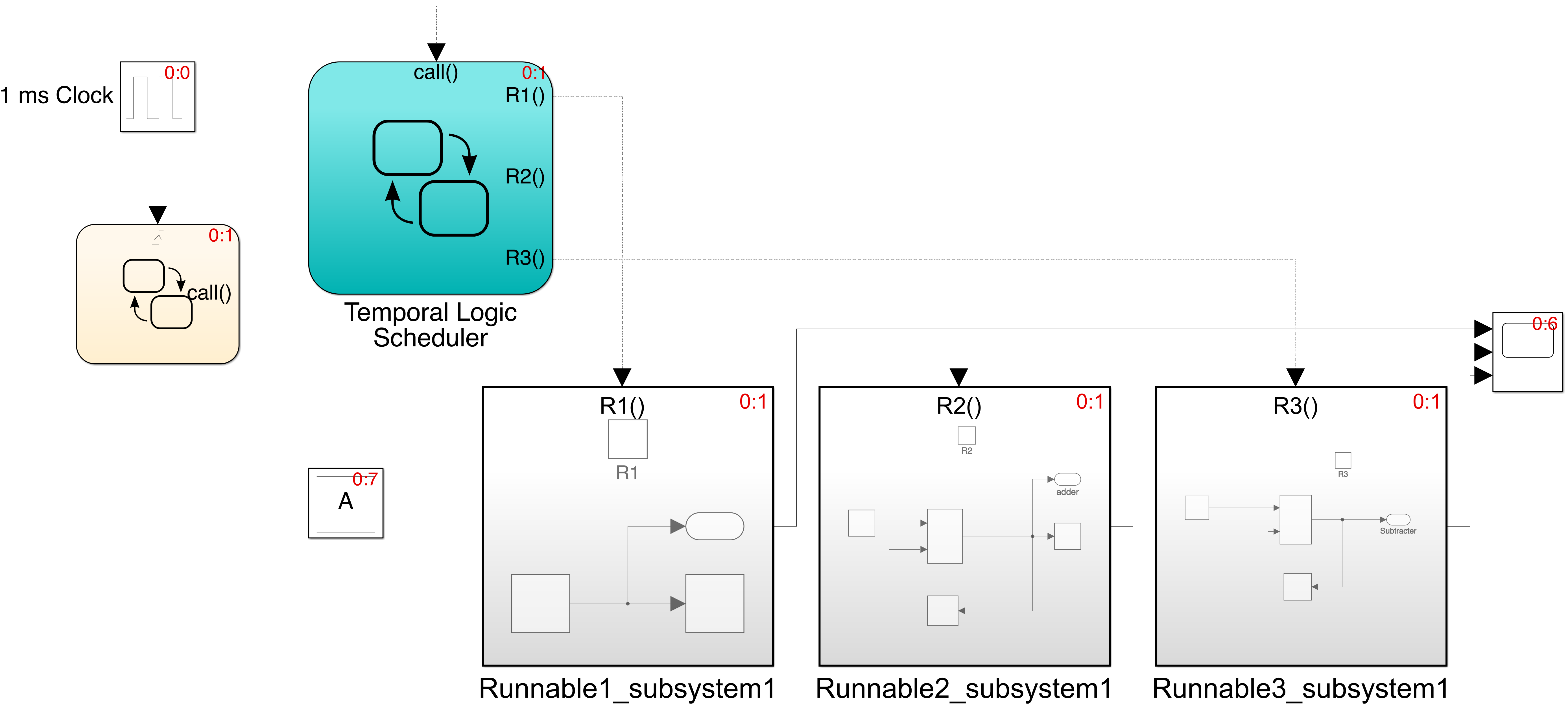} 
	\fcap{A simple example of using Stateflow to schedule AUTOSAR SW-Cs.}
	\label{fig:stateflowExecOrder_6}
\end{Figure}

\begin{Figure}
	\includegraphics[width=.99\columnwidth]{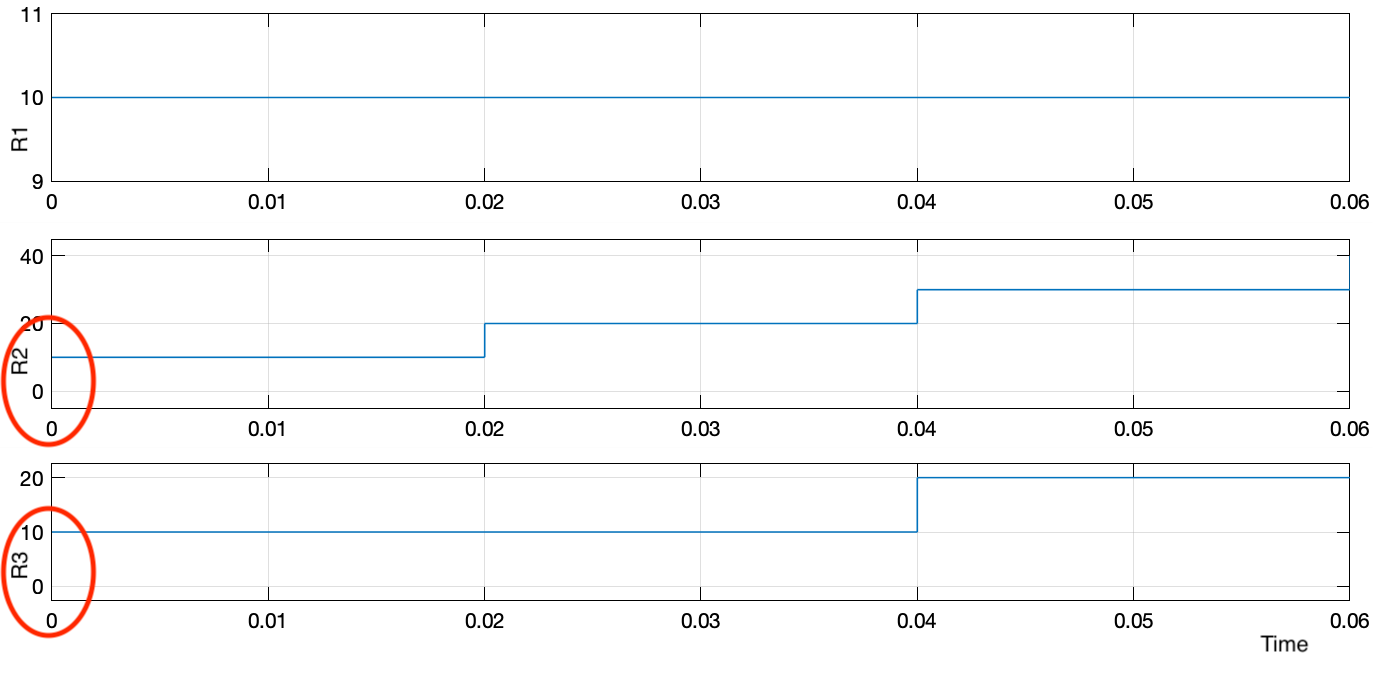}
	\fcap{Simple example output of Stateflow scheduler simulation.}
	\label{fig:sfoutput}
\end{Figure}

We use a Stateflow scheduler version of the simple example shown in Figure \ref{fig:stateflowExecOrder_6} to show the typical Stateflow scheduler simulation result, then compare it with the SimSched scheduler simulation result. The task parameters are all the same shown in Table \ref{tbl:simpleparameter_M}. We apply the same task configurations for both the Stateflow scheduler and SimSched models for simulation. Figure \ref{fig:sfoutput} shows the Stateflow scheduler simulation result, and Figure \ref{fig:SimSchedoutput} shows the SimSched simulation result. The result figures show the output value of each runnable. From the figure, we can see that $R_1$, $R_2$ and $R_3$ are all executed at time 0 in Figure \ref{fig:sfoutput}; $R_1$, is executed time $0$, $R_2$ is executed at $3ms$, and $R_3$ is executed at $6ms$ in Figure \ref{fig:SimSchedoutput}. $R_2$ and $R_3$ are executed later than the Stateflow scheduler simulation in Figure \ref{fig:sfoutput}. This is because that SimSched takes into account execution time, and each task must be executed until the previous task is completed on a single core platform.

\begin{Figure}
	\includegraphics[width=.99\columnwidth]{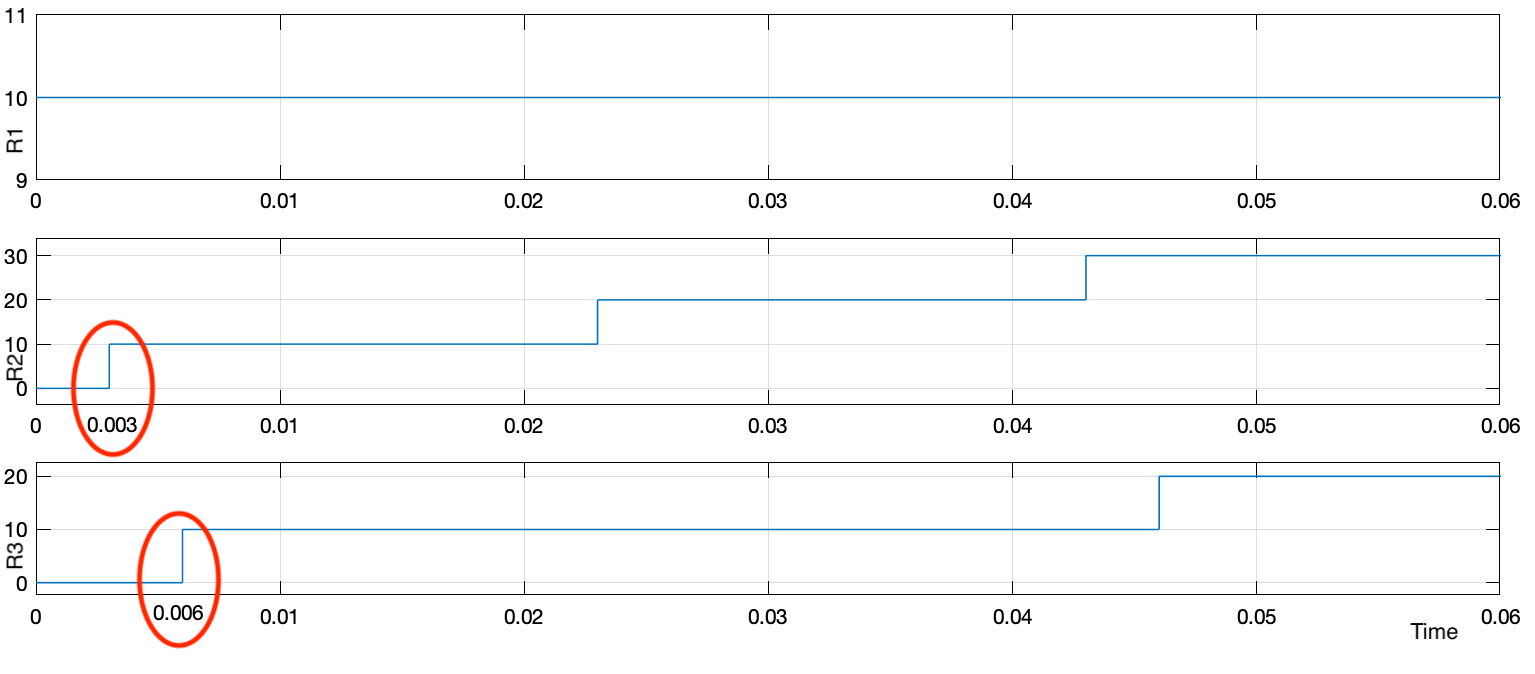}
	\fcap{Simple example output of SimSched simulation without applying any mutation operator.}
	\label{fig:SimSchedoutput}
\end{Figure}

\subsection{Offset mutation operators}

We first apply the \emph{mITO} mutation operator to the running example, let's say that increase $\delta_1=3ms$ for $T_1$ then we have the task execution timeline in Figure \ref{fig:ioffset}. We can see $T_2$ is executed first at time $0$, and $T_1$ preempts $T_2$ at $3ms$ in the first period due to the offset effect. After the first period, there is no preemption between $T_1$ and $T_2$. Then, we apply the \emph{mDTO} mutation operator based on the previous settings. We set $\delta_1=-1ms$ to $T_1$ then the offset for $T_1$ is $2ms$. Figure \ref{fig:doffset} shows the task execution timeline. $T_2$ is preempted by $T_1$ during the execution of the first period. Compared to the task execution Gantt chart of our running example shown in Figure \ref{fig:taskexecution} with no \emph{offset}, we can clearly see the preemption effect. 

\begin{Figure}
	\includegraphics[width=.99\columnwidth]{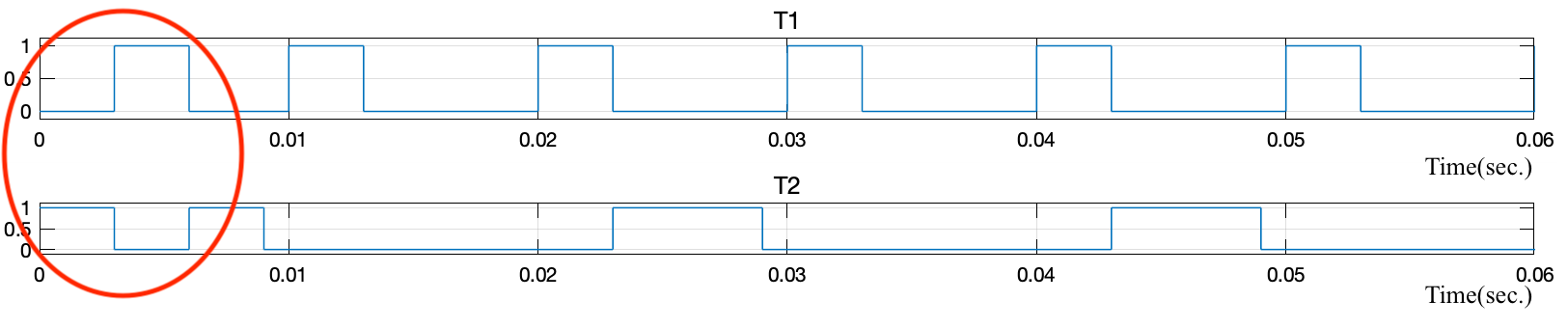}
	\fcaption{Task executions Gantt chart of the running example after increase offset mutation operator is applied.}
	\label{fig:ioffset}
\end{Figure}

\begin{Figure}
	\includegraphics[width=.99\columnwidth]{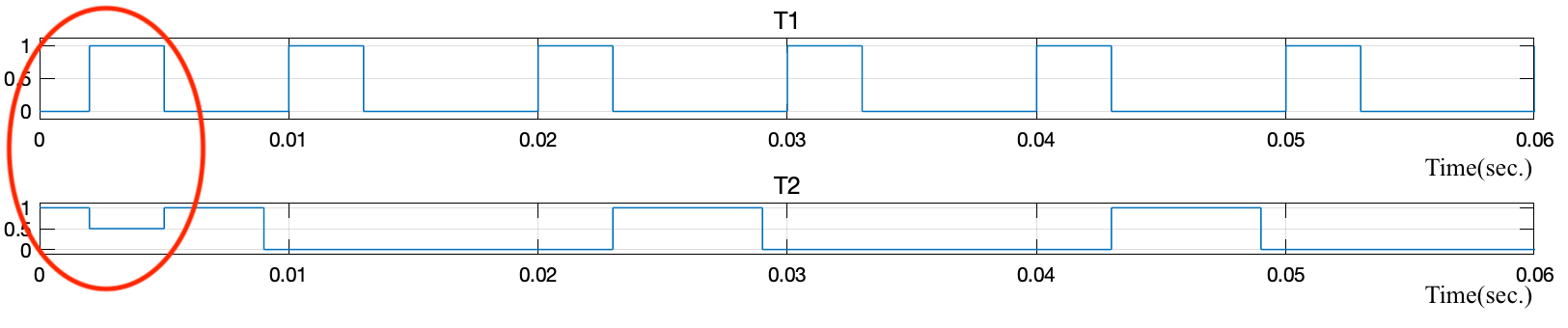}
	\fcaption{Task executions Gantt chart of the running example after decrease offset mutation operator is applied.}
	\label{fig:doffset}
\end{Figure}

The running example's output after applying the \emph{mITO} mutation operator is shown in Figure \ref{fig:outputofioffset} which is different from Figure \ref{fig:SimSchedoutput}. Because $T_1$ preempts $T_2$ at the first instance execution, the output of $R_3$ is from zero to ten then goes back to zero then goes up instead of always increasing value. The running example's output after applying the \emph{mDTO} mutation operator is the same as Figure \ref{fig:SimSchedoutput} because the offset operator only affects the initial execution of each task, and the preemption occurs before the first execution of $R_2$ instance completion. Inside our model scheduler program, we trigger each subsystem at the end of each execution time slot. Technically, the execution order of this example is still $R_1$, $R_2$, and $R_3$ so the output of the simulation keeps the same.

\begin{Figure}
	\includegraphics[width=.99\columnwidth]{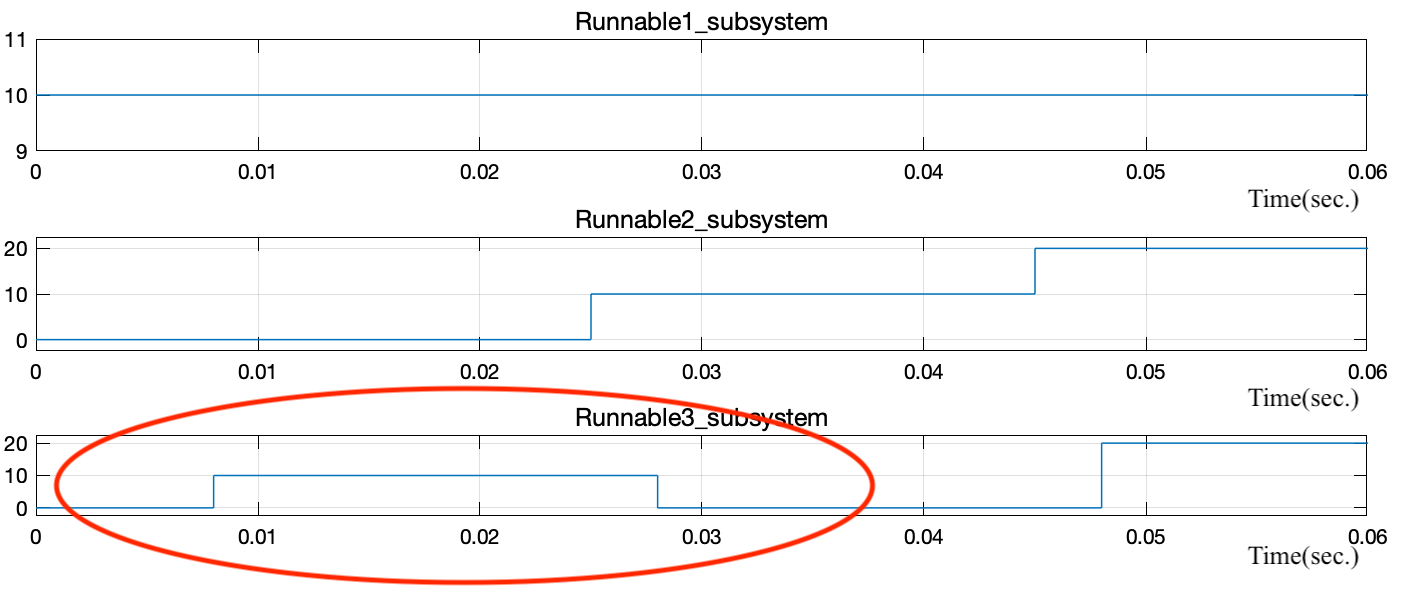}
	\fcap{Simple example output of SimSched simulation after applying \emph{mITO} mutation operator.}
	\label{fig:outputofioffset}
\end{Figure}

\subsection{Period mutation operators}
We apply the \emph{mITPER} operator to this example to increase the period of a task. We set $\delta_1=1ms$ to $T_1$ so the period of the task $T_1$ is $11ms$ now. Figure \ref{fig:iperiod} shows that $T_2$ is preempted  at the time of $22ms$, and the simulation yields a wrong result due to this preemption shown in Figure \ref{fig:outputiperiod}. The output of $R_3$ is an alternating value instead of an increasing value.

\begin{Figure}
	\includegraphics[width=.99\columnwidth]{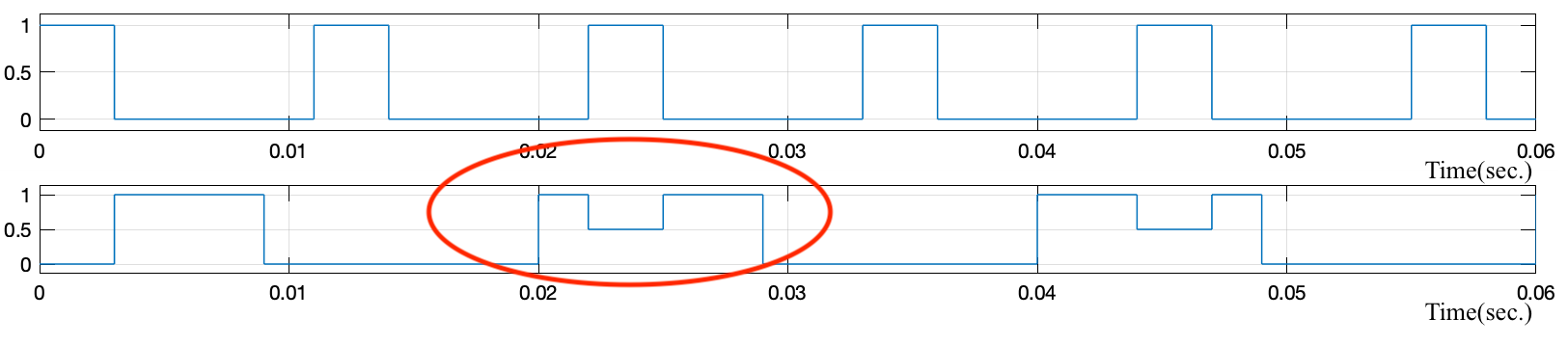}
	\fcap{Task executions Gantt chart of the running example\newline after increase period mutation operator is applied.}
	\label{fig:iperiod}
\end{Figure}

\begin{Figure}
	\includegraphics[width=.99\columnwidth]{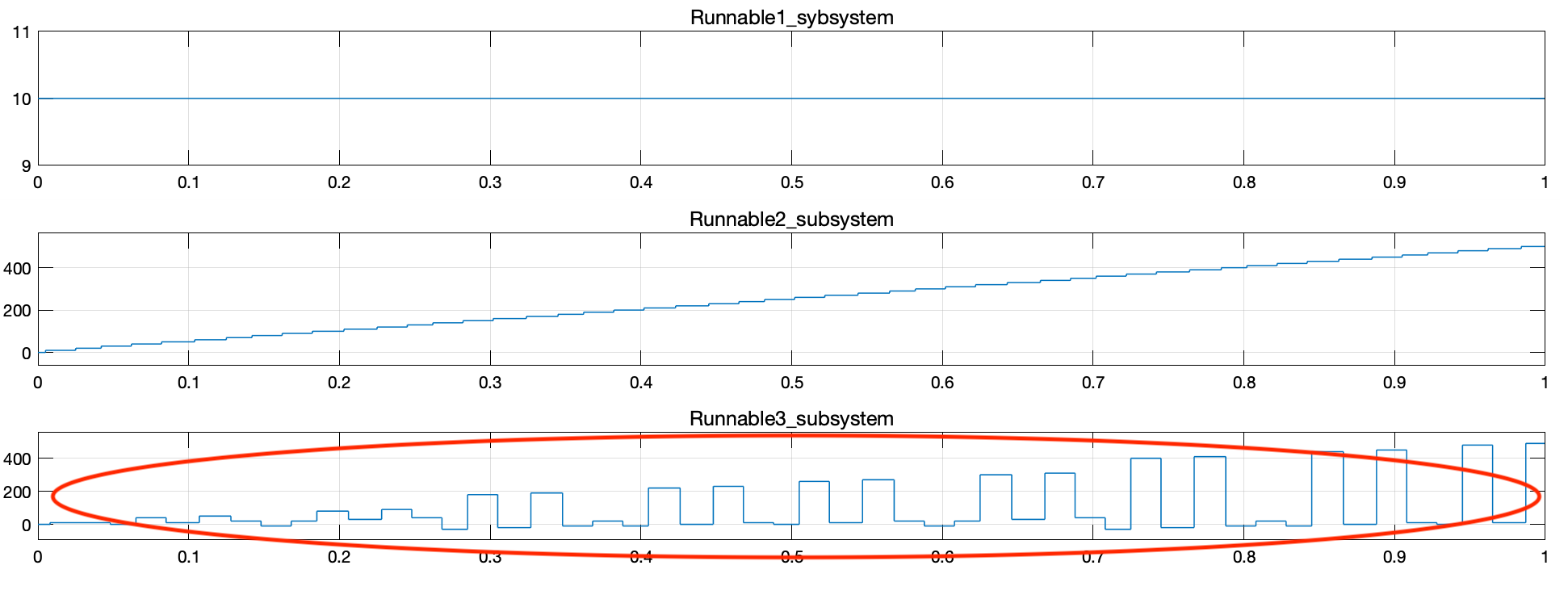}
	\fcap{Simple example output of SimSched simulation after applying \emph{mITPER} mutation operator.}
	\label{fig:outputiperiod}
\end{Figure}

We apply the \emph{mDTPER} operator to this example to decrease the period of a task. We set $\delta_1= -4ms$ to $T_1$ so the period of the task $T_1$ is $6ms$ now. Then, we run the simulation,  $T_2$ is preempted  by $T_1$ shown in Figure \ref{fig:dperiod} and it yields a wrong simulation result shown in Figure \ref{fig:outputdperiod}. The output of $R_3$ is either zero or ten instead of an increasing value.

\begin{Figure}
    \includegraphics[width=.99\columnwidth]{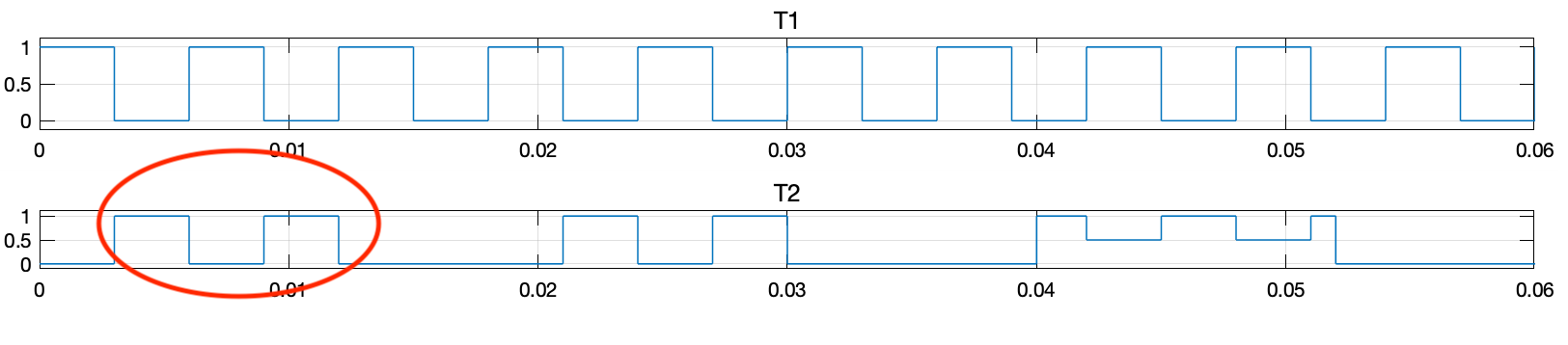}
	\fcaption{Task executions Gantt chart of the running example after decreasing period mutation operator is applied.}
	\label{fig:dperiod}
\end{Figure}

\begin{Figure}
	\includegraphics[width=.99\columnwidth]{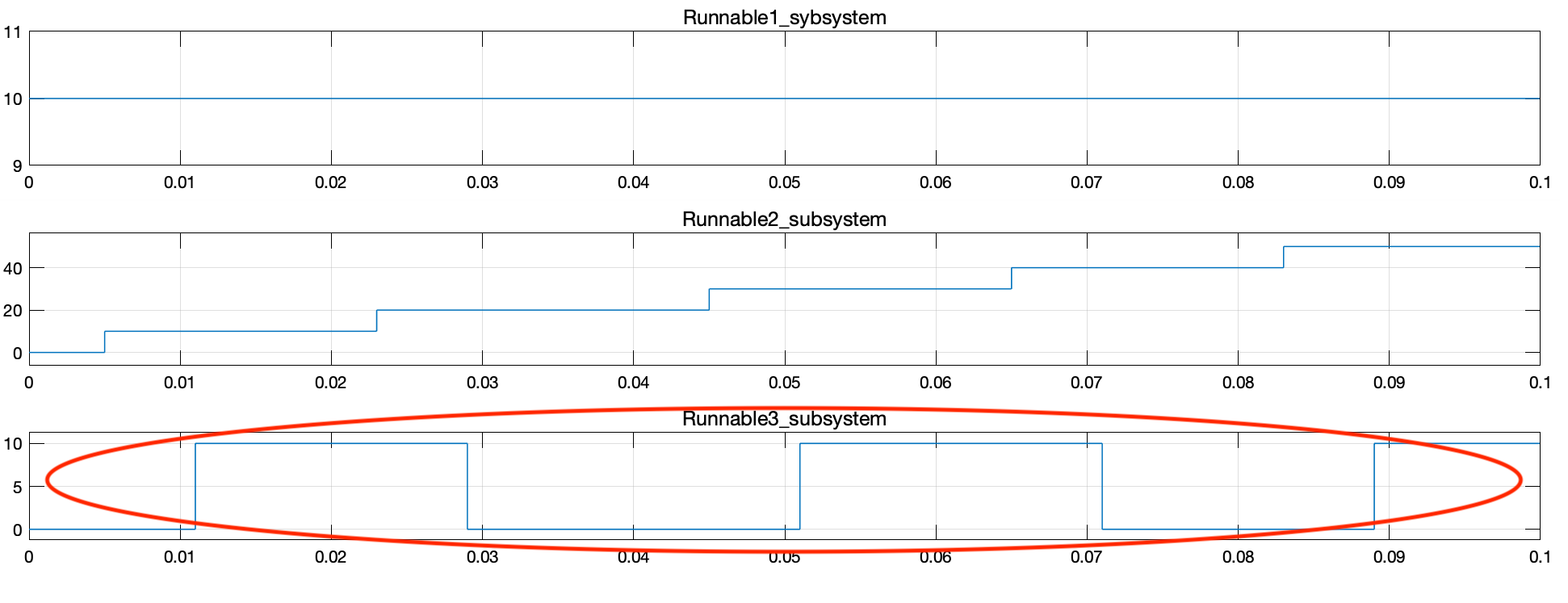}
	\fcaption{Simple example output of SimSched simulation after applying \emph{mDTPER} mutation operator.}
	\label{fig:outputdperiod}
\end{Figure}

\subsection{Execution time mutation operators}
We apply the \emph{mITET} operator to this example to increase the execution time of a task. We can specify any runnable to increase its execution time within a task.  For example, we set $\delta_1=4ms$ to $R_2$ in $T_2$ so the execution of $R_2$ is $7ms$ and $T_2$ takes $10ms$ to execute now. Figure \ref{fig:iexectime} shows that $T_2$ is preempted  at the time of $10ms$, and the simulation yields a wrong result due to this preemption. The wrong result is the same as the example of applying decreasing task period.  
We apply the \emph{mDTET} operator to this example to decrease the execution time of a task. We set $\delta_1= -1ms$ to $T_1$ so the execution time of the task $T_1$ is $2ms$ now. Then, we run the simulation, there is no preemption that occurs between these two tasks and the output is as expected as the original model. 
\begin{Figure}
	\includegraphics[width=.99\columnwidth]{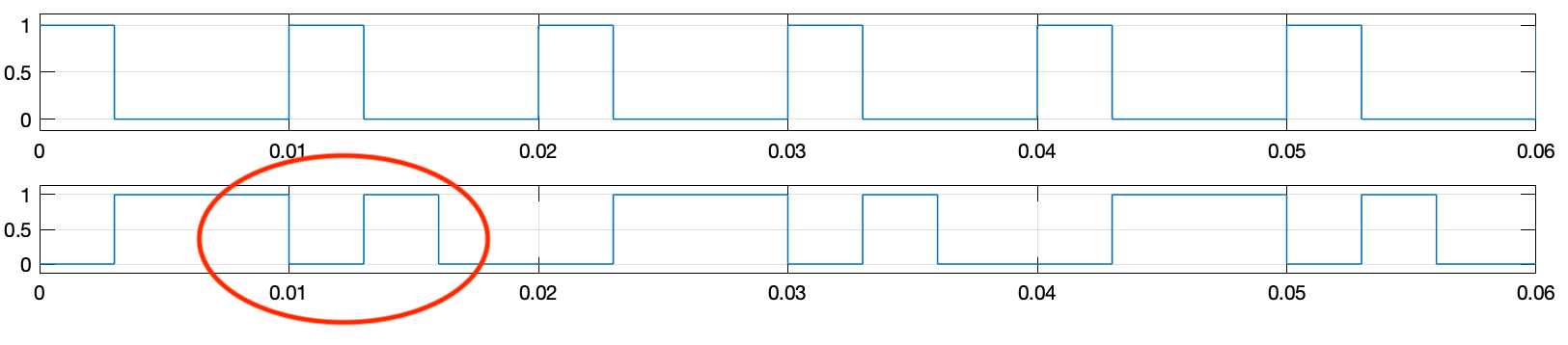}
	\fcaption{Task executions Gantt chart of the running example after increase execution time mutation operator is applied.}
	\label{fig:iexectime}
\end{Figure}

\subsection{Execution precedence mutation operators}
We introduce the second example as Table \ref{tbl:simpleparamete2} to explain the \emph{mATPREC} and \emph{{mRTPREC}} operators. Figure \ref{fig:example2} shows the task execution Gantt chart of this example. From the task execution chart, we can see the execution order of the tasks is $T_1, T_2, T_1, T_3$.
\begin{Table}
	\tcap{The simple example settings}
	\label{tbl:simpleparamete2}
	\begin{tabular}{ c c c c c}
		\hline
		Task & Period & Execution & Priority & Runnable \\ 
		 &  ($ms$) & Time($ms$) &  &  \\ \hline
		$T_1$   & 5 & 1 & 3 & $R_1$ \\
		$T_2$   & 10 & 4 & 2 & $R_2$ \\
		$T_3$   & 10 & 3 & 1 & $R_3$ \\
		\hline
	\end{tabular}
\end{Table}

\begin{Figure}
	\includegraphics[width=.99\columnwidth]{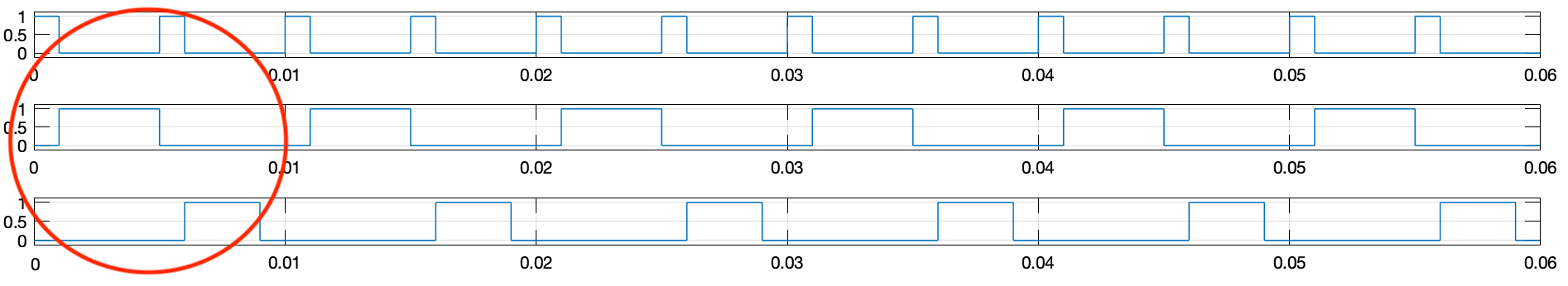}
	\fcaption{Task executions Gantt chart of example 2.}
	\label{fig:example2}
\end{Figure}

First, we assume there is no precedence relation among tasks so we use the \emph{mATPREC}  mutation operator to add a precedence relation $\tau_3$ to $prect_2$, which specifies that a new instance of $T_2$ cannot start unless $T_3$ has executed after the last instance of $T_2$. Hence, we set the execution order that $T_3$ is executed before $T_2$ in the setting dialogue. Figure \ref{fig:atprec} shows the execution result that $T_2$ is preempted by $T_1$. If $T_2$ is not a re-entrant function then this preemption may cause potential failure execution. 

\begin{Figure}
    \includegraphics[width=.99\columnwidth]{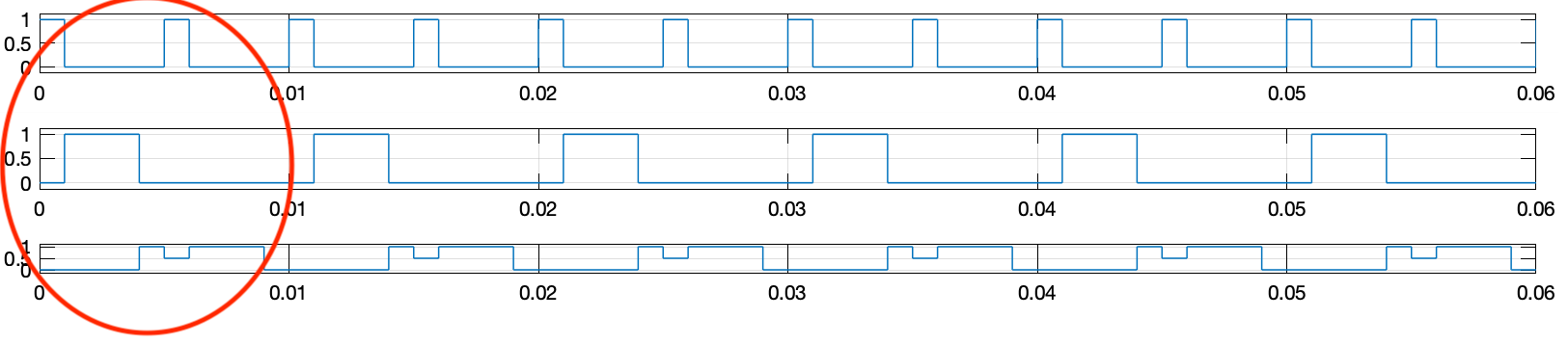}
	\fcaption{Task executions Gantt chart of example 2 after task precedence mutation operator is applied.}
	\label{fig:atprec}
\end{Figure}

Then, we assume there is a precedence relation between $T_1$ and $T_3$ and the task execution diagram is the same in Figure \ref{fig:example2}. We apply the \emph{mRTPREC}  mutation operator to remove the precedence relation $prect_3$ from $\tau_1$. The result is the same as shown in Figure \ref{fig:atprec}.

We add one runnable $R_4$ to the first example and assign it to $T_1$. This new task configuration is shown in Table \ref{tbl:simpleparamete2_prec}. $R_4$ writes a different constant value from $R_1$ to the global variable $A$. We apply \emph{mARPREC} mutation operator to this new example, which adds $\gamma_1$ to $precr_4$. $R_4$ requires $R_1$ execute first so $R_4$ overwrites the value written by $R_1$. The operator changes the execution order of runnables. 
\begin{Table}
	\tcap{Task configuration settings for runnable precedence mutation operators.}
	\label{tbl:simpleparamete2_prec}
	\begin{tabular}{ c c c c c}
		\hline
		Task & Period & Execution & Priority & Runnable \\ 
		 &  ($ms$) & Time($ms$) &  &  \\ \hline
		$T_1$   & 10 & 2 & 2 & $R_1$ \\
		$T_2$   & 20 & 2 & 1 & $R_2$ \\
		$T_3$   & 20 & 2 & 1 & $R_3$ \\
		$T_1$   & 10 & 2 & 1 & $R_4$ \\
		\hline
	\end{tabular}
\end{Table}

In example one, $T_2$ has two runnables $R_2$ and $R_3$ with a precedence relation between them. We apply \emph{mRRPREC} runnable remove precedence mutation operator to remove the precedence $\gamma_2$ from $precr_3$. We schedule $R_3$ runs before $R_2$ since no precedence constraint that turns out different than the original simulation. The original output of $R_3$ is an increasing value along with the execution instead of a value of either zero or a fixed value. The reason is that $R_3$ executes first and it reads $A$ before $R_2$ writes any new value to $A$. The bottom output line in Figure \ref{fig:rrprec} shows the execution result.
\begin{Figure}
    \includegraphics[width=.99\columnwidth]{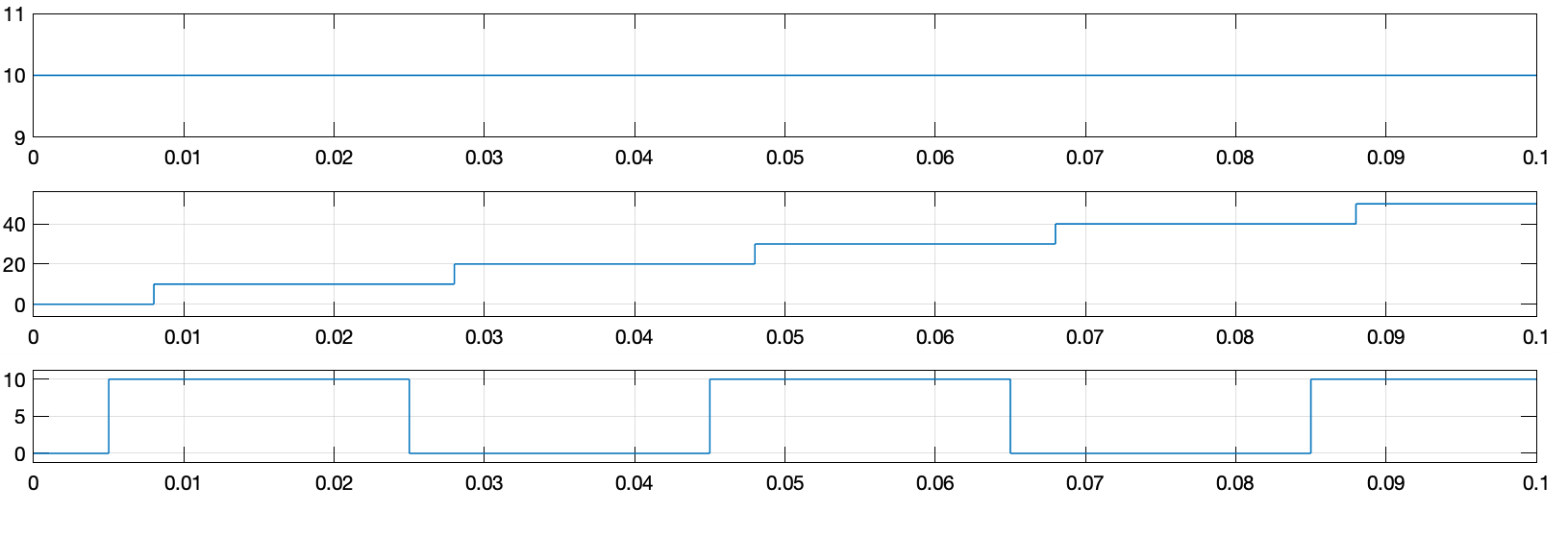}
	\fcaption{The outputs of example one three runnables.}
	\label{fig:rrprec}
\end{Figure}

\subsection{Priority mutation operators}

\begin{Table}
	\tcap{Priority mutation operator example settings}
	\label{tbl:priority}
	\begin{tabular}{ c c c c c}
		\hline
		Task & Period & Execution & Priority & Runnable \\ 
		 &  ($ms$) & Time($ms$) &  &  \\ \hline
		$T_1$   & 10 & 1 & 4 & $R_1$ \\
		$T_2$   & 10 & 2 & 3 & $R_2$ \\
		$T_3$   & 10 & 3 & 2 & $R_3$ \\
		\hline
	\end{tabular}
\end{Table}

We apply \emph{mITPRI} operator to the example in Table \ref{tbl:priority} to increase the priority of $T_3$. This mutation operator changes the priority of $prio_3$ to $proi_3+3$ so the $T_3$ has the highest priority 5 in this example, which results in $T_3$ being executed at first. Figure \ref{fig:itpriority} shows $T_3$ is triggered first in the task execution Gantt chart. This mutation alters the task execution order. 
\begin{Figure}
    \includegraphics[width=.99\columnwidth]{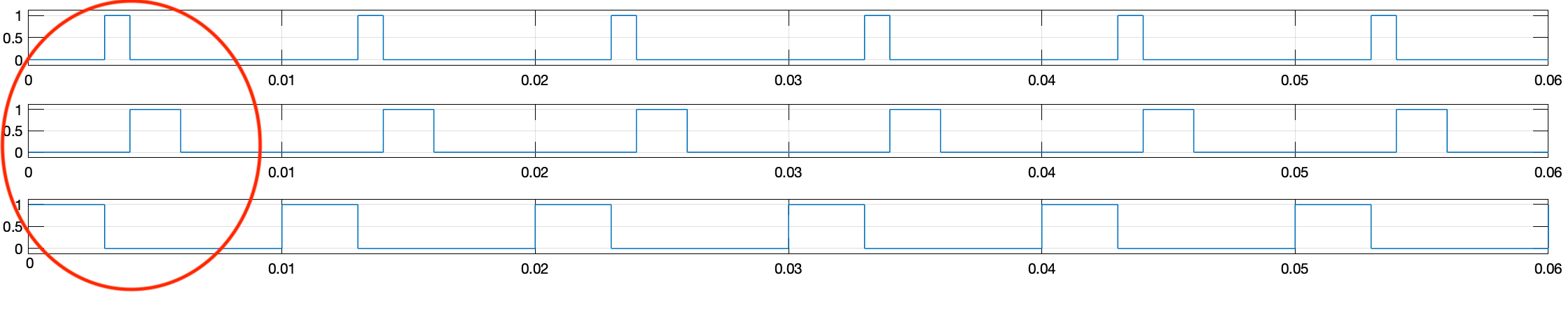}
	\fcaption{Task executions Gantt chart after applying increasing task priority mutation operator.}
	\label{fig:itpriority}
\end{Figure}
We apply \emph{mDTPRI} operator to decrease the priority of $T_1$. This mutation operator changes the priority of $prio_i$ to $proi_-3$ so the $T_1$ has the lowest priority 1 in this example, which results in $T_1$ being executed at last. The task execution Gantt chart is shown in Figure \ref{fig:dtpriority}.
\begin{Figure}
    \includegraphics[width=.99\columnwidth]{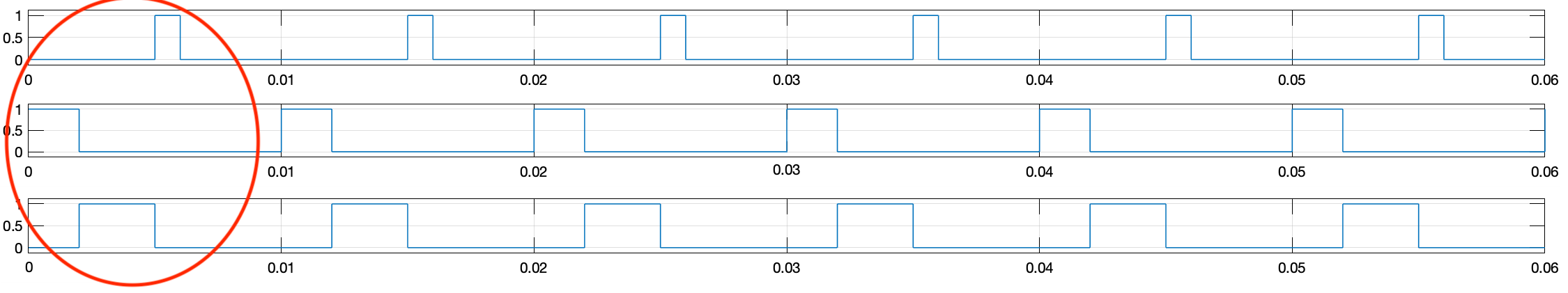}
	\fcaption{Task executions Gantt chart after applying decreasing task priority mutation operator.}
	\label{fig:dtpriority}
\end{Figure}

\subsection{Jitter mutation operators}
We apply \emph{mITJ} operator to increase a jitter time of a task. For example, let $\delta=2$, this mutation operator changes the real release time of the task to $jitter_1=0+2$. Figure \ref{fig:jitter} shows the execution of $T_2$ is preempted by $T_1$ caused by the jitter. 
\begin{Figure}
    \includegraphics[width=.99\columnwidth]{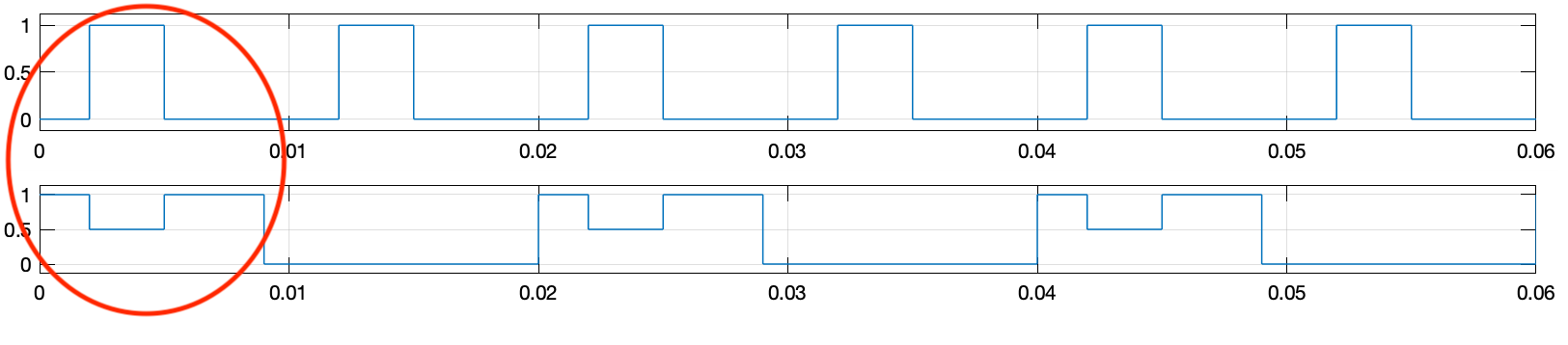}
	\fcaption{Task executions Gantt chart after applying increasing jitter mutation operator.}
	\label{fig:jitter}
\end{Figure}
Then \emph{mDTJ} mutation operator decreases the jitter time of a task. We apply this operator to the above example and let $\delta=-1$ so the task $jitter_1=2-1$. $T_2$ is preempted by $T_1$ during the simulation phase.

\subsection{Shared memory mutation operators} 
In this shared memory category, we introduce five mutation operators.  The first one is \emph{mDSM}, and this operator assigns a new value to the memory store before a read. For our example, we add a Data Store Write block right before the Data Store Read execution so that the Data Store Write block defines a new value to the variable, and we chose the initial value of this variable as the default new value. The mutant using \emph{mUDSM} operator is shown in Figure \ref{fig:dsm}, which only shows the changes of Runnable2\_subsystem. We add a constant block and a Data Store Write block at the top left corner.
\begin{Figure}
	\includegraphics[width=.99\columnwidth]{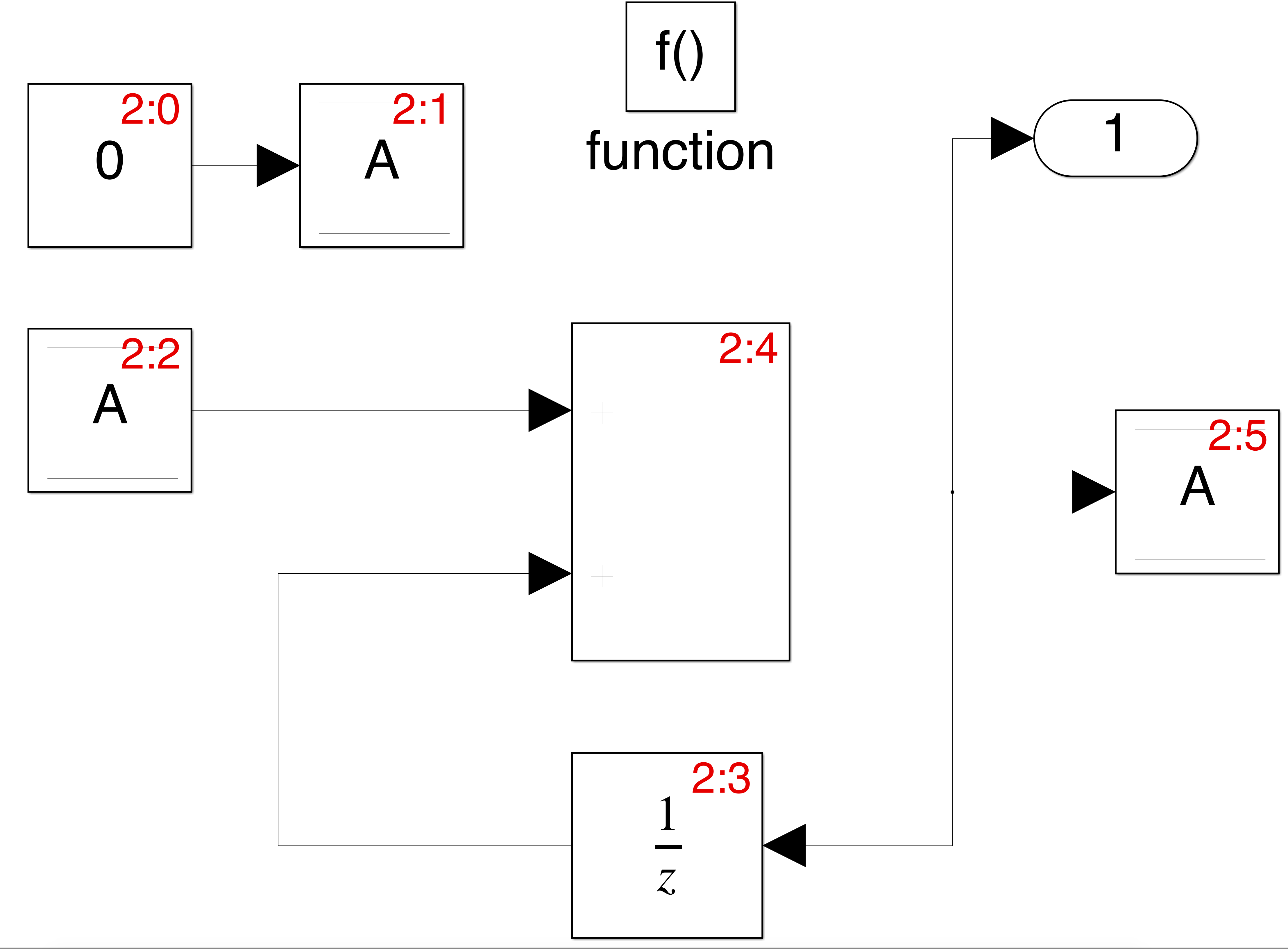} 
	\fcaption{A simple example of DSM mutant.}
	\label{fig:dsm}
\end{Figure}

The second mutant operator is \emph{mUDSM}, and this operator disregards a write to a Data Store block. For our example, we remove the Data Store Write block. Figure \ref{fig:udsm} shows the \emph{mUDSM} mutant that the Data Store Write has been removed.

\begin{Figure}
	\includegraphics[width=.99\columnwidth]{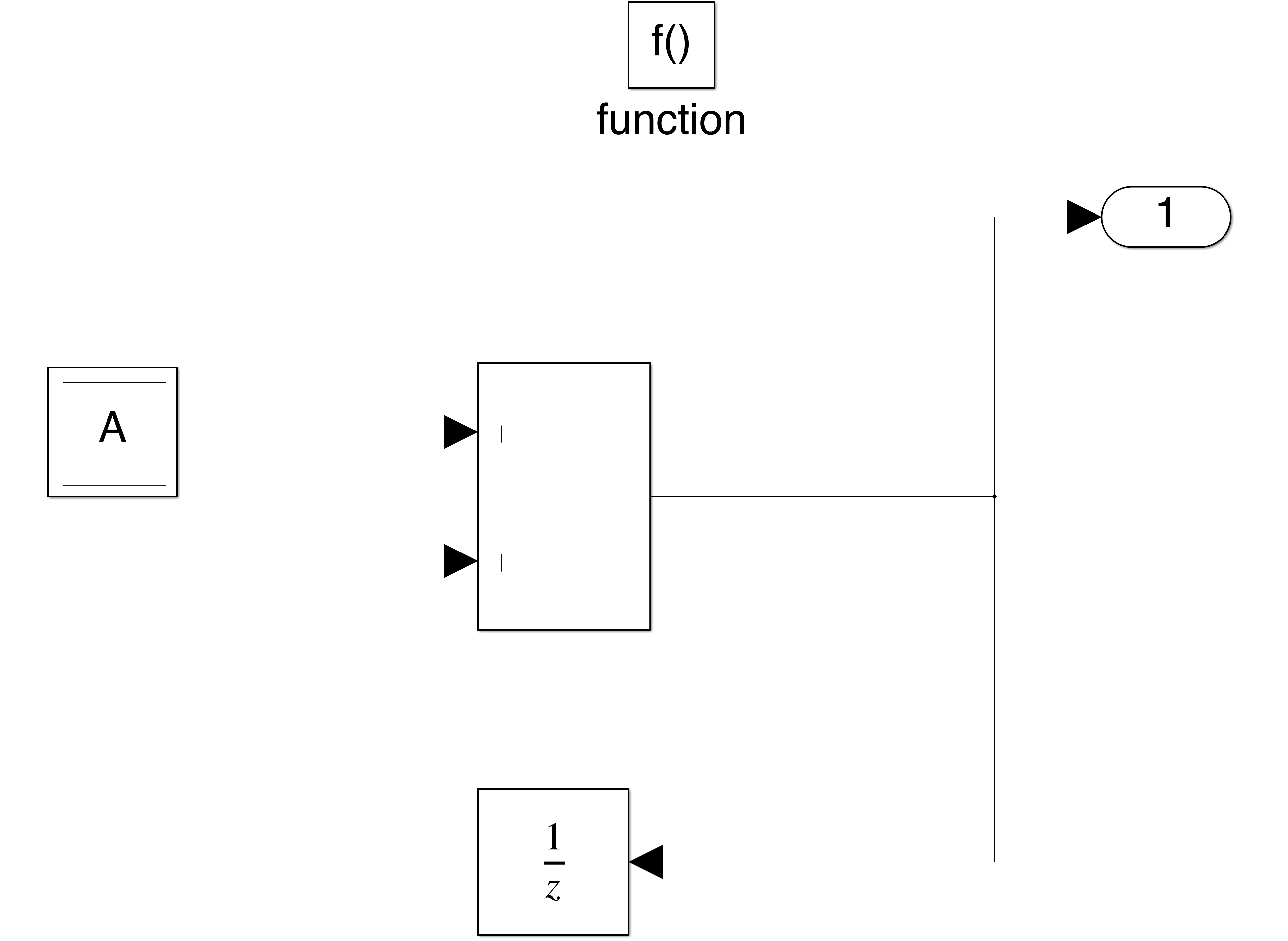} 
	\fcaption{A simple example of UDSM mutant.}
	\label{fig:udsm}
\end{Figure}

The third mutant operator is \emph{mRDSM}, and this operator removes an initialization value to a Data Store memory. In many programs, variables require an initial value before they can use properly. For our example, Runnable1\_subsystem is such a process of initializing Data Store A; then, we remove the Data Store Write block in Runnable1\_subsystem. The Simulink model can still run simulations without any issues however the output of the simulation only yields a single value. 

Mutant operator $mRSM$ adds a new reference to shared memory. Figure \ref{fig:mrsmexample} shows the block diagrams of a Simulink model with three subsystems and they are mapped to two tasks. The model has a DSM block $A$ in the root-level system. There is a Data   Store Write block inside subsystems $Task\_B1$ and a Data Store Read block in $Task\_B2$. The period of $Task\_A$ is  $5ms$ and the period of $Task\_B$ is $10ms$. To implement the $mRSM$, we add a Data Store Read block to the $Task_A$ subsystem which shows in Figure \ref{fig:mrsm}.  In the original example, $Task\_A$ executes first then $Task\_B1$ writes $A$ and $Task\_B2$ reads $A$. The mutant program has the same execution order as the original model. However, when the Data Store Read block in $Task\_A$ executes, the block reads data from an uninitialized data store or a previous instant of $Task\_B1$ as $Task\_B$ has not executed yet or has been executed previously.

\begin{Figure}
	\includegraphics[width=.99\columnwidth]{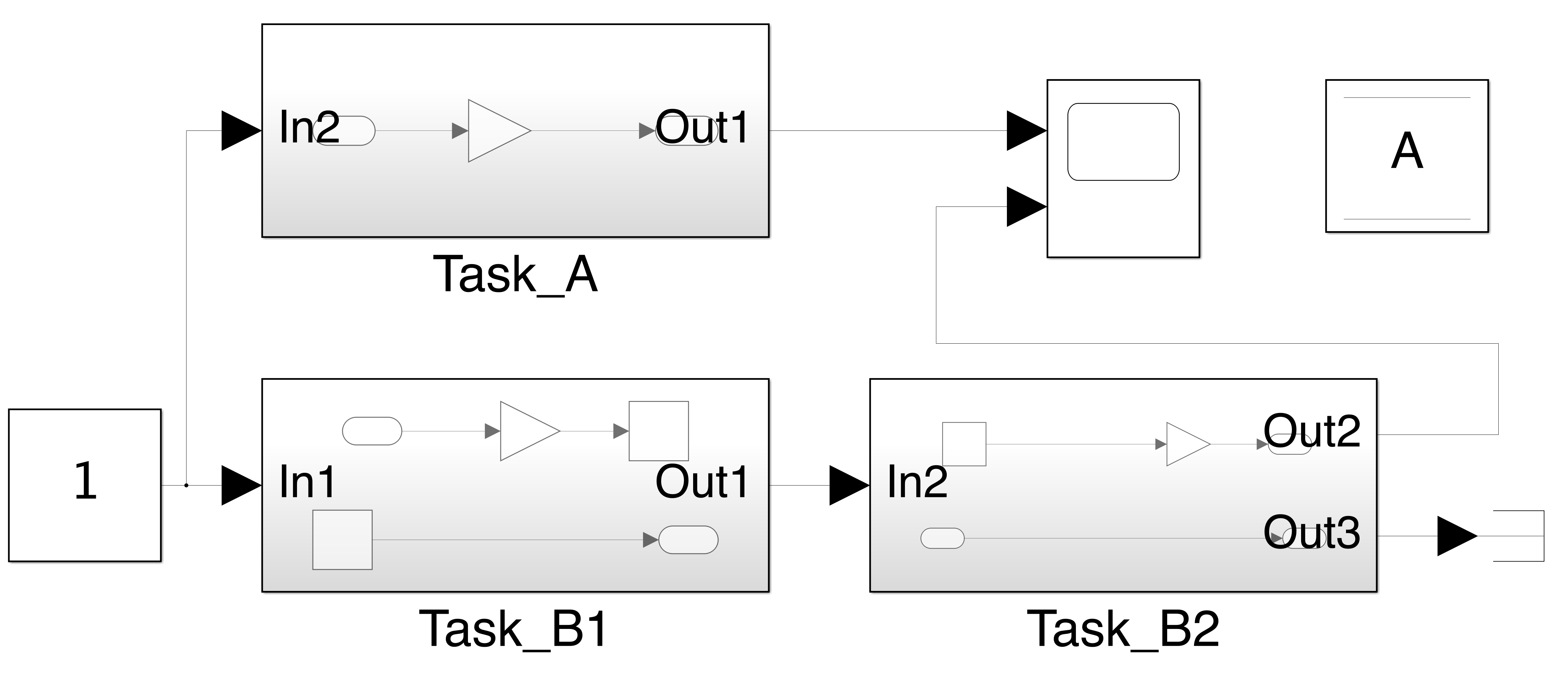} 
	\fcaption{A simple Simulink model.}
	\label{fig:mrsmexample}
\end{Figure}

\begin{Figure}
	\includegraphics[width=.99\columnwidth]{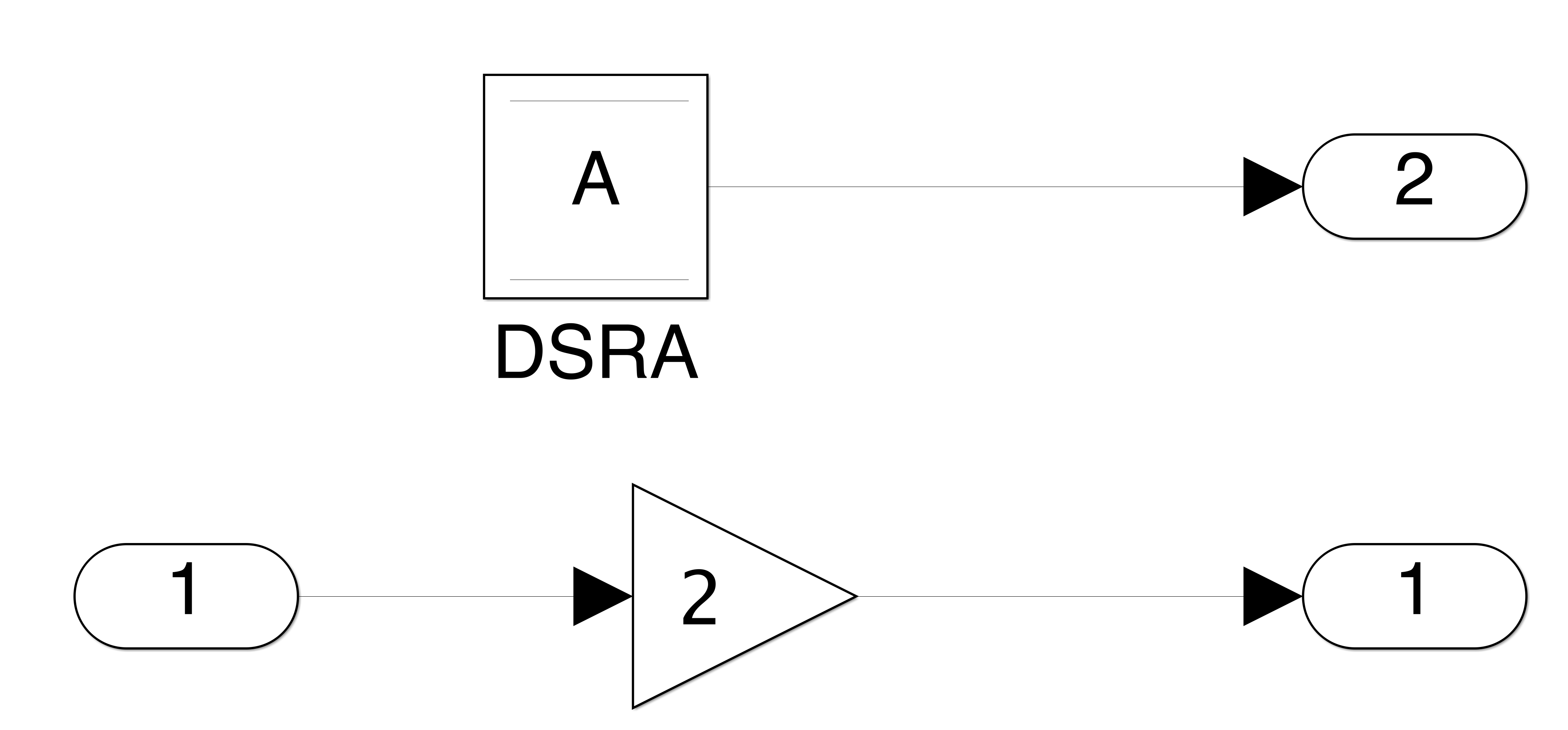} 
	\fcaption{ An example of mRSM mutant operator. Adding a Data Store Read block to Task\_A block.}
	\label{fig:mrsm}
\end{Figure}

Mutant operator $mRMSMR$ deletes a reference to shared memory. In Figure \ref{fig:mrsmexample}, $Task\_B2$ has a reference to a DSM block $A$ in the root-level system. To implement the $mRMSMR$, we delete the Data Store Read block in the $Task_B2$ subsystem. In the mutant program, $Task\_B2$ has a constant output value of zero since there is no reference.

\section{Evaluation Phase}

In the previous section, we describe how a model scheduler SimSched can validate the real-time context during a simulation, and we utilize mutation testing to evaluate SimSched. In this section, we perform experiments to demonstrate the use of our mutation testing framework to evaluate the quality of SimSched and Stateflow schedulers in scheduling tasks in real-time systems.

\subsection{Evaluation Process}
To validate the proposed mutation operators, we apply them to ML/SL models. We separate the evaluation process into two parts base and extension, according to the ability of ML/SL. We apply the first-order mutants (FOMs) \cite{Jia2011} to ML/SL models to generate a mutant, which means we generate a mutant by using a mutation operator only once.

\subsubsection{Base Case}
In the base case, we examine the simulation results of the original models and the SimSched models and their mutants. An original model $\mathcal{M}$  is an ML/SL model scheduled by Stateflow scheduler; A SimSched model $\mathcal{M'}$ is an original model scheduled by SimSched; The mutants ($\mathcal{M_\mu}$ or $\mathcal{M'_\mu}$) are either original model or SimSched models mutated by one of our mutation operators. Figure \ref{fig:modelmutation} shows the schematic diagram of our mutants generation process. We use the simulation result of $\mathcal{M}$ as a comparison baseline, and then we compare the baseline with every other simulation result of $\mathcal{M_\mu}$, and $\mathcal{M'_\mu}$. We examine the comparison result to see if the result reaches a verdict failure during model simulation. We say a mutant is killed if a verdict of failure is reached.

    \begin{Figure}
        \includegraphics[width=.6\columnwidth]{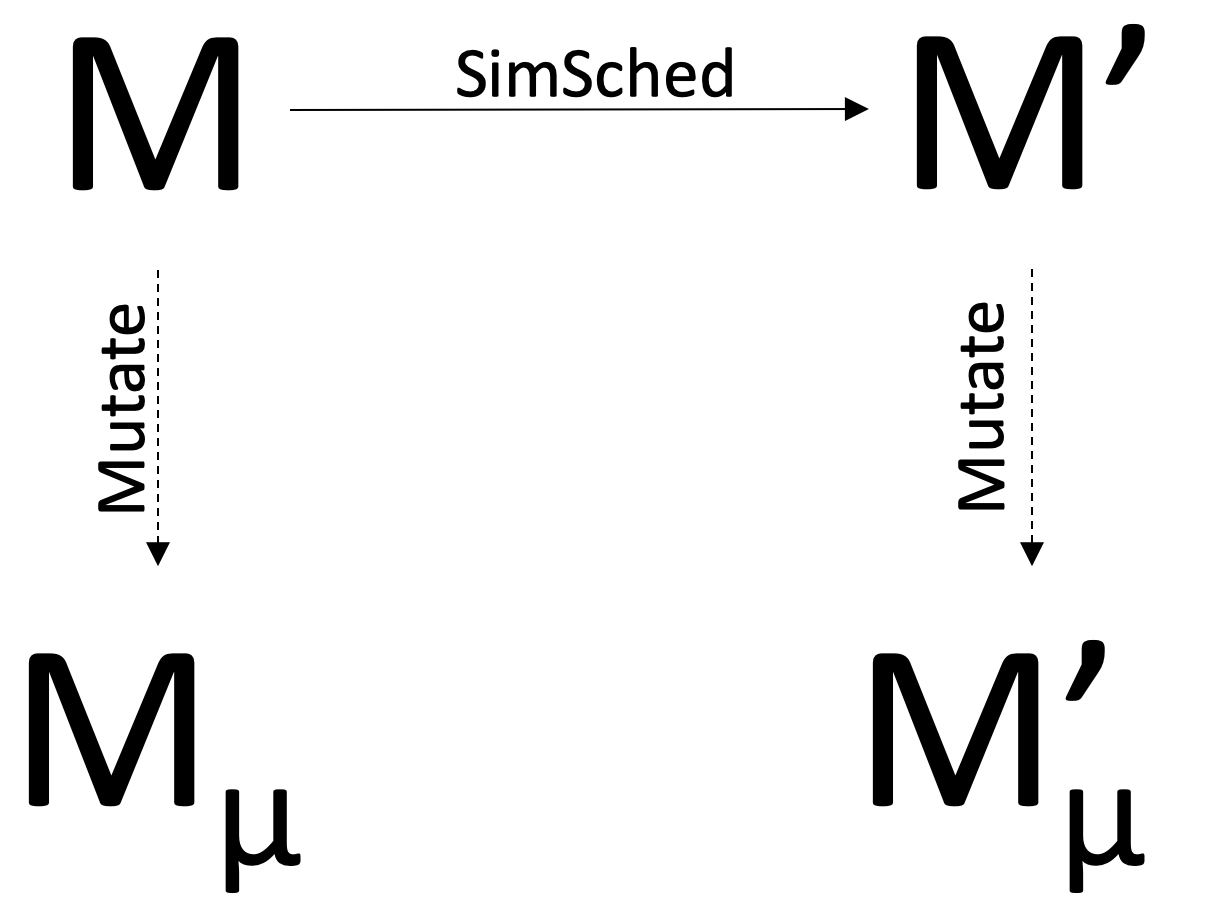} 
	    \fcaption{Schematic diagram of the model mutants generation process.}
	    \label{fig:modelmutation}
    \end{Figure}

    \begin{Figure}
        \includegraphics[width=.99\columnwidth]{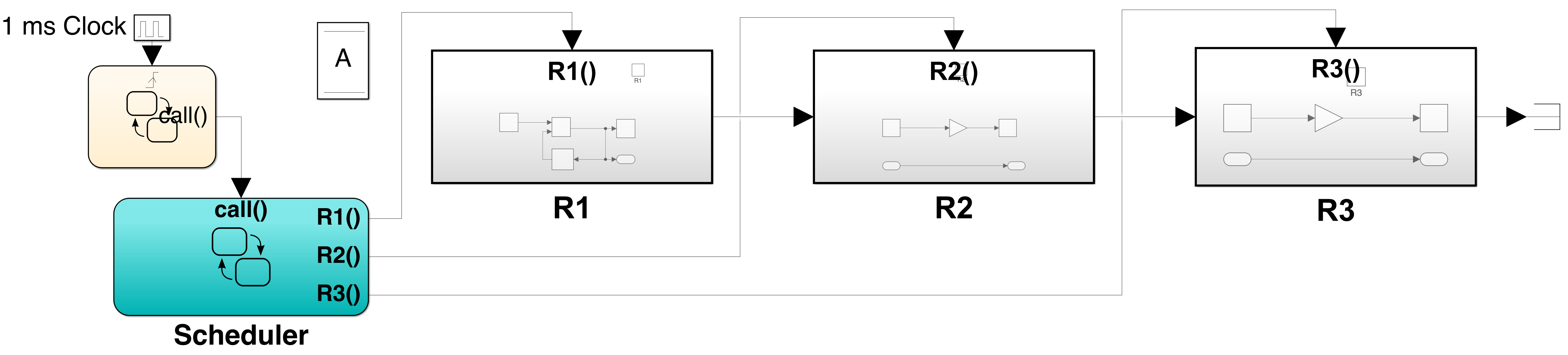} 
	    \fcaption{Simple evaluation example scheduled by Stateflow scheduler.}
	    \label{fig:evaluation}
    \end{Figure}

    We use a simple example shown in Figure \ref{fig:evaluation} to explain the base case evaluation process. This example is an original model. We replace the Stateflow scheduler with a SimSched scheduler to form a SimSched model. We generate mutants for both the original and SimSched models by a specific mutation operator, e.g., mDTPER, to decrease the task period. Then we run the simulation for both mutants and analyzed the results to see if there is any errors. If the simulation result of $\mathcal{M_\mu}$ or $\mathcal{M'_\mu}$ is different from the original model and shows a verdict failure, then we say the mutant is killed. 
    
    In this example, we have three runnables $R_1, R_2, R_3$ and they are mapped to two tasks $T_1, T_2$. $R_1$ is mapped to $T_1$ and $R_2,R_3$ are mapped to $T_2$. The period of $T_1$ is $3ms$ and The period of $T_2$ is $6ms$. The execution time of each runnable is $1ms$. The simulation result of the $\mathcal{M}$ is shown in Figure \ref{fig:a} and it shows each runnable output is a rising non-interlaced polyline.   We apply the \emph{mDTPER} mutation operator as decreasing $1ms$ to both the original model and SimSched model to generate mutants. The task $T_1$ in the mutants has a period of $2ms$. The simulation result of these simulations is shown in Figure \ref{fig:b} and Figure \ref{fig:c}. The simulation result of $\mathcal{M'_\mu}$ is different from the result of $\mathcal{M}$, and it shows the output of $R_2$ and $R_3$ are two rising interlaced polylines because SimSched can simulate the execution time and preemption. $T_1$ preempts $T_2$ in the SimSched mutant model to yield an alternative execution trace, and we say a verdict fail is reached. However, the simulation result of $\mathcal{M_\mu}$ is similar to the result of $\mathcal{M}$. Thus, the \emph{mDTPER} mutant is killed to the $\mathcal{M'_\mu}$ and is alive to the $\mathcal{M_\mu}$. We can not apply this means to all mutation operators due to the nature of ML/SL. We combine this method and the following method to evaluate the mutation operators.  

    \begin{Figure}
        \includegraphics[width=.6\columnwidth]{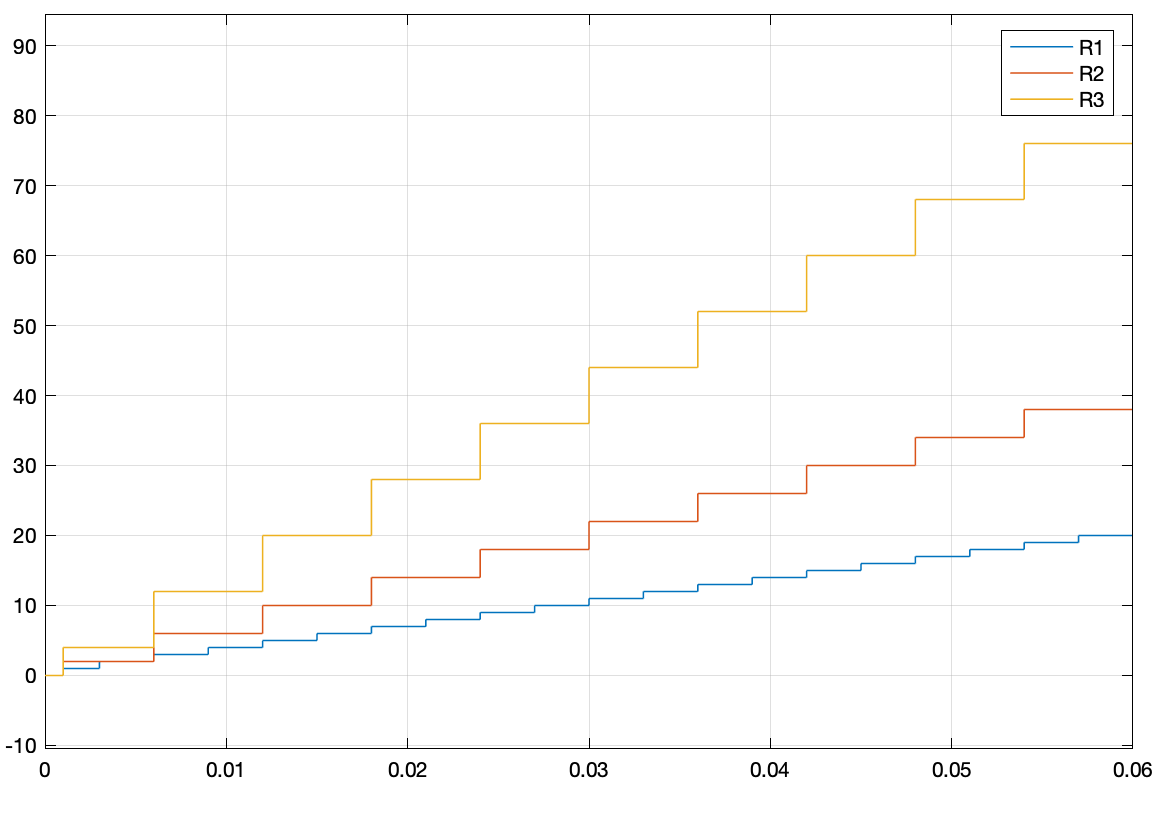}
        \fcaption{$\mathcal{M}$ simulation result.}
        \label{fig:a}
    \end{Figure}
    \begin{Figure}
        \includegraphics[width=.6\columnwidth]{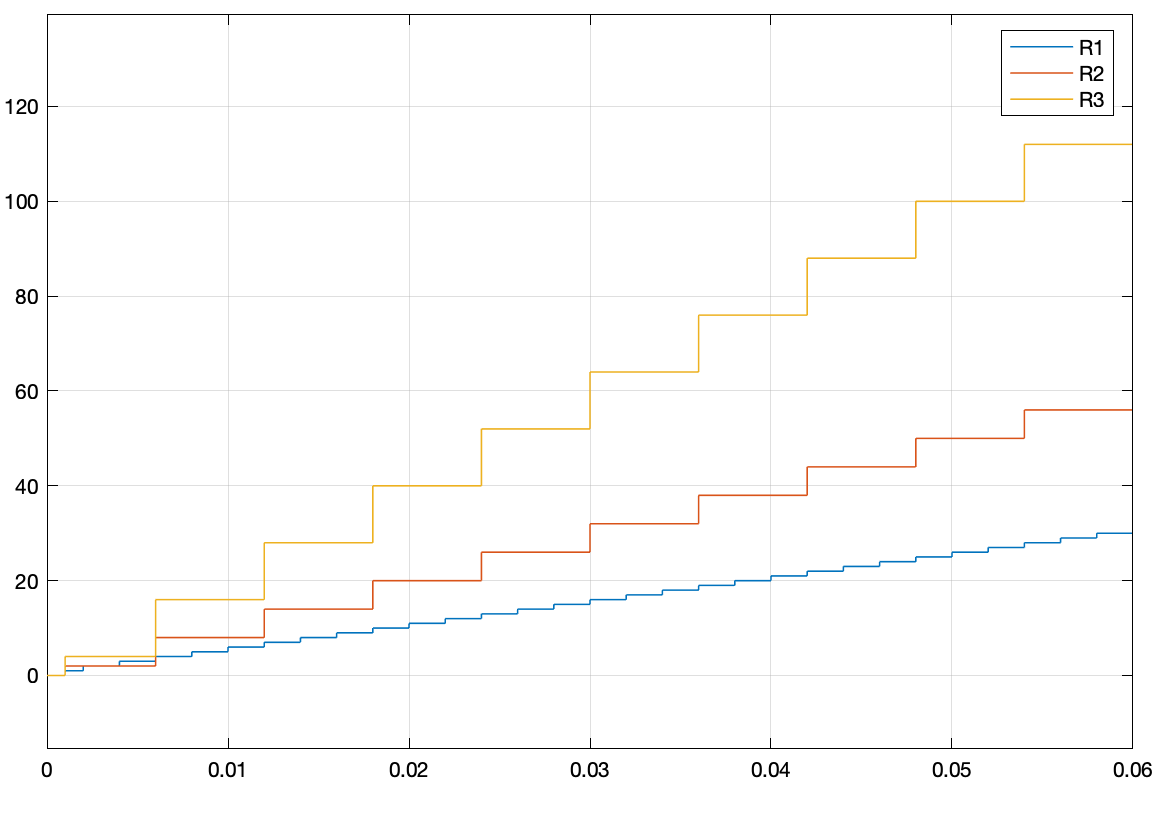}
        \fcaption{$\mathcal{M_\mu}$ simulation result.}
        \label{fig:b}
    \end{Figure}
    \begin{Figure}
        \includegraphics[width=.6\columnwidth]{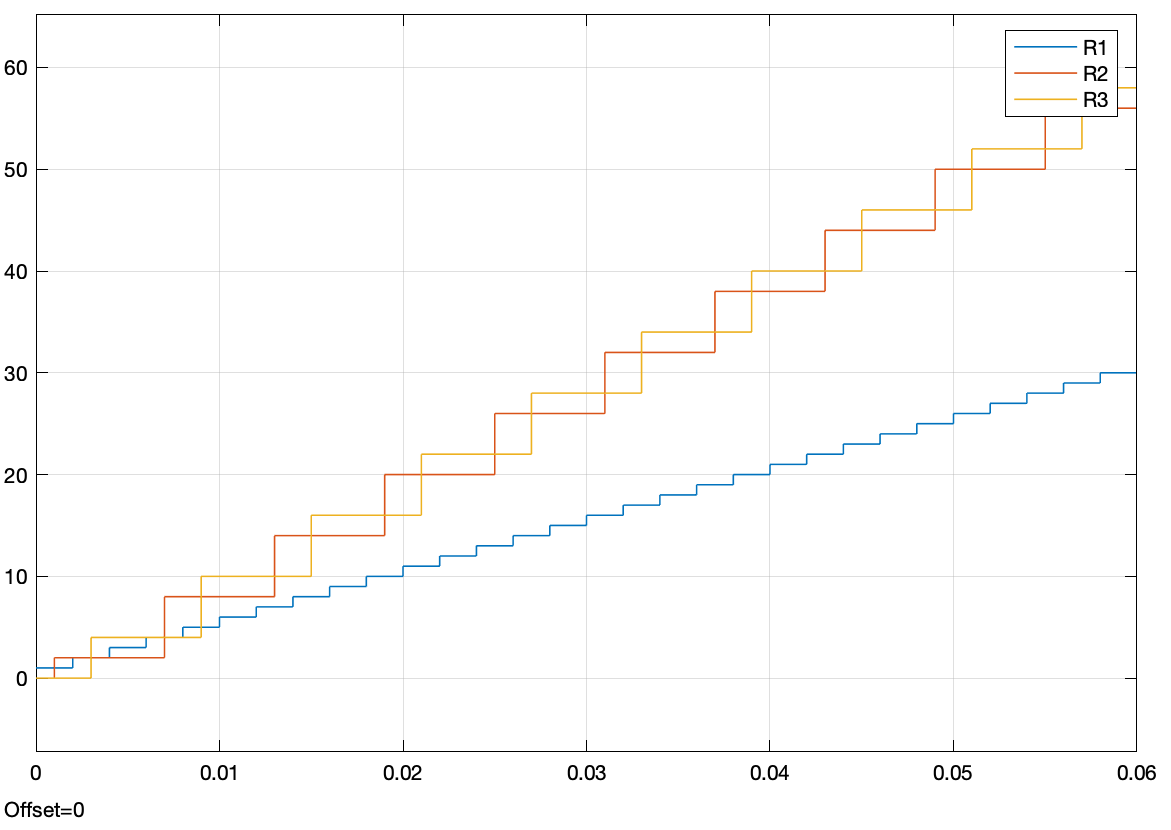}
        \fcaption{$\mathcal{M'_\mu}$ simulation result}
        \label{fig:c}
    \end{Figure}
    
    \subsubsection{Extension}
    To evaluate the rest of the mutation operators, we implement a mutation generator with additional functionalities to assist the validation process. One feature is to check the mutant model's schedulability at the given set of tasks configuration to decide if all task deadlines are met. The other function is to check the data access sequence. If there is a DataStore block in the mutated model, every read or write to this DataStore block is recorded. Then we use this mutated model data access sequence to compare with the original model data access sequence. The mutation generator is implemented as a Matlab script written in m-file.   

    The validation process takes a Stateflow scheduled ML/SL model and a test specification as input. The test specification specifies which mutation operator to use. A mutant generator applies the specified mutation operator to the ML/SL model via SimSched and generates a mutant. The mutant generator then executes the simulation both for the original model and the mutated model using the additional functionalities to analyze the simulation.  If the analysis shows at least one task misses its deadline in a mutated model, then we say a mutant is killed. Or at least one variable comparison result of the DataStore access sequence is unmatching, and then we say a mutant is killed; otherwise, we report the mutant is benign. 

    \begin{Figure}
		\includegraphics[width=0.9\textwidth]{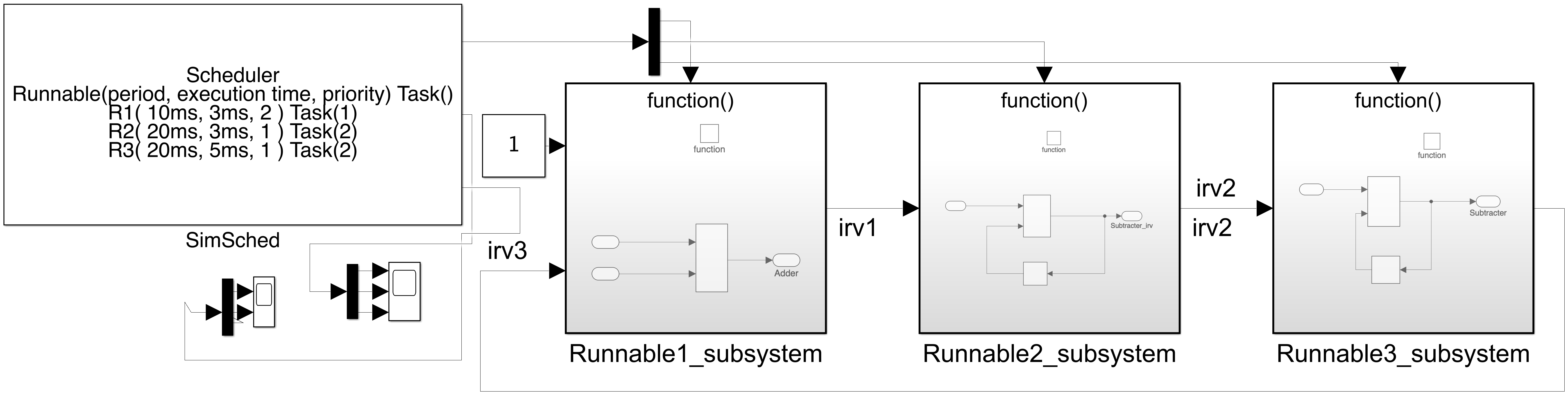}
		\fcaption{A simple example of using Model Scheduler to schedule AUTOSAR SW-Cs.}
		\label{fig:msSimpleExample}
	\end{Figure}

    We use an example shown in Figure \ref{fig:msSimpleExample} to explain the validation process. It has three runnables and is mapped to two tasks, $R_1$ map to $T_1$, $R_2$, and $R_3$ map to $T_2$. The period of task $T_1$ is $10 ms$, and $T_2$ is $20 ms$, every runnable's execution time is $3 ms$. There is a DataStore block named $A$ as a shared variable in this example model. If we apply the period mutation operator \emph{mDTPER} $\rho_i-\delta$ where $i-1$ and $\delta=6$ to this model to decrease the period of $T_1$ and generate a mutant, run it. The analysis result shows the $T_2$ missed deadline, then we say this mutant is killed. If we apply the execution time mutation operator \emph{mITET} $c_i+\delta$ where $i=1$ and $\delta=3$ to this model to increase the execution time for $T_1$ and generate a mutant. The DataStore access sequence of the original model is a pattern of $WRWR$ where $ W $ represents a \emph{write} to the shared variable, and $ R $ represents a \emph{read} to the shared variable. The mutant generates a different sequence, which is $WRWWR$. It is because the $T_1$ has a longer execution time than the original model, and it preempts $T_2$ during the execution of $T_2$. Hence, there is one more $W$ in the DataStore access sequence.

\subsection{Experiments}
We employ two examples to demonstrate the use of our mutation testing framework. We first explain the two examples in detail. We then apply the mutation operators to the two models scheduled by both the Stateflow scheduler and SimSched.

	\begin{Figure}
	    \includegraphics[width=.99\columnwidth]{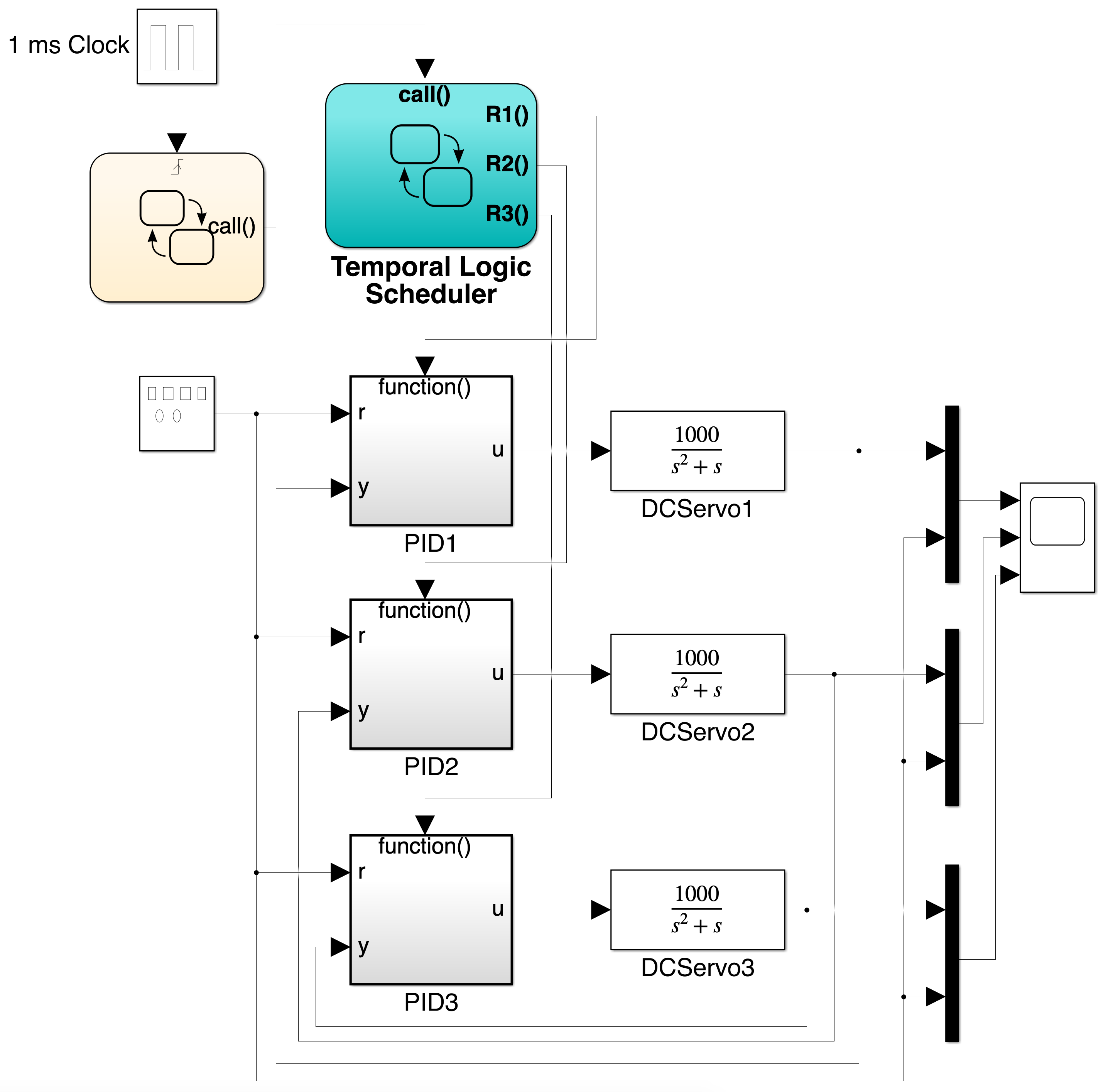} 
    	\fcaption{ The three-servo example adapted from \cite{Cremona2015} with Stateflow scheduler.}
    	\label{fig:MSthreeServossf}
    \end{Figure}

\subsubsection{Three Servos Model}
We adapt an example from the TrueTime \cite{Henriksson2003} example library, which shows a possible implementation of a three-servo PID control system. The example is shown in Figure \ref{fig:MSthreeServossf} with a Stateflow scheduler. In this example, three DC servos are modeled by a continuous-time system, and three PID controllers are implemented as three subsystems. We map three controller subsystems to three runnables $R_1$, $R_2$, and $R_3$ then they are mapped to tasks $T_1$, $T_2$, and $T_3$. The task periods are $T_1$=4 , $T_2$ = 5 and $T_3$ = 6 $ms$ respectively. Each task has the same execution time as 1$ms$. Task settings are shown in Table \ref{tbl:servoparameter_E_1}. The simulation result is shown in Figure \ref{fig:MSthreeServosoutputsf} based on the above task settings. The three graphs show the output of the motors using the three PID controllers when the corresponding task parameters are assigned accordingly. In the graph, the square wave is the reference input signal for the motors, where the computation delays are not taken into account. Three PID controllers are all smooth output signals as expected.

    \begin{Table}
	\tcap{Three Servo example settings.}
	\begin{tabular}{ c c c c c}
		\hline
		Task & Period & Execution  & Priority & Runnable \\ 
		 & ($ms$) & Time($ms$) &  &  \\ \hline
		$T_1$   & 4 & 1 & 3 & $R_1$ \\
		$T_2$   & 5 & 1 & 2 & $R_2$ \\
		$T_3$   & 6 & 1 & 1 & $R_3$ \\
		\hline
	\end{tabular}
	\label{tbl:servoparameter_E_1}
    \end{Table}

    \begin{Figure}
    	\includegraphics[width=.99\columnwidth]{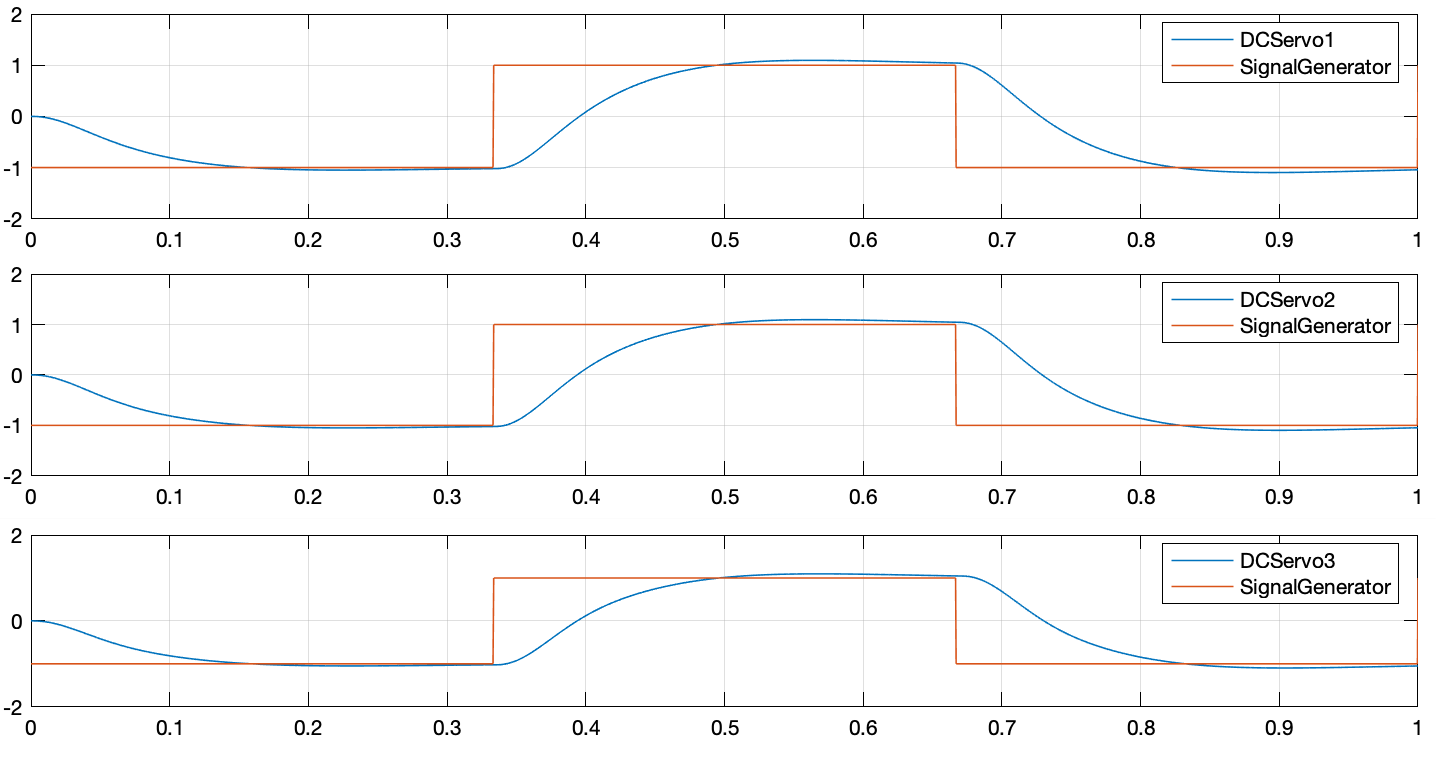} 
    	\fcaption{ The three servos example output with Stateflow scheduler.}
    	\label{fig:MSthreeServosoutputsf}
    \end{Figure}

  We replace the Stateflow scheduler with the SimSched scheduler, and the updated example is shown in Figure \ref{fig:MSthreeServos_E}. In this example, three DC servos have the same task setting as the Stateflow scheduler example. Each runnable has the same execution time as 1$ms$. The simulation result is the same as the Stateflow scheduler example based on the above task settings. There is no deadline missing for any task, so the simulation result shows every task has smooth control. Figure \ref{fig:MSthreeServosoutput1ms} shows the task active chart generated by SimSched. Every task has been executed within its own deadline.
 
     \begin{Figure}
    	\includegraphics[width=.99\columnwidth]{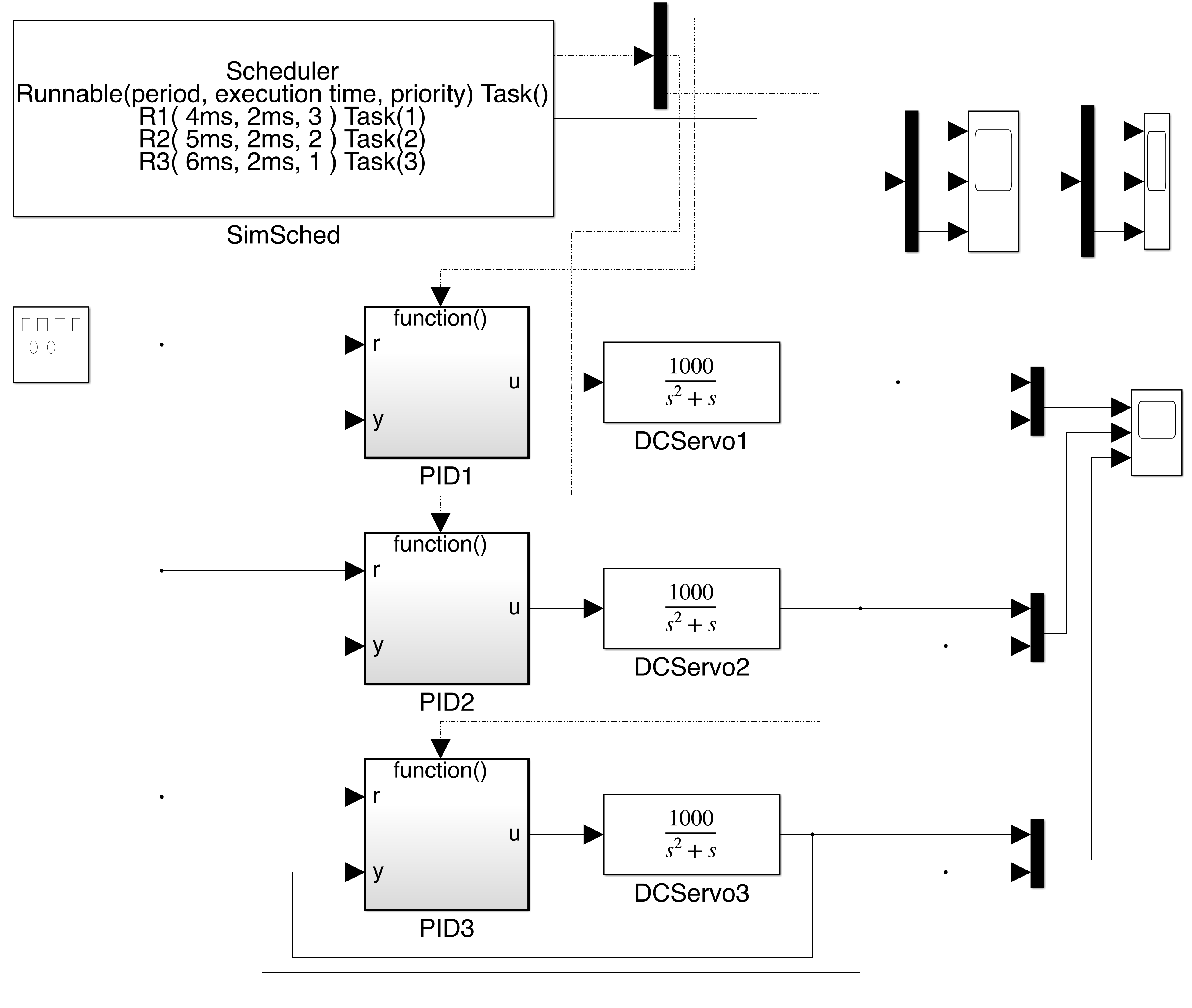} 
    	\fcaption{ The three servos example adapted from \cite{Cremona2015}.}
    	\label{fig:MSthreeServos_E}
    \end{Figure}

    \begin{Figure}
        \includegraphics[width=.99\columnwidth]{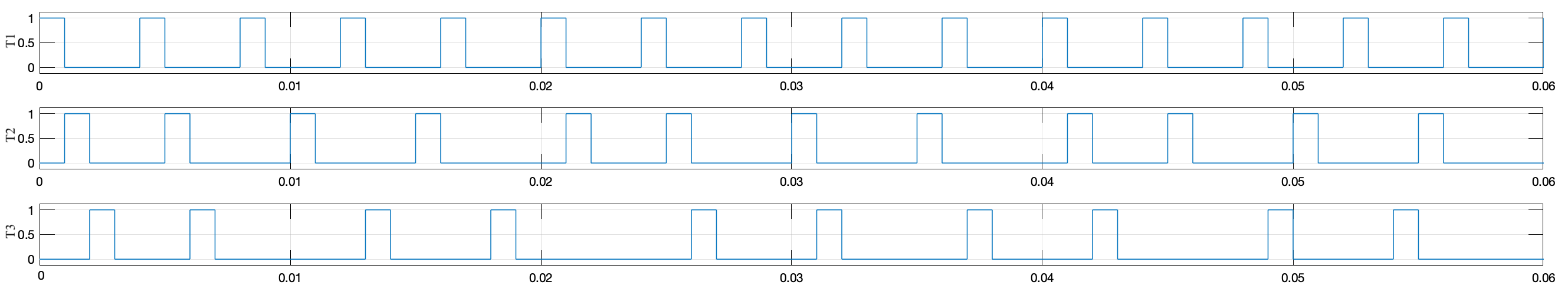} 
    	\fcaption{ The three servos example task active chart generated by SimSched.}
    	\label{fig:MSthreeServosoutput1ms}
    \end{Figure}
    
     \begin{Figure}
        \includegraphics[width=.6\columnwidth]{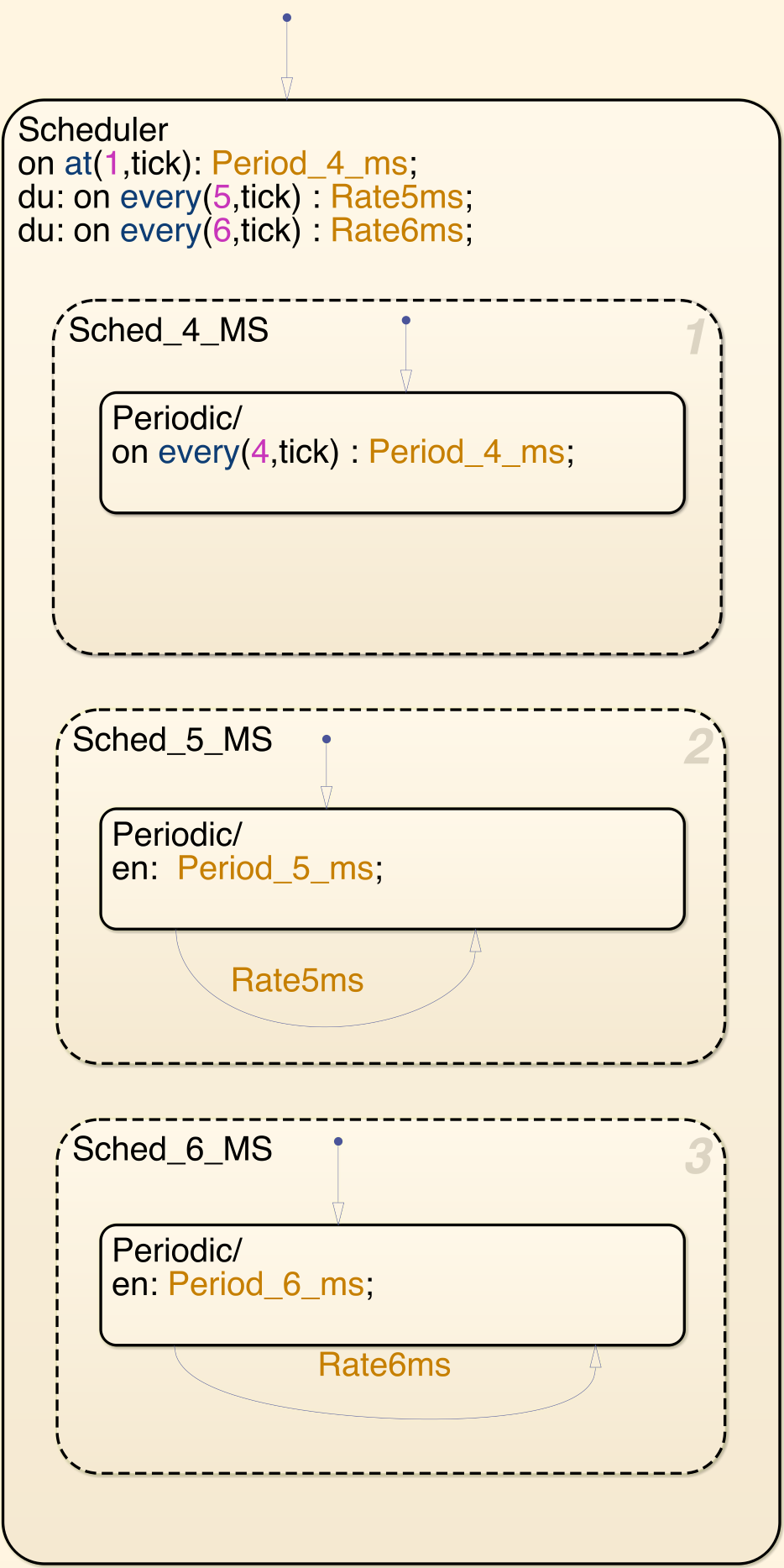} 
    	\fcaption{The adjusted Stateflow scheduler for \emph{mITO} mutation operator to increase $offset$ as 1$ms$ for DCServo1.}
    	\label{fig:Offsetscheduler}
    \end{Figure}
 
 Next step, we apply $mITO$, $mDTO$, $mITPER$, $mDTPER$,  $mARPREC$, $mRRPREC$,  $mITJ$, $mDTJ$ mutation operators to both Stateflow scheduler and SimShced examples to generate two versions of mutants with the same mutation operators. To apply some of the mutation operators to evaluate the Stateflow scheduler, we need to adjust the Stateflow scheduler so that it can be used on the generated mutants. Figure \ref{fig:Offsetscheduler} shows an example that is adjusted for the \emph{Offset} mutation operator. This example uses a temporal logic operator $at$ in the state to set the Offset parameter to generate a mutant for PID1 which runs at the period of $4ms$ in this example. This mutant increases \emph{Offset} as 1$ms$ for DCServo1 controlled by PID1. The mutant of the SimSched version can be easily generated by our model scheduler SimSiched.
    
    \begin{Figure}
        \includegraphics[width=.99\columnwidth]{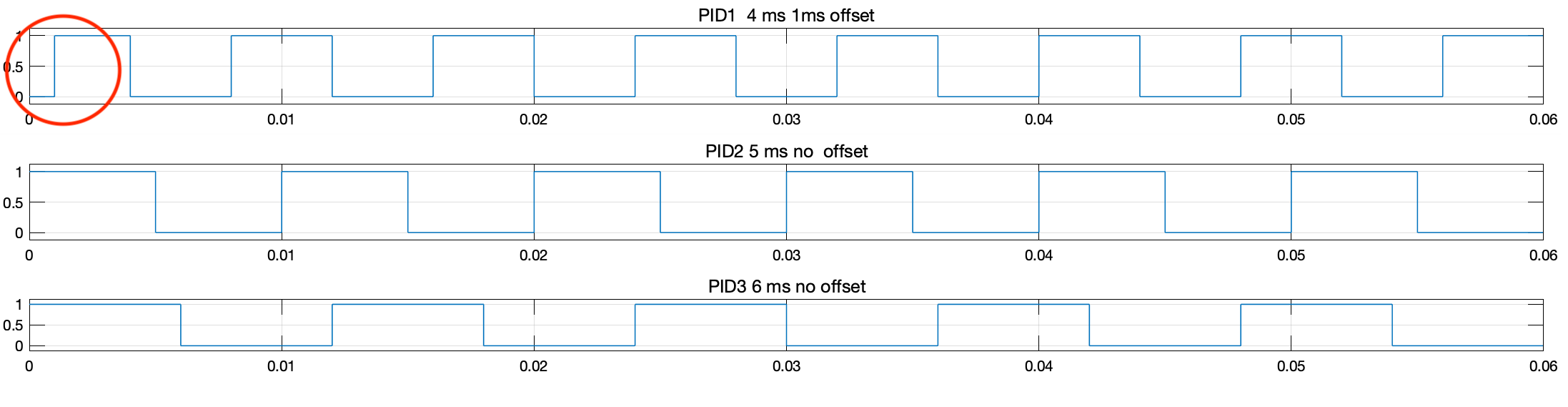} 
    	\fcaption{The Stateflow scheduled three-servo example task active chart after applying \emph{mITO} mutation operator to increase $offset$ as 1$ms$ for DCServo1.}
    	\label{fig:sf_offset}
    \end{Figure}
    
  We run simulations for both versions of the mutants generated by \emph{Offset} mutation operator. Both mutant versions' output of three servos is the same as shown in Figure \ref{fig:MSthreeServosoutputsf}. The only difference occurs at the beginning of the simulation but it does not affect the smooth control of DCServos. We can see the difference from the following comparison. Figure \ref{fig:sf_offset} shows the Stateflow scheduled task active chart after applying \emph{mITO} mutation operator to increase \emph{offset} as 1$ms$ for DCServo1. Before applying the mutation operator, every task is released at time 0. After applying the offset mutation operator, Task 1 is delayed by $1ms$ shown on the top of the figure. Task 2 and Task 3 are both released at time 0. Figure \ref{fig:ss_offset} shows the SimSched scheduled three servos example task active chart after applying \emph{mITO} mutation operator to increase \emph{offset} as 1$ms$ for DCServo1. There are three output signals representing three tasks from top to bottom T1, T2, and T3. As T1 has a 1$ms$ offset, Task 2 is executed first as shown in the figure the second line starts at time 0. Because the SimSched scheduler has the execution time parameter, Task 2 is executed at time 1 and Task 3 at time 2 respectively. 

    \begin{Figure}
        \includegraphics[width=.99\columnwidth]{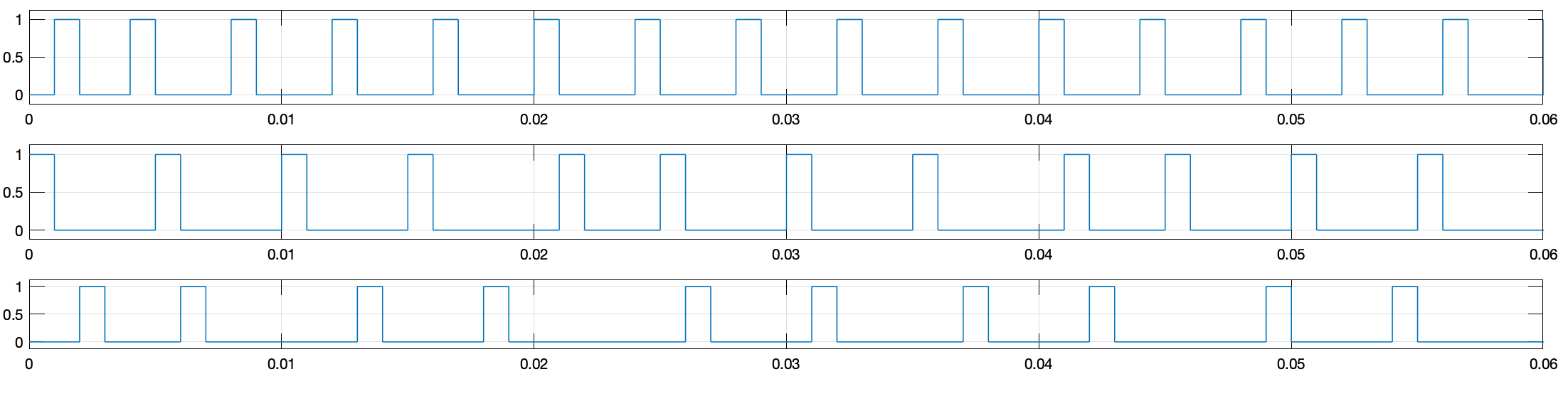} 
    	\fcaption{The SimSched scheduled three servos example task active chart after applying \emph{mITO} mutation operator to increase \emph{Offset} as 1$ms$ for DCServo1.}
    	\label{fig:ss_offset}
    \end{Figure}

    We use a similar approach to apply \emph{Period} mutation operator to the three-servo example and generate mutants for both the Stateflow scheduler model and SimSched model. We use two mutation configurations to show the similarities and differences between the two schedulers. The first configuration is [1,5,6]. It means Task 1 has a period of 1$ms$ and the period of Task 2 and Task 3 keep the same as 5$ms$ and 6$ms$, respectively. Figure \ref{fig:sf_period} shows the Stateflow scheduler mutant simulation result. Because Task 1 has a 1$ms$ period, it has too many times of calculations, and the output value is out of the chart. Task 2 and Task 3 keep the same output as before. Figure \ref{fig:ss_period} shows the SimSched mutant simulation result. The output of Task 1 is the same as the Stateflow scheduler mutant. However, Task 2 and Task 3 are different from the Stateflow one. Because the SimSched takes the execution time into account, Task 1 has the highest priority and has an execution time of 1$ms$, and Task 1 always runs during the simulation. Task 2 and Task 3 always are preempted by Task 1 because they have a lower priority than Task 1.  
    
    \begin{Figure}
        \includegraphics[width=.99\columnwidth]{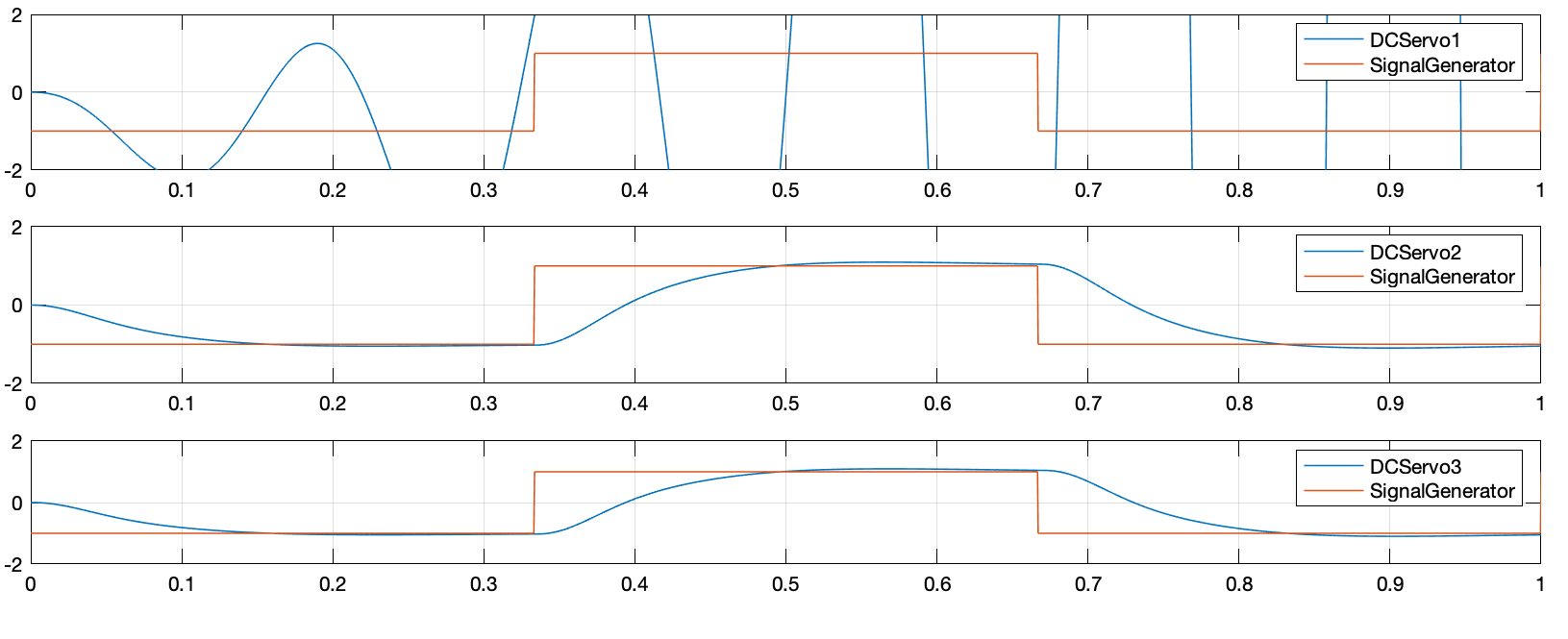} 
    	\fcaption{The Stateflow scheduled three servos example output after applying \emph{mDTPER} mutation operator to decrease $period$ as 1$ms$ for DCServo1.}
    	\label{fig:sf_period}
    \end{Figure}

    \begin{Figure}
        \includegraphics[width=.99\columnwidth]{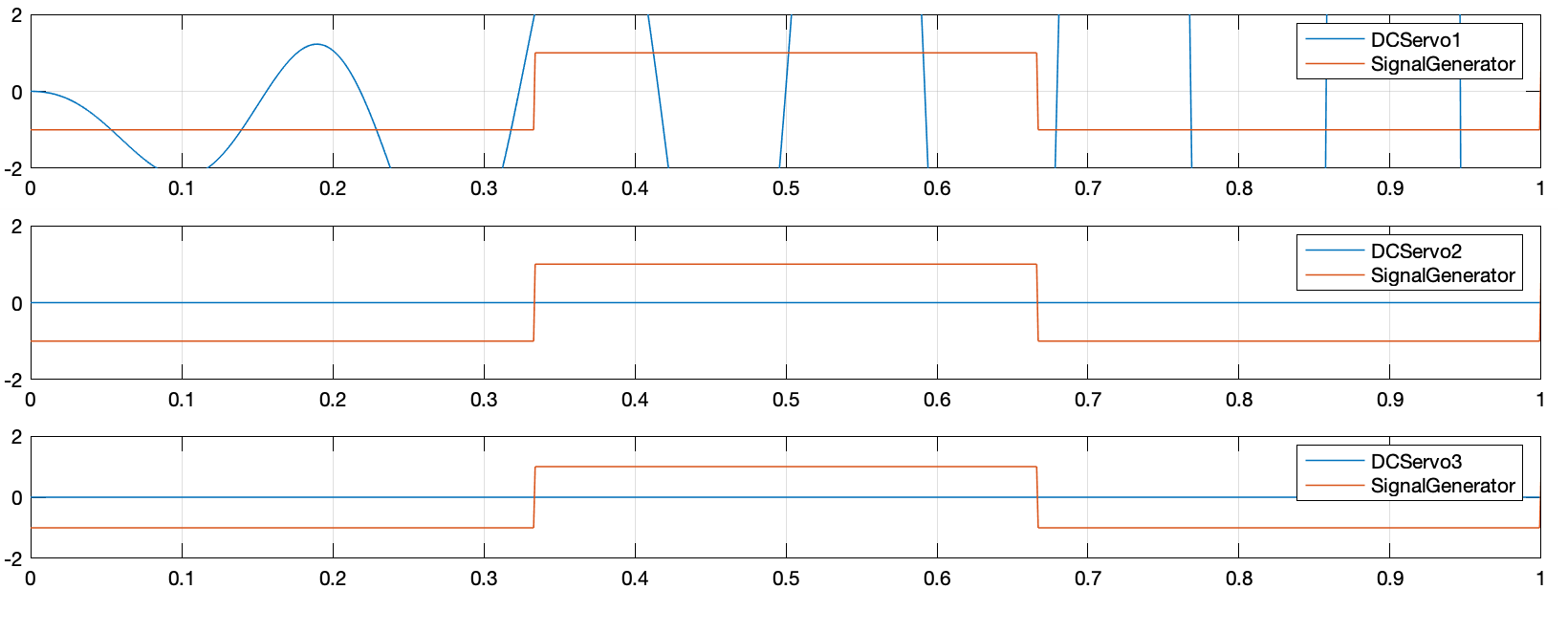} 
    	\fcaption{The SimSched scheduled three servos example output after applying \emph{mDTPER} mutation operator to decrease $period$ as 1$ms$ for DCServo1.}
    	\label{fig:ss_period}
    \end{Figure}
    
    The second configuration is [13,5,6]. It means Task 1 has a period of 13$ms$ and the period of Task 2 and Task 3 keep the same as 5$ms$ and 6$ms$, respectively. Figure \ref{fig:sfss_period} shows the Stasflow scheduler mutant simulation result and the SimSched mutant has the same output as shown in the figure. Because Task 1 has a 13$ms$ period, it has fewer computations than the original model. Although the output behavior looks like Task 1 misses its deadline in the figure, every execution of Task 1 meets its deadline and it is executed as scheduled.
    
    \begin{Figure}
        \includegraphics[width=.99\columnwidth]{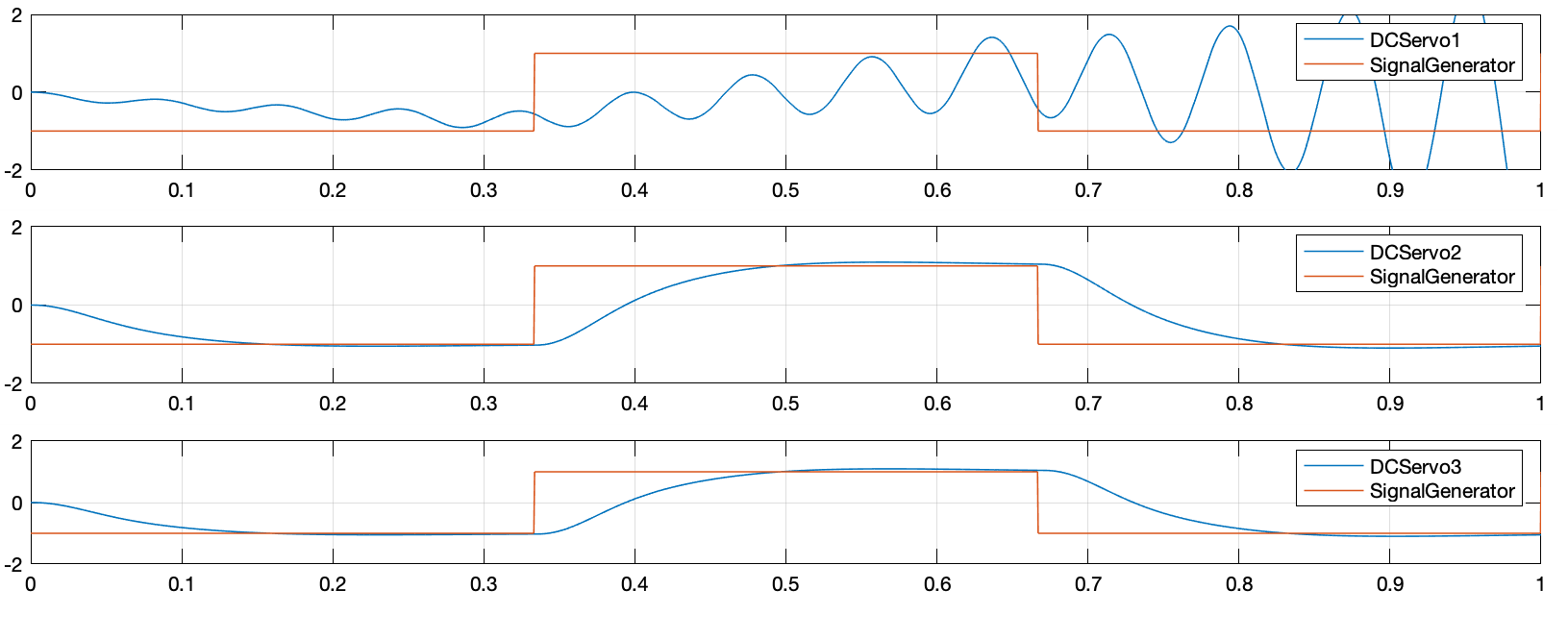} 
    	\fcaption{The Stateflow scheduler three servos example output after applying \emph{mITPER} mutation operator to increase $period$ as 13$ms$ for DCServo1.}
    	\label{fig:sfss_period}
    \end{Figure}

    We generate mutants for $mARPREC$ and $mRRPREC$ operators by setting each parallel state's execution order in the Stateflow scheduler model and configuring the parameters and connections for the SimSched model.  The two mutants' simulation results are the same as the original model, except that each task's execution order is different from the original model.
    
    We generate mutants for $mITJ$ and $mDTJ$ operators by adapting the Stateflow scheduler in the model and configuring the parameters in the SimSched model. We set the configuration as [1,0,0]. It means only Task 1 has a jitter as 1$ms$. Figure \ref{fig:ss_jitter} shows the SimSched scheduler three servos example task active chart after applying \emph{mITJ} mutation operator to increase $jitter$ as 1$ms$ for DCServo1. Because Task 1 has 1$ms$ jitter, Task 2 is executed first then Task1 and Task3.  
    
    \begin{Figure}
        \includegraphics[width=.99\columnwidth]{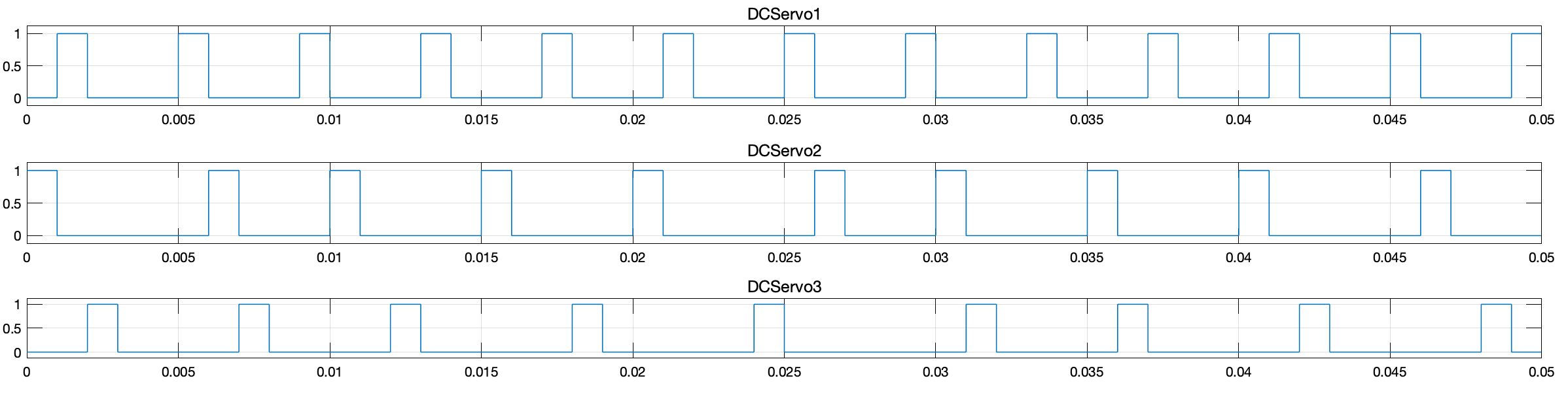} 
    	\fcaption{The SimSched scheduler three servos example task active chart after applying \emph{mITJ} mutation operator to increase $jitter$ as 1$ms$ for DCServo1.}
    	\label{fig:ss_jitter}
    \end{Figure}

 We only apply the execution time operator to the SimSched model due to the lack of support for the Stateflow scheduler. Figure \ref{fig:ss_executiontime} shows the effect output of $mITET$ mutation operator. We set $c_1=3ms$ using the $mITET$ mutation operator for Task 1 to generate a mutant. The output of Task 3, shown as DCservo 3 at the bottom of the figure, is a curly wave. It is an unstable control due to the preemption by $T_1$ and $T_3$ missing its deadline. Figure \ref{fig:ss_ettask} shows the task preemption effect. Task 1 takes 3$ms$ to execute and Task 2 takes 1$ms$ to execute. After the execution of Task 1 and Task 2, Task 3 should be executed; however, it is the time that Task 1 is scheduled to run. Task 1 has a higher priority, so Task 3 is preempted by Task 1. The first instance of Task 3 is executed at 19$ms$ so Task 3 does not have a smooth control signal output as the other tasks. Although some task preemptions occur in Task 2, it does not miss enough deadlines to significantly affect the output. Task 2 still has a smooth output signal as shown in Figure \ref{fig:ss_executiontime} as the second chart.

    \begin{Figure}
        \includegraphics[width=.99\columnwidth]{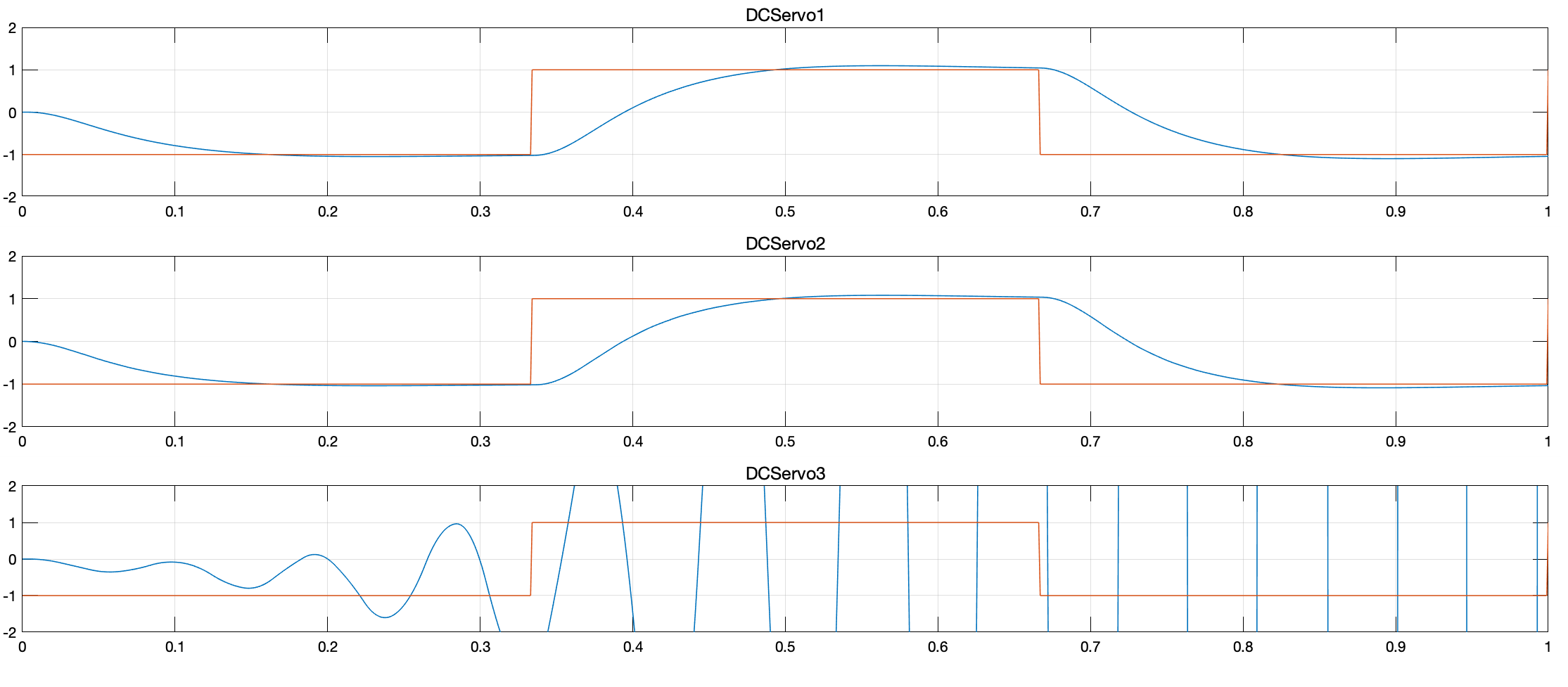} 
    	\fcaption{The SimSched scheduler three servos example signal output after applying \emph{mITET} mutation operator to increase $execution time$ to 3$ms$ for DCServo1.}
    	\label{fig:ss_executiontime}
    \end{Figure}
 
     \begin{Figure}
        \includegraphics[width=.99\columnwidth]{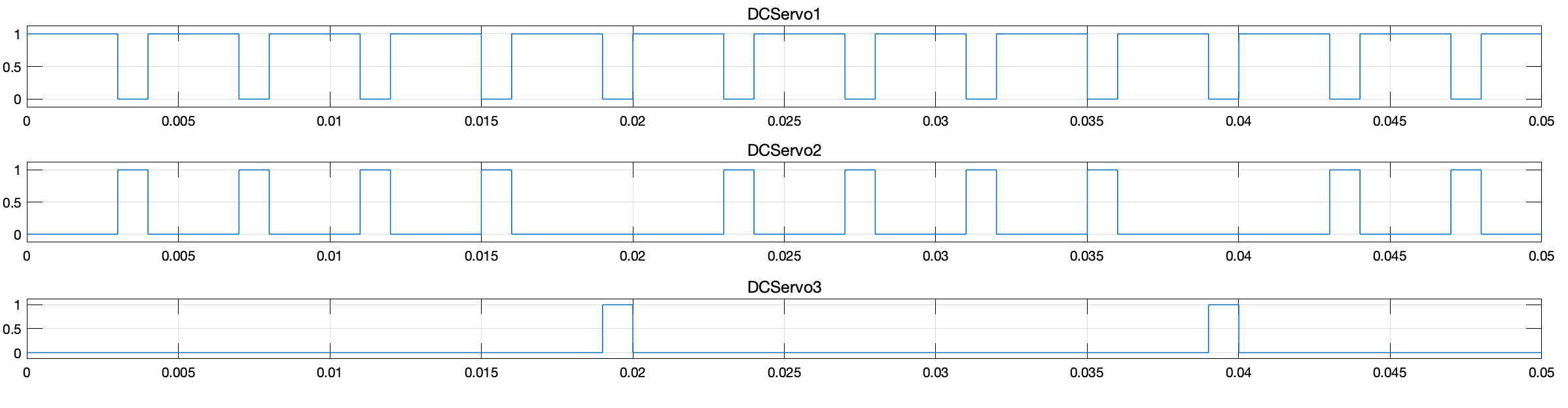} 
        \fcaption{The SimSched scheduler three servos example task active chart after applying \emph{mITET} mutation operator to increase $execution time$ to 3$ms$ for DCServo1.}
        \label{fig:ss_ettask}
    \end{Figure}

    \subsubsection{Throttle Position Control Model}

    We adopt an AUTOSAR software component Simulink model from Mathworks shown in Figure \ref{fig:casestudy}. It implements a throttle position control system for an automobile and contains three sensors, one monitor, one controller, and one actuator. They are implemented as six subsystems and mapped to six runnables $TPSSecondary$, $Monitor$, $Controller$, $Actuator$,  $APPSnsr$ and $TPSPrimary$ then they are mapped to tasks $T_1$, and $T_2$. The task periods are $T_1$=5$ms$ and $T_2$ = 10 $ms$ respectively. Each runnable has the same execution time of 1$ms$. Task settings are shown in Table \ref{tbl:throttleparameter}. This example uses seven DataStoreMemory blocks to access the shared resources.

    \begin{Figure}
    	\includegraphics[width=.99\columnwidth]{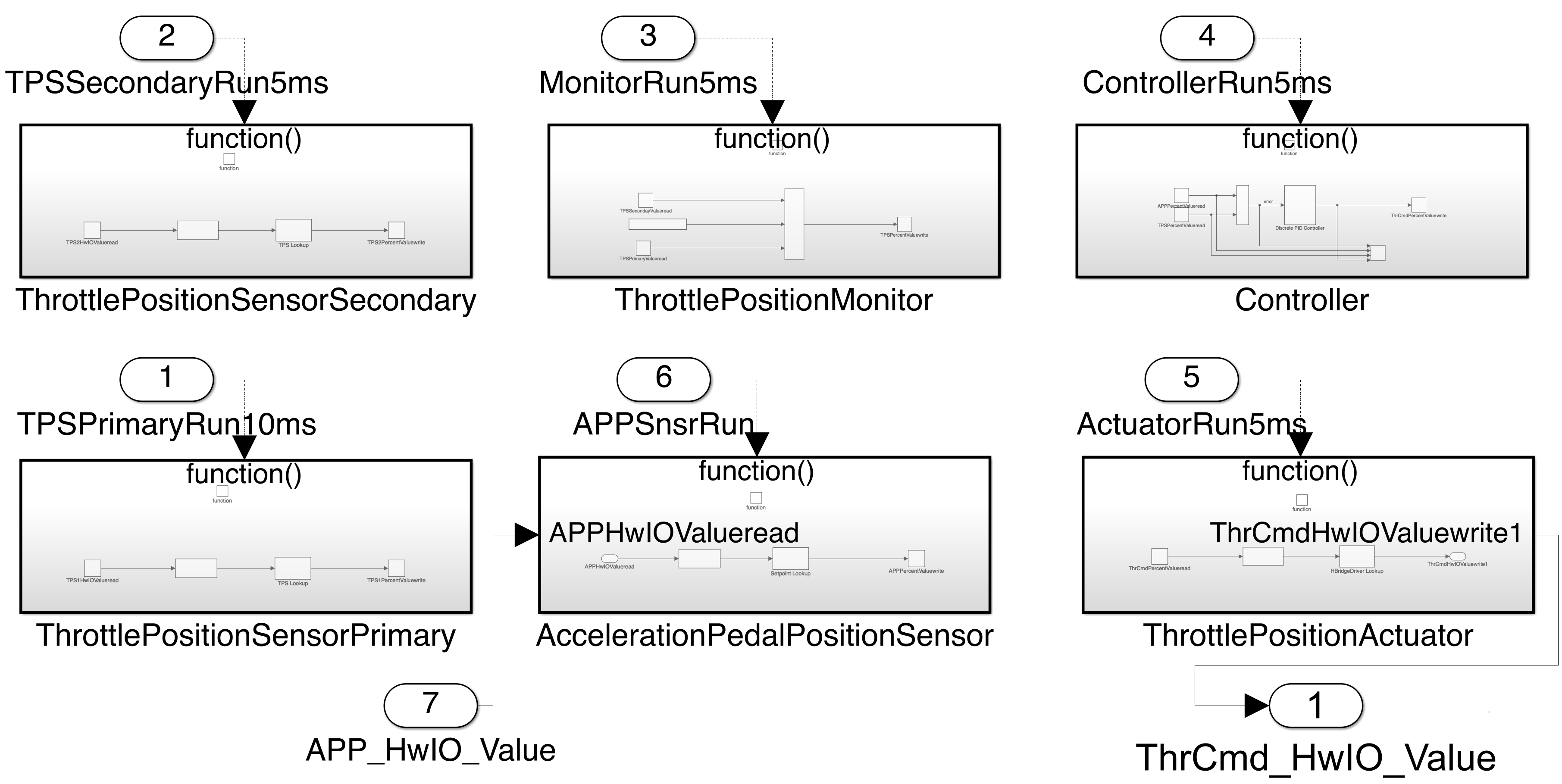} 
    	\fcaption{ Throttle position control Simulink model contains six runnables.}
    	\label{fig:casestudy}
    \end{Figure}

\begin{Table} 
	\tcap{Throttle control example settings.}
	\begin{tabular}{ c c c c c}
		\hline
		Task & Period & Execution & Priority & Runnable \\
		&   ($ms$)& Time($ms$) &          &                  \\ \hline
		$T_2$   & 10 & 1 & 1 & TPSPrimary  \\
		$T_1$   & 5 & 1 & 2 &  TPSSecondary \\
		$T_1$   & 5 & 1 & 2 & Monitor  \\
		$T_1$   & 5 & 1 & 2 &  Controller \\
		$T_1$   & 5 & 1 & 2 &  Actuator \\
		$T_2$   & 10 & 1 & 1 & APPSnsr \\
		\hline
	\end{tabular}
	\label{tbl:throttleparameter}
    \end{Table}

 Figure \ref{fig:tpc_original} shows the simulation result, which is generated by a Stateflow scheduler. The square wave in the figure is the simulated pedal input, and the curly wave is the output of the throttle body, representing the current throttle position. The Stateflow scheduler simulates the throttle control controller's process well and simulates the entire control process.

    \begin{Figure}
        \includegraphics[width=.99\columnwidth]{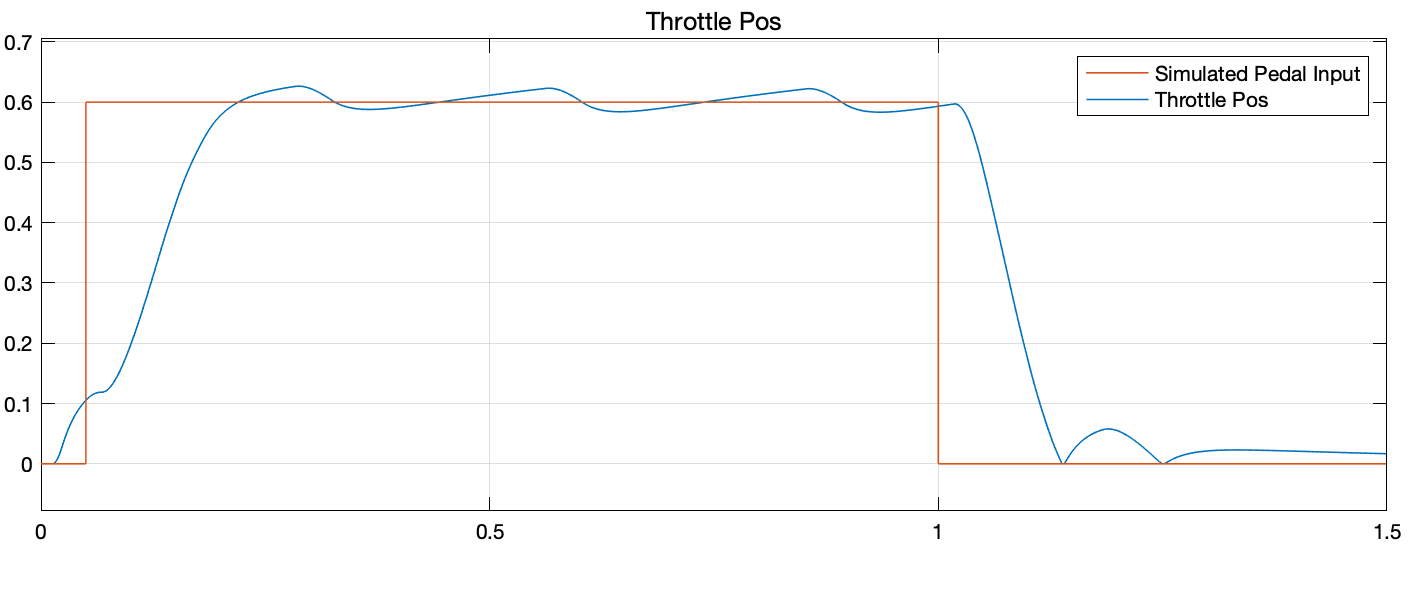}
    	\fcaption{The simulated throttle position of the throttle position control model scheduled by the Stateflow scheduler.}
    	\label{fig:tpc_original}
    \end{Figure}

 Figure \ref{fig:sf_tpc_r} shows the runnable active chart scheduled by the Stateflow scheduler. All runnables are scheduled and executed at time 0. Runnable \emph{TPSPrimary} and \emph{APPSensor} are mapped to $T_2$ and they are scheduled and executed every 10$ms$ and the top and bottom chars shown in the figure are their active charts. The active charts of runnable \emph{TPPSSendary}, \emph{Monitor}, \emph{Controller}, and \emph{Actuator} are the four charts in the middle of the figure. They are scheduled and executed every 5$ms$.
 
    \begin{Figure}
        \includegraphics[width=.99\columnwidth]{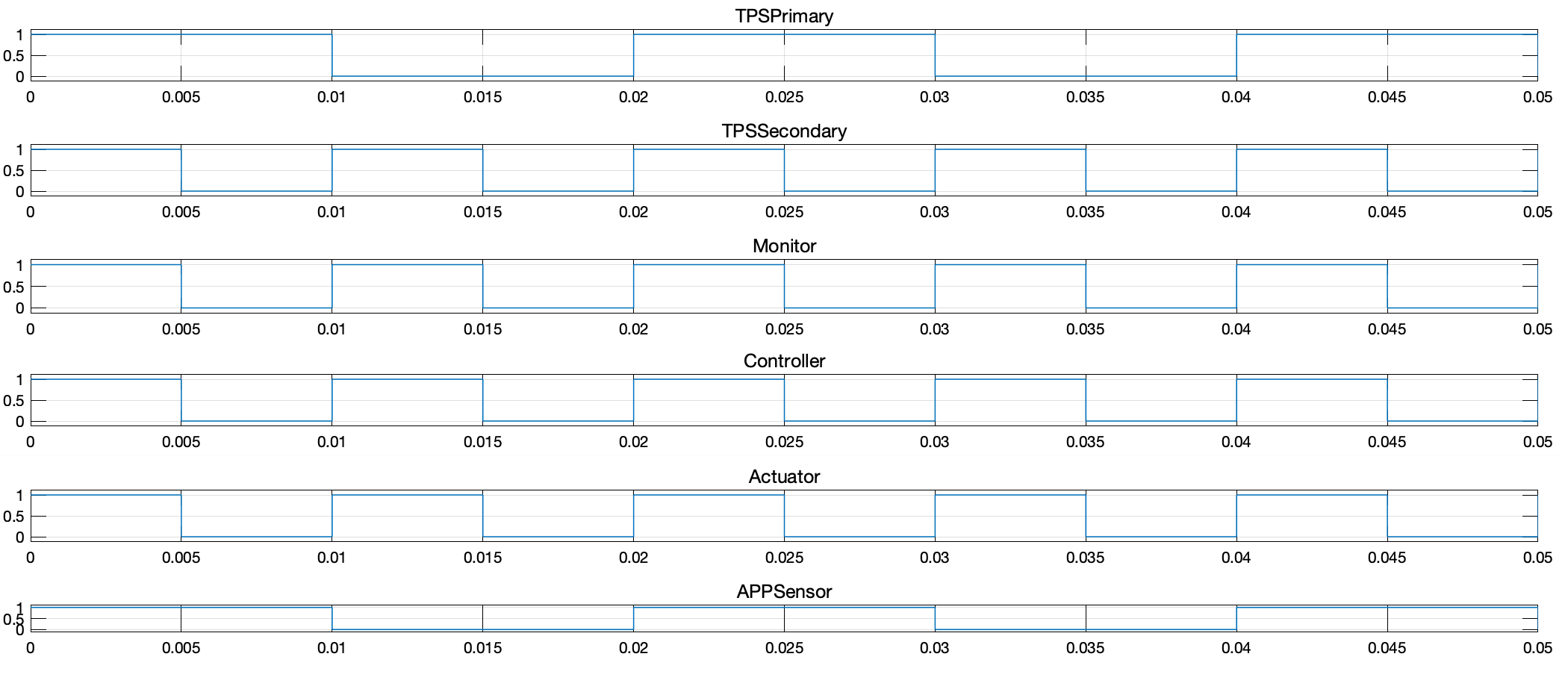}
    	\fcaption{The runnable active chart of the throttle position control model scheduled by the Stateflow scheduler.}
    	\label{fig:sf_tpc_r}
    \end{Figure}

    We apply SimSched to the Stateflow scheduler model, and we can get a similar simulation result as the Stateflow scheduler example based on the above task settings. Figure \ref{fig:ss_tpc_t} shows the task active chart generated by SimSched. Every task has been executed within its own deadline. Both task $T_1$ and $T_2$ are scheduled at time 0 but only $T_1$ is executed at time 0 due to its higher priority. $T_1$ takes up 4$ms$ to run. After the first instance of $T_1$ is finished, $T_2$ is executed. The second instance of $T_1$ arrives at 5$ms$, which is during the middle of the execution of $T_2$. $T_2$ is preempted by $T_1$ and resumes at the completion of the second instance of $T_1$.

    \begin{Figure}
        \includegraphics[width=.99\columnwidth]{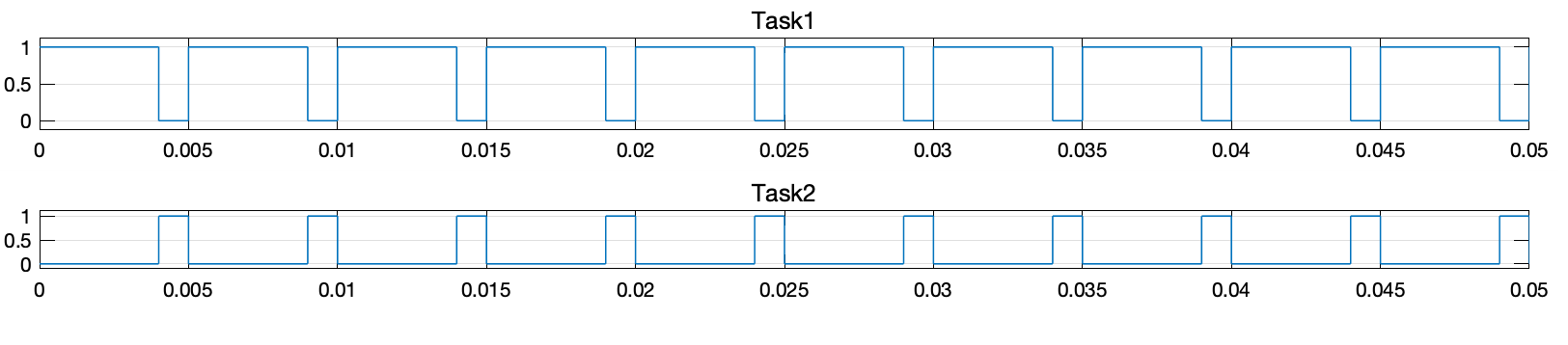}
    	\fcaption{The task level active chart of the throttle position control model scheduled by the SimSched scheduler.}
    	\label{fig:ss_tpc_t}
    \end{Figure}
    
    Figure \ref{fig:ss_tpc_r} shows the runnable level active chart of the throttle position control model scheduled by the SimSched scheduler. From this figure, we can clearly see the activity of each runnable. It exactly shows the execution order of each runnable.  Runnable \emph{TPPSSendary}, \emph{Monitor}, \emph{Controller}, and \emph{Actuator} are executed one after another followed by \emph{TPSPrimary}.  Runnable \emph{APPSensor} is executed after the second instance of $T_1$ and it is the preemption point of $T_2$. 
    
    \begin{Figure}
        \includegraphics[width=.99\columnwidth]{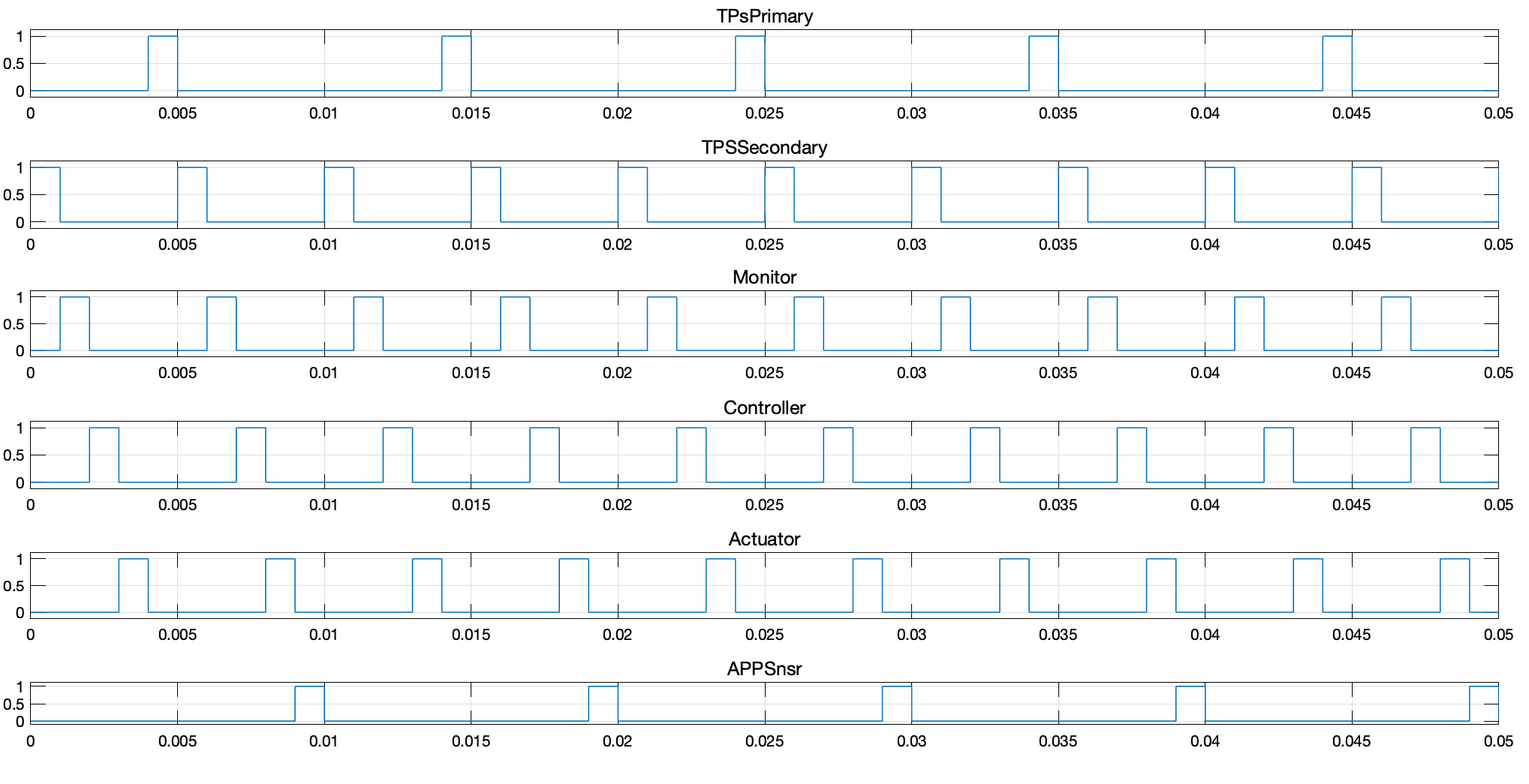}
    	\fcaption{The runnable active chart of the throttle position control model scheduled by the SimSched scheduler.}
    	\label{fig:ss_tpc_r}
    \end{Figure}

    We use the same means applied to three servos example to apply it to the throttle position control model. We adopt the Stateflow schedule and replace it with SimSched to generate mutants for both the Stateflow scheduler and SimSched for the experiments.
    
    Figure \ref{fig:ss_tpc_offset_t1_1} shows the runnable active chart of the Throttle Position Control model scheduled by SimSched after applying $mTIO$ mutation operator as increasing 2$ms$ offset for $T_1$. The runnable Actuator active chart shown in the second bottom chart in the figure is missing its first execution. $T_1$ takes 4$ms$ to run, and it also has 2$ms$ offset, so the total execution time of $T_1$ exceeds its period 5$ms$. On the other hand, the mutant generated by the Stateflow scheduler can not simulate this overrun situation due to the lack of execution time simulation support. 
    
    \begin{Figure}
        \includegraphics[width=.99\columnwidth]{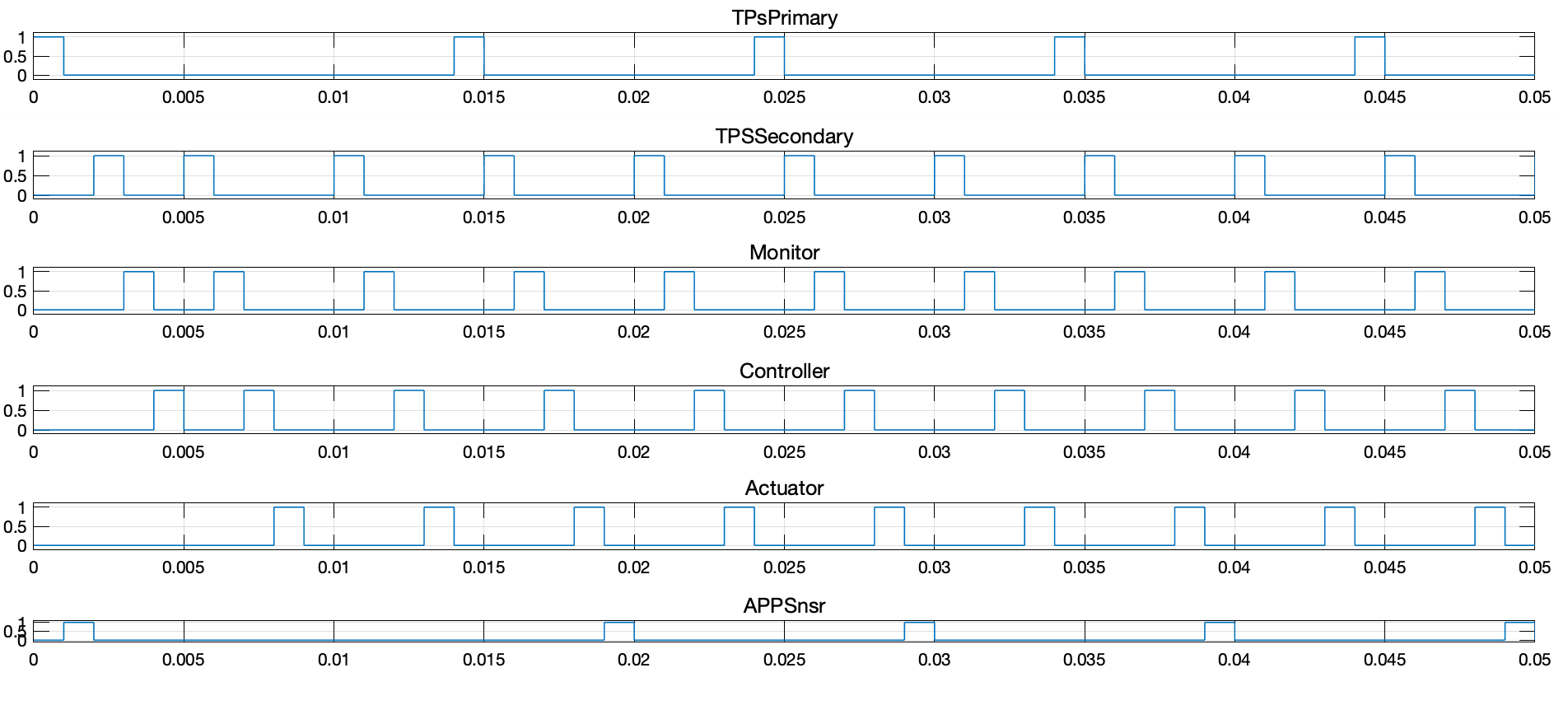} 
    	\fcaption{The runnable active chart of Throttle Position Control model scheduled by SimSched after applying $mTIO$ mutation operator as increasing 2$ms$ offset for $T_1$.}
    	\label{fig:ss_tpc_offset_t1_1}
    \end{Figure}
    
    Figure \ref{fig:ss_tpc_itper_t1} shows the simulated throttle position of the throttle position control model scheduled by SimSched after applying \emph{mITPER} mutation operator for $T_1$ at 100$ms$. The Stateflow scheduler also can output the same figure. The $mITPER$ mutation operator can reduce the computation times of a task at the same amount of time, which results in unstable control as shown in the figure. 
    \begin{Figure}
        \includegraphics[width=.99\columnwidth]{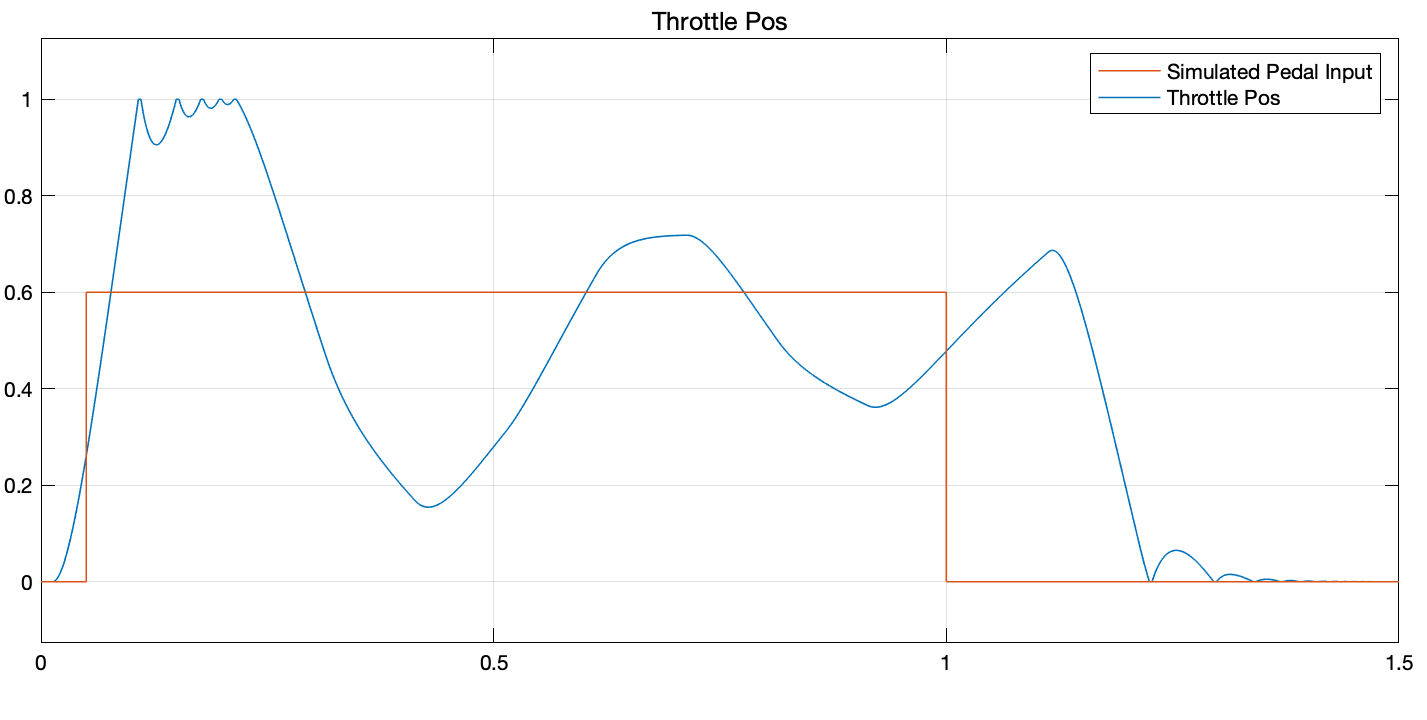} 
    	\fcaption{The simulated throttle position of the throttle position control model scheduled by the Stateflow scheduler after applying \emph{mITPER} mutation operator.}
    	\label{fig:ss_tpc_itper_t1}
    \end{Figure}
    
    Figure \ref{fig:ss_tpc_dtper_t1} shows the simulated throttle position of the throttle position control model scheduled by SimSched after applying \emph{mDTPER} mutation operator for $T_1$ at 4$ms$. The $mDTPER$ can result in no output signal for $T_2$ because a higher rate task increases the computation times, and the lower rate task does not get an execution. In this example, we set $\rho_1=4ms$ using the $mDTPER$ mutation operator, then the $T_2$ is always preempted by $T_1$ and does not have a chance to execute. 
    
    \begin{Figure}
        \includegraphics[width=.99\columnwidth]{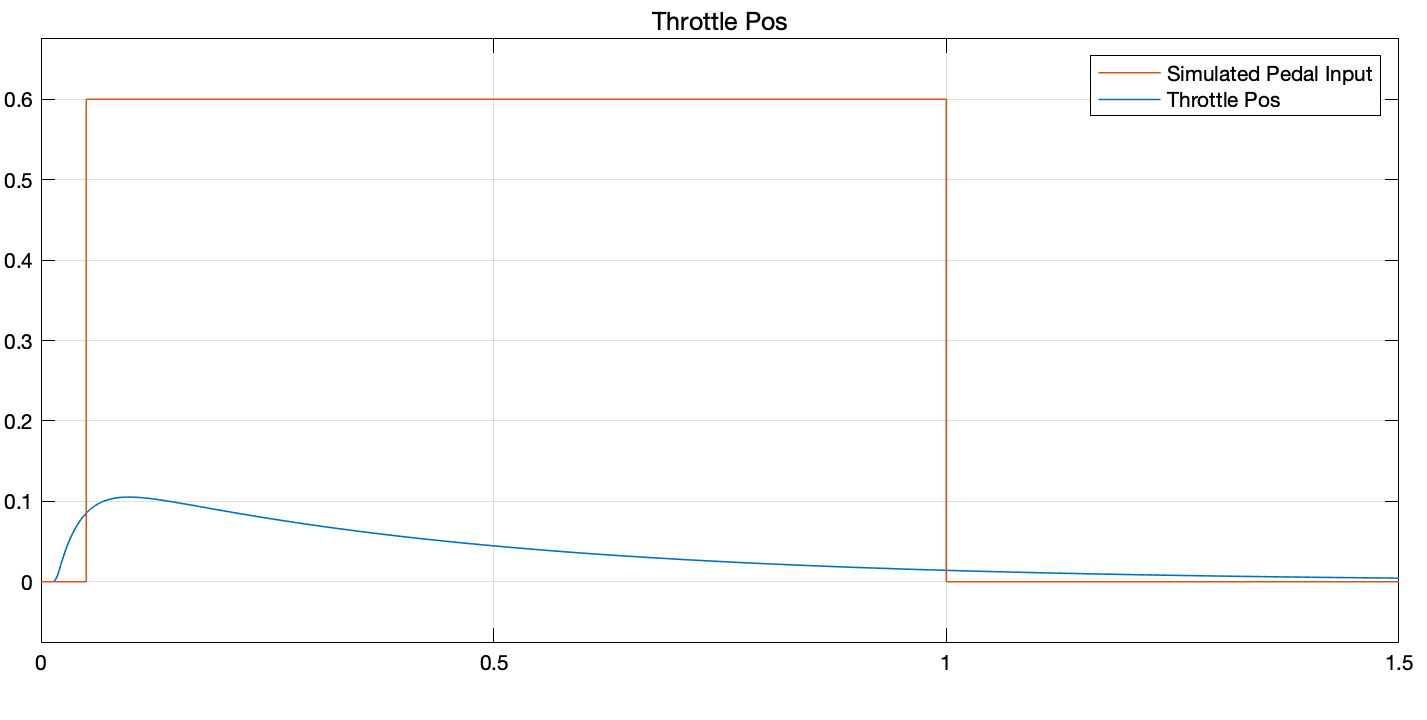} 
    	\fcaption{The simulated throttle position of the throttle position control model scheduled by SimSched after applying \emph{mDTPER} mutation operator.}
    	\label{fig:ss_tpc_dtper_t1}
    \end{Figure}
    
    We only apply the execution time operator to SimSched models. We set $c_1=5ms$ using $mITET$ mutation operator for $T_1$ to generate a mutant. This mutant just outputs the same throttle position as shown in Figure \ref{fig:ss_tpc_dtper_t1}. Since $T_1$ has increased its execution time by 1$ms$, it just takes up all the time slots in its period. $T_2$ is preempted by $T_1$ during the simulation process.

   We apply $mARPREC$ and $mRRPREC$ mutation operators to this throttle position control example. First, there is no precedence between runnable $APPSnsr$ and $TPSPrimary$, and we use a $mARPREC$ mutation operator to add precedence $\gamma_{APPSnsr}$ to $precr_{TPSPrimary}$ to generate mutants for both schedulers. Runnable $Controller$ consumes the values produced by $APPSnsr$ and $TPSPrimary$ to calculate the throttle percent value for the throttle actuator. The changes in the simulation results of both mutants are trivial.   Figure \ref{fig:casestudyprec} shows the simulation result comparison between the original model and the SimSched mutant. 

    Second, there is precedence between $Controller$ and $Actuator$, and $Controller$ is executed before $Actuator$. We remove the precedence from the pair of runnables, so the $Actuator$ (destination) runs before $Controller$ (source), which changes the data dependency and delays the data. The difference in simulation is similar to Figure \ref{fig:casestudyprec}.
    \begin{Figure}
        \includegraphics[width=.99\columnwidth]{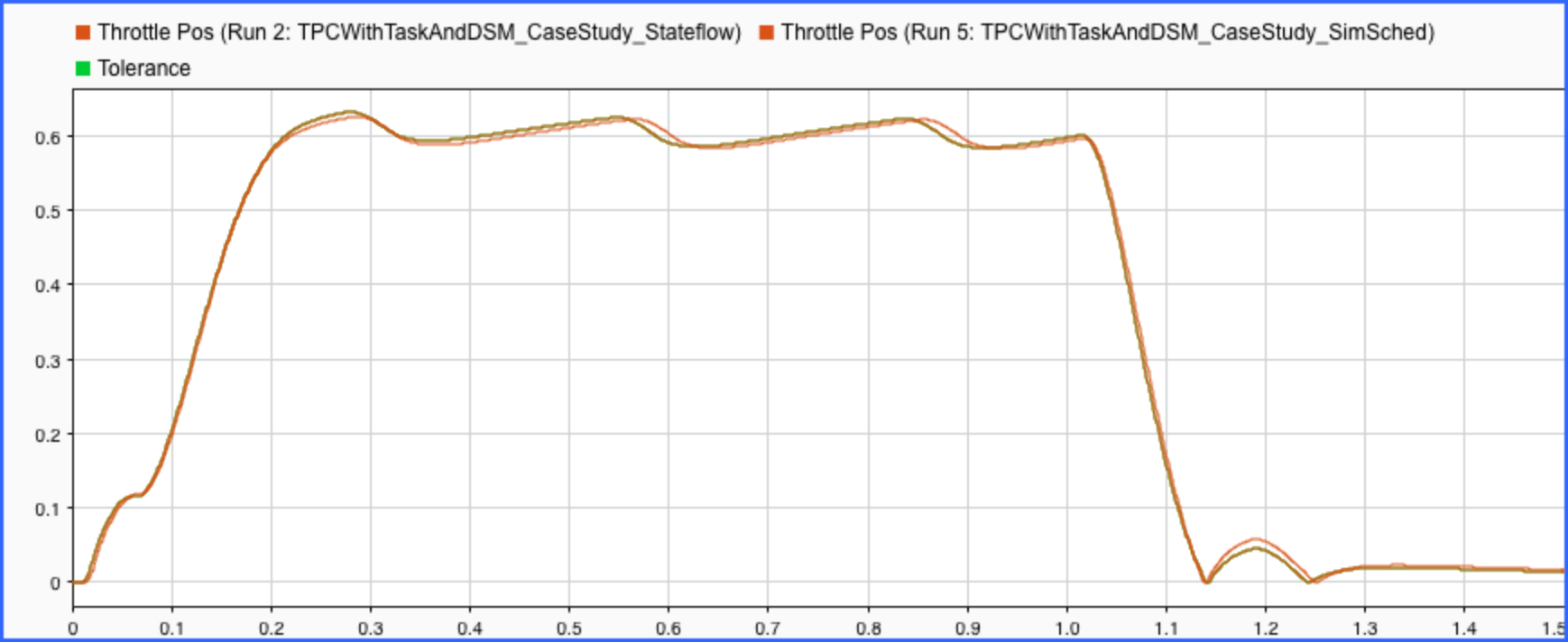}
    	\fcaption{The difference of simulation result between the original model and the mutant with $mARPREC$ mutation operator scheduled by SimSichde.}
    	\label{fig:casestudyprec}
    \end{Figure}

    We apply $mDSM$, $mUDSM$, $mRDSM$, $mRSM$, $mRMSMR$, and $mRSMR$ mutation operators to this example and generate accordingly mutants for both the Stateflow scheduler and SimSched. Interestingly, since the task time properties have not changed for the shared memory mutation operators, the simulation results are consistent between the two types of mutants. For example, we apply the \emph{mDSM} mutation operator to variable $ThrCmdPercentValus$ and set the variable to a new constant value. The simulation result of both types of mutants is the same shown in Figure \ref{fig:mdsm_E}. Since this variable is an input to a Lookup table, there is always an output that matches the input value and yields a corresponding value to the output. The input value is constant, so the throttle position's output is a smooth curve.

    \begin{Figure}
        \includegraphics[width=.99\columnwidth]{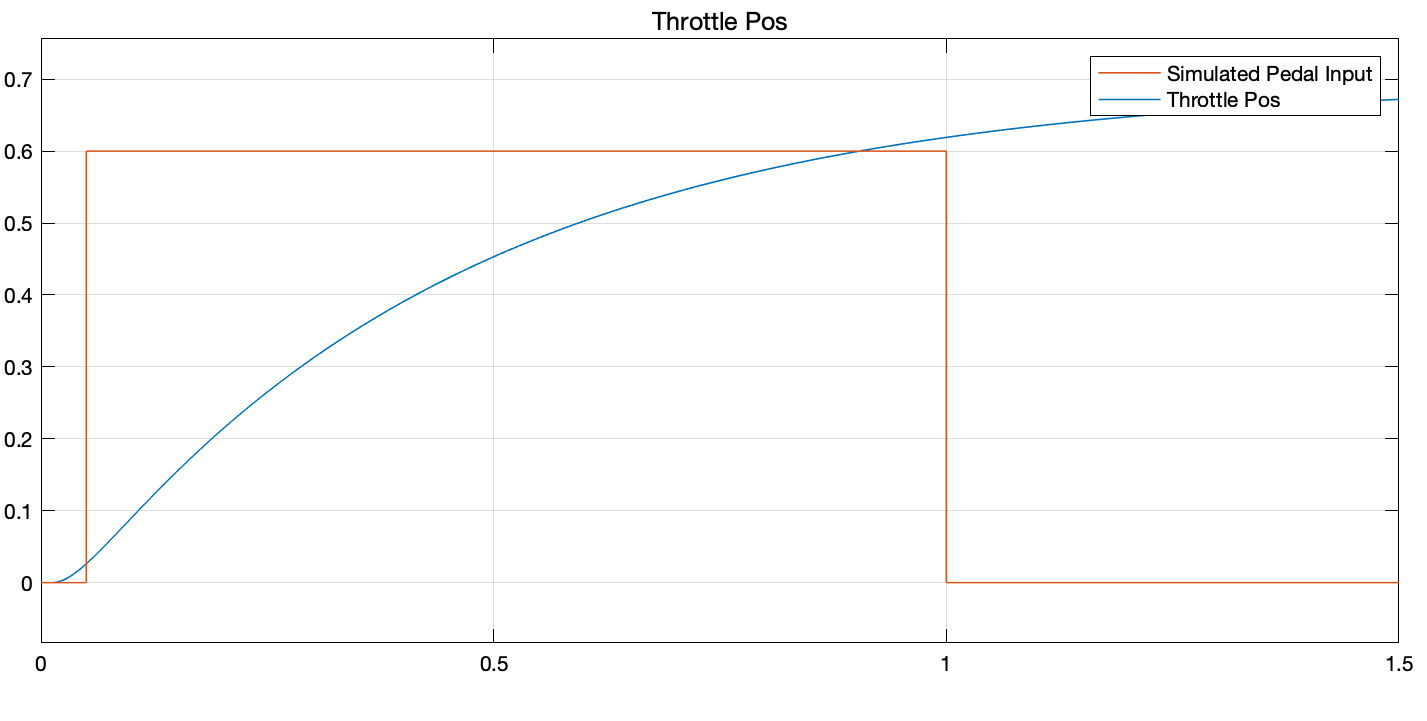}
    	\fcaption{The throttle position output after applying \emph{mDSM} mutation operator to variable $ThrCmdPercentValus$ scheduled. }
    	\label{fig:mdsm_E}
    \end{Figure}

\subsection{Evaluation Result}
    We apply the evaluation process to the example models to investigate the mutation operator's effectiveness to kill mutants. Table \ref{tbl:mutationanalysis_threeservo} and \ref{tbl:mutationanalysis_tpc} summarize the results of the efficacy of mutation operators, where each row provides the number of mutants and the mutation score of our mutation operators. The mutation score is a measure that gives the percentage of killed mutants with the total number of mutations.

    \begin{Table}
	\tcap{Mutation analysis of mutation operators for three servos example. }
	\begin{tabular}{|c|c|c|c|c|}
    \hline
    \multicolumn{1}{|l|}{}                & \multicolumn{2}{c|}{Stateflow Scheduler} & \multicolumn{2}{c|}{SimSched} \\ \hline
    Operator                              & Mutants              & Kills             & Mutants        & Kills        \\ \hline
    Offset                                & 36                   & 24                & 36             & 27           \\ \hline
    Period                                & 39                   & 25                & 39             & 25            \\ \hline
    Execution                         & N/A                  & N/A               & 36             & 31           \\ 
    Time                        &                &              &             &          \\ \hline
    Precedence                            & 5                    & 0                 & 5              & 0            \\ \hline
    Priority                              & 18                   & 0                 & 18             & 0            \\ \hline
    Jitter                                & 36                   & 24                & 36             & 27           \\ \hline
    Mutation                         & \multicolumn{2}{c|}{54.48\%}             & \multicolumn{2}{c|}{64.71\%}  \\ 
    Score                        & \multicolumn{2}{c|}{}             & \multicolumn{2}{c|}{}  \\ \hline
    \end{tabular}
	\label{tbl:mutationanalysis_threeservo}
    \end{Table}

For the three servo example, we generate 134 mutants for the Stateflow scheduler models and 170 mutants for the SimSched models. We achieve a mutation score of 54.48\% for the Stateflow scheduler model and 64.71\% for the SimSched models. Evaluation results show that the time-related mutation operators have the most effect on the mutation testing, such as the \emph{Offset}, \emph{Period}, \emph{Execution Time}, and \emph{Jitter} mutation operator. We observe that the \emph{Precedence} and \emph{Priority} mutation operators kill zero mutant because, in this example, each controller individually controls a motor. There is no connection between them, so the change of precedence and priority does not cause any simulation changes. However, the three controllers run on a single CPU, and one task execution time's length affects other tasks. We also observe the \emph{Offset} mutation operator only affects each task's initial execution, and tasks miss the deadline. Still, each mutant's simulation results show each controller can have stable control of each servo.

    \begin{Table}
    	\tcap{Mutation analysis of mutation operators for throttle position control example. }
	    \centering
        \begin{tabular}{|c|c|c|c|c|}
            \hline
            \multicolumn{1}{|l|}{}              & \multicolumn{2}{c|}{Stateflow Scheduler} & \multicolumn{2}{c|}{SimSched} \\ \hline
            Operator                            & Mutants              & Kills             & Mutants        & Kills        \\ \hline
            Offset                              & 19                   & 7                 & 19             & 15           \\ \hline
            Period                              & 30                   & 6                 & 30             & 19           \\ \hline
            Execution                       & N/A                  & N/A               & 34             & 34           \\ 
            Time                      &                   &               &              &            \\ \hline
            Precedence                          & 23                   & 10                & 23             & 10           \\ \hline
            Priority                            & 11                   & 0                 & 11             & 0            \\ \hline
            Jitter                              & 19                   & 12                & 19             & 15           \\ \hline
            Shared                        & 72                   & 46                & 72             & 46           \\
            Memory                       &                    &                 &              &            \\ \hline
            Mutation                       & \multicolumn{2}{c|}{49.39\%}             & \multicolumn{2}{c|}{70.2\%}  \\ 
            Score                      & \multicolumn{2}{c|}{}             & \multicolumn{2}{c|}{}  \\ \hline
        \end{tabular}
        \label{tbl:mutationanalysis_tpc}
    \end{Table}
    
    For the throttle position control example, we generate 164 mutants for the Stateflow scheduler models and 198 mutants for the SimSched models. We achieve a mutation score of 49.39\% for the Stateflow scheduler model and 70.2\% for the SimSched models. Evaluation results are similar to the previous example. The time-related mutation operators have the most effective for mutation testing.  We observe that the \emph{Shared Memory} mutation operators have the same kills for both $\mathcal{M_\mu}$ and $\mathcal{M'_\mu}$.  In this example, task $T_2$ has two Runnable \emph{TPSPrimary} and \emph{APPSnsr} each has a shared variable to update at each execution, and there is no direct relation between them. Though SimSched can simulate the preemption of $T_2$ to interrupt its execution, shared memory mutants' model behaviors are the same for both $\mathcal{M_\mu}$ and $\mathcal{M'_\mu}$. 
    
   From the above two examples, we can see the mutation operators are application-dependent, and SimSched can achieve a higher mutation score at the time-related mutation operators. For example, the \emph{Precedence} mutation operator kills zero mutants for the three servo example, but it kills ten mutants for the throttle position control example for both Stateflow Scheduler and SimSched. The three-servo example does not require any precedence at all; each task only controls itself. However, the throttle position control example requires precedence. A runnable consumes data from a previous runnable execution. If we alter the precedence,  the execution order is different from the original model execution; it produces a  different data flow.  For example, in the \emph{Period} mutation operator both Stateflow scheduler and SimSched kill the same mutants for the three serve example, but SimSched kills more mutants than the Stateflow Scheduler in the throttle position control example. Because the throttle position control example has four runnables in $T_1$ and each runnable has $1ms$ execution time. We use a $mDTPER$ to $T_1$ and generate a mutant that the period of $T_1$ is $4ms$ instead of $5ms$, then $T_1$ will occupy all the execution time slots, and $T_2$  will not be executed for the SimSched. However, the Stateflow Scheduler does not consider the execution time, so both $T_1$ and $T_2$ are executed as scheduled.   
	
	\section{Related Work}
	Our work aligns with the MBMT, and it has been applied to various models. Trakhtenbrot \cite{Trakhtenbrot2010}  proposes mutation testing on statechart-based models for reactive systems. This approach mainly deals with the conformance between specific semantics of statechart models and the model's implementation. Mutation testing has been applied to feature models to test software product lines \cite{Henard2013}. The feature models represent the variability of software product lines and configurable systems. El-Fakih \etal \cite{El-Fakih2008} develop a technique of mutation-based test case generation toward extended finite state machines (EFSM) that examines the EFSM under test agrees to user-defined faults. Belli \etal \cite{Belli2016} have surveyed MBMT approach a great deal and detailed the approach applied to graph-based models, including directed graphs, event sequence graphs, finite-state machines, and statecharts. A recent mutation testing survey\cite{papadakis2019mutation} presents up-to-date advanced approaches that use mutants to support software engineering activities to model artifacts.
	
	Our work also fits in the timed system testing, which requires a real-time environment. Researchers \cite{Aichernig2013,Nilsson2004,Nilsson2007} have utilized the most studied formalisms TA to inject faults to the timed system and reveal that the time-related errors are unable to find by using randomly generated test suites. Nilsson \etal \cite{Nilsson2004} first proposed a set of extended TA mutation operators based on TA with Tasks (TAT) \cite{Norstrom1999} to test real-time systems which depend on the execution time and execution order of individual tasks. The mutation operators are interested in the \emph{timeliness} of a task to meet its deadlines. Aichernig \etal \cite{Aichernig2013} propose a mutation testing framework for timed systems, where they define eight mutation operators to mutate the model and its mutants are expressed as a variant of TA in Uppaal specification format. The authors also develop a mutation-based test case generation framework for real-time, where they use symbolic bounded model checking techniques and incremental solving. Cornaglia \etal \cite{8727843} presents an automated framework MODELTime that facilitates the study of target platform-dependent timing during the model-based development of embedded applications using MATLAB/Simulink simulations.
	
	ML/SL is one of the most popular formalisms to model and simulate embedded systems, and many researchers have explored the various type of mutation operators applied to ML/SL models. Hanh \etal \cite{Hanh2012} propose a set of mutation operators based on investigating common faults in ML/SL models to validate test suites and present a process of mutation testing for ML/SL models. They provide twelve mutation operators and divide them into five categories. Stephan \etal \cite{Stephan2014a} utilize the mutation testing technique to compare model-clone detection tools for ML/SL models. They present a taxonomy of ML/SL and prose a set of structural mutation operators based on three clone types. The mutation operators are used to evaluate the model-clone detectors. Using the mutation-based technique to generate test cases for ML/SL models automatically has been studied \cite{Brillout2010,5981757}. They can effectively generate small sets of test cases that achieve high coverage on a collection of Simulink models from the automotive domain. A recent work SLEMI \cite{9283988} has applied mutation techniques to the Simulink compiler and uses tools to generate mutants of the seed model and found 9 confirmed bugs in Simulink models.
	
	Our work intends to exploit mutation analysis to identify potential time-related errors in ML/SL models. Roy and Cordy \cite{Roy2009a,Roy2009b} first propose using mutation analysis to assist the evaluation of software clone detection tools. They develop a framework for testing code-clone detectors based on mutation. Stephan \etal \cite{Stephan2013,Stephan2014b} proposed a framework that can objectively and quantitatively evaluate and compare model-clone detectors using mutation analysis. Their work is based on a structural mutation method for ML/SL model mutation. Our mutation operators are based on a timed system task model, whereas, there are no relevant existing studies that directly integrated the ML/SL models in the timed systems in the MIL phase; thus, we carry out the work presented in this paper.
	
    Co-simulation\cite{gomes2018co} is a widely used technique in model-based testing to verify as much of the system functionality, among subsystems, as possible. Composing the simulations of sub-simulators can achieve a joint simulation of a coupled system. Many different languages and tools are used for other purposes in the model-based design domain, either designing continuous plants or discrete controllers.  A relatively recent open standard functional mock-up interface (FMI) is developed for exchange simulation models in a  standardized format, including support for co-simulation. Gomes \etal \cite{gomes2020application} propose an approach to facilitate the implementation of the Functional Mock-up Interface standard. They use the MBT methodology to evaluate the tools that export Functional Mock-up Units (FMUs). Hence, they can root out the ambiguities and improve conformance to the FMI standard.  Garro  \etal \cite{bouskela2021formal} employs FMI to perform co-simulation to verify the system requirements based on the FOrmal Requirements Modeling Language and the Modelica language. Zafar \etal  \cite{zafar2021towards} present a systematic tool-supported MBT workflow to facilitate the simulation-based testing process of an embedded system. The workflow expends from the requirements phase,  and generation of executable test scripts, to the execution of generated test scripts on simulation levels.
	
	\section{Conclusion and future work}
	In this paper, we proposed a set of timed mutation operators for the ML/SL model that is primarily intended to integrate the timed task model in the ML/SL model to better support MIL simulation using mutation analysis. Moreover, testing at an earlier stage during the development process reduces development costs since earlier changes and fixing errors are much more manageable. We introduce a timed task model and present a set of mutation operators for the ML/SL based on this task model. We implement a mutation analysis framework that can apply mutation operators to the simple ML/SL models. We demonstrate the approach on several ML/SL models. The results validate that mutation analysis can reveal time-related faults. We intend to automate the mutation testing process for the ML/SL environment and improve the mutation operators to expose defects in the future. We will further validate our mutation analysis method to more industrial complex ML/SL model sets.

\section*{Acknowledgments}This work was supported in part by the Natural Sciences and Engineering Research Council of Canada (NSERC), as part of the NECSIS Automotive Partnership with General Motors, IBM Canada, and Malina Software Corp.

\bibliographystyle{plain}
\bibliography{paper}
\end{multicols}
\end{document}